\documentclass[aps,prd,preprint,floats,epsf,nofootinbib]{revtex4}
\pdfoutput=1
\usepackage{textcomp}
\usepackage{amssymb}
\usepackage{anysize}
\usepackage{bold-extra}
\usepackage{enumerate}
\usepackage{graphicx}
\usepackage{lscape}
\usepackage{mathrsfs}
\usepackage{multirow}
\usepackage{hhline}
\usepackage{makecell}
\usepackage[final]{pdfpages}
\usepackage{rotating}
\usepackage{subfig}
\usepackage{url}
\usepackage{wrapfig}
\usepackage{amsmath}
\usepackage{color}
\usepackage{fancyhdr}
\usepackage{verbatim}
\usepackage{latexsym}
\usepackage{epstopdf}
\usepackage[section]{placeins}
\usepackage[normalem]{ulem} 
\usepackage[colorlinks=true,
            linkcolor=red,
            urlcolor=blue,
            citecolor=blue]{hyperref}
\def\nn{\nonumber}
\def\bea{\begin{eqnarray}}
\def\eea{\end{eqnarray}}
\def\ba{\begin{eqnarray}}
\def\ea{\end{eqnarray}}
\def\be{\begin{equation}}
\def\ee{\end{equation}}
\def\beq{\begin{equation}}
\def\eeq{\end{equation}}

\newcommand{\gev}{~\text{GeV}}
\newcommand{\tev}{~\text{TeV}}
\newcommand{\fb}{~\text{fb}}
\newcommand{\pb}{~\text{pb}}

\newcommand{\abi}{~\text{ab}^{-1}}
\newcommand{\fbi}{~\text{fb}^{-1}}

\begin{document}
\preprint{ACFI-T19-11}

\title{Triply charged Higgs bosons at a 100~TeV $pp$ collider}
\author{Junxing Pan$^{1}$\footnote{panjunxing2007@163.com}}
\author{Jung-Hsin Chen$^{2}$\footnote{lewis02030405@gmail.com}}
\author{Xiao-Gang He$^{1,2,3,4}$\footnote{hexg@phys.ntu.edu.tw}}
\author{Gang Li$^{2,5}$\footnote{ligang@umass.edu}}
\author{Jhih-Ying Su$^{2}$\footnote{b02202013@ntu.edu.tw}}

\affiliation{$^{1}$School of Physics and Information Engineering, Shanxi Normal University, Linfen 041004, China}
\affiliation{$^{2}$Department of Physics, National Taiwan University, Taipei 10617, Taiwan}
\affiliation{$^{3}$Tsung-Dao Lee Institute, and School of Physics and Astronomy, Shanghai Jiao Tong University, Shanghai 200240, China}
\affiliation{$^{4}$Physics Division, National Center for Theoretical Sciences, Hsinchu 30013, Taiwan}
\affiliation{$^{5}$Amherst Center for Fundamental Interactions, Department of Physics,
University of Massachusetts Amherst, MA 01003, USA}

\date{\today}

\begin{abstract}
The neutral Higgs boson predicted from spontaneous breaking of electroweak symmetry in the standard model (SM) has been discovered. Precision test of the Higgs boson properties is one of the promising ways to study new physics beyond SM. Some of the most important information to know is whether there are additional Higgs bosons, neutral, singly charged and even multi-charged ones. In this work, we study the potential to search for triply charged Higgs bosons in the final state with at least three same-sign leptons. A detailed collider analysis of the SM backgrounds and signals at a 100~TeV $pp$ collider is performed with $5\sigma$ discovery prospects being obtained, which are expressed as a function of the vacuum expectation value (VEV) $v_\Delta$ or the mass splitting $\Delta m$. We also revisit the sensitivity at the Large Hadron Collider (LHC) by projecting that at a 100~TeV $pp$ collider. For a comparison, two benchmark values $v_\Delta=10^{-6}~\text{GeV}$ and $5\times 10^{-3}~\text{GeV}$ are taken. We find that for the triply charged Higgs boson mass below 1~TeV $5\sigma$ discovery significance can be reached at a 100~TeV $pp$ collider with $3.3~\text{fb}^{-1}$ and $110~\text{fb}^{-1}$ of data, respectively, the sensitivity of which is much better than that at the LHC.

\end{abstract}

\maketitle
\section{Introduction}
\label{sec:intro}

The neutral Higgs boson predicted from spontaneous symmetry breaking of electroweak interaction in the standard model (SM) has been discovered. This is a great success of the SM. Is there new physics beyond the SM is one of the most asked questions after the discovery of the Higgs boson. There are many theoretical arguments which support the existence of new physics beyond the SM. However, only experimental data can answer this important question.
Precision test of the Higgs boson properties is one of the promising ways to study new physics beyond the SM. 
More directly, one can study if there are additional Higgs bosons, neutral, singly charged and even multi-charged ones in Nature. The Large Hadron Collider (LHC) has not detected any signals beyond the SM. With more data becoming available from LHC and future colliders, we will know more what lay ahead of us. Before that we should keep an open mind about different possibilities. In this work we study the possibility of discovering multi-charged Higgs bosons at a 100 TeV $pp$ collider which will extend the kinematic region beyond the LHC.

There are many highly motivated theoretical models, in which there exist new Higgs bosons beyond the neutral Higgs boson in the SM that has been discovered at the LHC. The two-Higgs-doublet~\cite{Branco:2011iw}, minimal SUSY~\cite{Martin:1997ns}, and multi-Higgs doublet~\cite{Bento:2017eti,Bento:2018fmy} models are some of the most studied models. In these models, there are not only new neutral Higgs bosons, but also singly charged Higgs boson. Electrically multi-charged Higgs bosons also exist in some well-motivated models, such as the doubly charged Higgs boson in the Type-II seesaw model~\cite{Perez:2008ha,Aoki:2011pz,Yagyu:2012qp,kang:2014jia}, and even higher (multiple) charged Higgs boson in minimal dark matter models~\cite{Cirelli:2005uq}. The discovery of any of these color signlet Higgs bosons will be evidence of new physics beyond the SM. To this end, one needs to know how  such color singlet multi-charged Higgs bosons can be produced and detected at various experimental facilities. 

Before going to some detailed discussions, let us briefly discuss the main mechanism of producing multi-charged Higgs bosons and how they can be detected. We will indicate a color singlet higher dimensional Higgs boson $H_n$ transforming under $SU(2)_L\times U(1)_Y$ as $(n, Y)$. We will take values of $Y$ such that the resulting electric charges of the Higgs bosons are zero or integers and write the component fields as $h^Q_n$ with electric charge given by $Q=m+Y_n$. Here $m$ is the third component of isospin $I_n$.
The kinetic and interaction terms of the Higgs multiplet $H_n$ and the SM Higgs doublet $H$ are given by
\begin{eqnarray}
\label{eq:kinetic}
&&\mathcal{L}_{\text{int}} = (D^\mu H)^\dagger D_\mu H + (D^\mu H_n)^\dagger D_\mu H_n\;,\nonumber\\
&&D_\mu = \partial_\mu + i g T^a W_\mu^a +i g^\prime Y B_\mu\nonumber\\
&&\hspace{0.65cm}=\partial_\mu +i {g\over \sqrt{2}}(T_+ W^+_\mu + T_- W^-_\mu) +i e (T_3+Y) A_\mu +i {g\over c_W} (T_3 - (T_3+Y) s^2_W)Z_\mu\;,
\end{eqnarray}
where $T^a$ is the $SU(2)_L$ generator for $n$-dimensional representation with the normalization $\text{Tr}(T^aT^b) = \delta^{ab}/2$.
$T_\pm$ are the raising and lowering operators, the operator $T_3h^{Q} = m h^Q_n$, $Yh^Q_n= Y_n h^Q_n$ and $e = gg^\prime/\sqrt{g^2+g^{\prime 2}}$ with $g^\prime$ and $g$ being the $SU(2)_L$ and $U(1)_Y$ gauge couplings. The abbreviations $c_W\equiv \cos\theta_W$ and $s_W\equiv \sin\theta_W$ with $\theta_W$ being the weak mixing angle are used. The gauge interactions are the main interactions responsible for production of the multi-charged Higgs bosons.

The neutral components of $H$ and $H_n$ can be decomposed as $(v_{H}+h^0+iI^0)/{\sqrt{2}}$ and $(v_{n}+h_n^0+iI_n^0)/{\sqrt{2}}$. If $H$ and $H_n$ have non-zero vacuum expectation of values (VEVs) $v_H/\sqrt{2}$ and $v_n/\sqrt{2}$, the $Z$ and $W^\pm$ will receive masses and the $SU(2)_L\times U(1)_Y$ will break down to $U(1)_{\text{em}}$.  Certain linear combinations of components in $H$ and $H_n$ will become the would-be Goldstone bosons $G_Z$ and $G_W^+$``eaten'' by the $Z$ and $W^\pm$ bosons. When discussing detecting physical Higgs bosons, these would-be Goldstone bosons should be separated and counted as longitudinal components of the $Z$ and $W^\pm$ bosons. We provide details by expanding Eq.~\eqref{eq:kinetic} in Appendix~\ref{sec:app1}.

After the would-be Goldstone bosons are removed, one can identify the physical degrees of freedom for the Higgs bosons and discuss their production. If there is only one Higgs multiplet $H_n$. 
The production of a multi-charged Higgs boson $h^{|Q|\pm}$~\footnote{$h^{|Q|\pm}$ means a multi-charged Higgs boson with the electric charge being $\pm |Q|$.} at a $pp$ collider can happen in the following fashions: (1) the Drell-Yan type through the $s$-channel exchange of a virtual $\gamma$, $Z$ or $W^\pm$ boson in $pp\to \gamma^*,\;Z^* \to h^{|Q|+}_n + h^{|Q|-}_n$ or $pp \to W^{\pm*} \to  h^{|Q|\pm}_n h^{|Q-1|\mp}_n$; and (2) two vector boson fusion type through the $pp$ collision to produce pair $\gamma \gamma$, $\gamma Z$, $Z Z$, $(\gamma, Z) W^\pm$ or $W^+ W^-$ followed by $\gamma \gamma, \gamma Z, Z Z, W^{+} W^{-} \to h^{|Q|+}h^{|Q|-}$, or
$(\gamma, Z) W^{\pm} \to h^{|Q|\pm}_n h^{|Q-1|\mp}_n$. Here the vector bosons are virtual, but the photons can be almost real at the LHC~\cite{Han:2007bk,Drees:1994zx,Babu:2016rcr}. It is found that the production cross section is dominated by the Drell-Yan process for $|Q|=1,2$~\cite{Babu:2016rcr}; while for $|Q|\geq 3$ the cross section of $\gamma\gamma$ fusion becomes more significant and even comparable to the Drell-Yan cross section~\cite{Ghosh:2017jbw}.

The multi-charged Higgs boson $h_n^{|Q|+}$ (similarly for $h_n^{|Q|-}$) produced can be detected by the decays $h_n^{|Q|+} \to h_n^{|Q-1|+} W^+\to \cdots\to h_n^{2+ }\underbrace{W^+\cdots W^+}_{|Q-2|}$, where the multi-charged Higgs boson $h_n^{q+}$ $(2\leq q \leq |Q-1|)$ and $W^+$ boson may be on shell or off shell. The decay of the doubly charged Higgs boson $h_n^{2+}$ is model dependent: it can decay into $h_n^+ W^+$ or $W^+W^+$. In the former case, $h_n^+$ can be detected by $h_n^+\to \bar{f}f^\prime$ with $f$ and $f^\prime$ denoting lighter fermions, $W^+h^0$ followed by the decay of neutral Higgs boson $h_n^0$ into SM particles, or $W^+ Z$ if there exists at least one $H_n$ representation with $n\geq 3$~\cite{Grifols:1980uq,Gunion:1989we}.  If we consider the couplings of doubly charged Higgs boson to charged leptons, as in the Type-II seesaw model~\cite{Mohapatra:1980yp,Schechter:1980gr,Cheng:1980qt,
Lazarides:1980nt,Wetterich:1981bx}, the decay channel $h_n^{2+} \to \ell^+\ell^+$ can also be utilized. The decay $h_n^{|Q|+} \to h_n^{|Q-1|+} W^+\to \cdots\to h_n^{0}\underbrace{W^+\cdots W^+}_{|Q|}$ depends on the mass splitting between the charged Higgs bosons with $\Delta Q=\pm 1$ and is independent of the VEV $v_n$. The widths of $h_n^+\to \bar{f}f^\prime$, $W^+Z$ and $h_n^{2+}\to W^+W^+$ however are proportional to $v_n^2$, while that of $h_n^{2+}\to \ell^+\ell^+$ is proportional to $1/v_n^2$. 

If there exist the SM doublet $H$ and Higgs multiplet $H_n$ simultaneously, the singly charged Higgs boson $h_n^+$ and neutral Higgs boson $h_n^0$ above are not the mass eigenstates , see Eqs.~\eqref{sing_charged}~\eqref{real_neutral}. However, since the VEV $v_n$ is much smaller than $v_H$ constrained by the $\rho$ parameter~\cite{Tanabashi:2018oca}, $h_n^+$, $h_n^0$ are almost the same as the mass eigenstates if $H_n$ is in the real representation.

There have been plenty phenomenological studies of singly, doubly and triply charged Higgs boson searches at the LHC. A review of thorough studies of singly charged Higgs boson in the two-Higgs-doublet models (2HDMs) can be found in Ref.~\cite{Akeroyd:2016ymd}. Apart from the 2HDMs, singly charged Higgs bosons in models with weak singlet charged scalar and triplet models have also been investigated, which are characterized by sizable couplings to the first two generation fermions~\cite{Cao:2017ffm,Cao:2018ywk} and tree-level coupling to $W^+Z$~\cite{Cen:2018okf}, respectively. Doubly charged Higgs bosons have been studied in the Type-II seesaw model in $h_3^{2+}\to \ell^+\ell^+, h_3^+ W^+, W^+W^+$ channels~\cite{Perez:2008ha,Aoki:2011pz,
Du:2018eaw,Han:2007bk} and in the Georgi-Machacek model~\cite{Georgi:1985nv,Chanowitz:1985ug} in the $h_3^{2+}\to h_3^+ W^+,W^+W^+$ channels~\cite{Chiang:2012dk,Chiang:2015amq}. For the triply charged Higgs boson, it has not been discussed as much as the charged Higgs bosons with smaller electric charges. To this end, we will take a specifically non-trivial model with a Higgs quadruplet $(n=4, Y_n = 3/2)$ to show how a triply charged Higgs boson can be detected in the following sections.

Triply charged Higgs boson in the model with a Higgs quadruplet was firstly investigated in Ref.~\cite{Babu:2009aq}, in which a mechanism for generating tiny neutrino masses at tree level via dimension-7 operators was proposed. The detailed phenomenology of triply charged Higgs boson in this model at the LHC were discussed in Refs.~\cite{Bambhaniya:2013yca,Ghosh:2017jbw,Ghosh:2018drw} with same-sign leptons. Thanks to the high charge of the triply charged Higgs boson, at least three same-sign charged leptons can be produced in the final state, which is distinctive from the same-sign dilepton and multiple lepton searches in various new physics or SM studies~\cite{CMS:2019nig,CMS:2019see,Aaboud:2017dmy}. At the $13\tev$ or $14\tev$ LHC, triply charged Higgs bosons in the same-sign trilepton (SS3L) signature have been investigated~\cite{Bambhaniya:2013yca,Ghosh:2017jbw,Ghosh:2018drw} with only a few benchmark values of the quadruplet VEV, which controls the decay of the triply charged Higgs boson, and an incomplete list of SM backgrounds being considered. Besides, it was shown~\cite{Ghosh:2018drw} that for the Higgs quadruplet VEV being $5\times 10^{-3}\gev$, a triply charged Higgs boson with mass above about 600~GeV cannot be discovered even at the High-Luminosity LHC with the integrated luminosity of $3\abi$~\cite{Apollinari:2017cqg}. As we will discuss in Sec.~\ref{sec:prod_decay}, the production cross section of triply charged Higgs boson increases with the collider energy, so it is natural to ask how future facilities can help to search for triply charged Higgs bosons. From experimental searches for the final states containing a pair of same-sign leptons or multiple leptons at the LHC~\cite{CMS:2019nig,CMS:2019see,Aaboud:2017dmy}, we obtain a more complete list of SM backgrounds for the final state with at least three same-sign leptons at $pp$ colliders, some of which are sizable but missed in previous studies~\cite{Bambhaniya:2013yca,Ghosh:2017jbw,Ghosh:2018drw}, see also Refs.~\cite{Mukhopadhyaya:2010qf,Agarwalla:2018xpc}. For the signals, we generate them according to their dependence on the quadruplet VEV $v_\Delta$ and the mass splitting $\Delta m$, and obtain the discovery prospects with a function of $v_\Delta$ or $\Delta m$.

When going beyond the LHC, there may be greater chance to discover multi-charged Higgs bosons. In this work we will concentrate on  the study of the discovery potential for the triply charged Higgs boson at a 100~TeV $pp$ collider. To compare with the sensitivity at the LHC, we consider two benchmark values of $v_\Delta$ and show the discovery prospects in the plane of $m_{\Delta^{\pm\pm\pm}}$ and $\Delta m$ as in Ref.~\cite{Ghosh:2018drw}. However, the discovery significance in Ref.~\cite{Ghosh:2018drw} was evaluated using $n_s/\sqrt{n_s+n_b}$, which underestimates the significance by several times, so the discovery contours in Ref.~\cite{Ghosh:2018drw} cannot be used for our comparison. After investigating the kinematic distributions at a 100~TeV $pp$ collider and the LHC, we find that it is possible to project the results at a 100~TeV $pp$ collider to the LHC without repeating the collider simulation. Similar analyses can be extended to higher multi-charged Higgs bosons following what outlined earlier. 

This paper is organized as follows: in Sec.~\ref{sec:model}, details of the model with a Higgs quadruplet and vector-like triplet leptons will be given. In Sec.~\ref{sec:constraint}, current (indirect) constraints on this model are discussed. In Sec.~\ref{sec:prod_decay}, we will discuss the production and decay of the triply charged Higgs boson. The production in three processes are included. The dependence of the total width and decay branching ratios on the mass splitting $\Delta m$ and quadruplet VEV $v_\Delta$ are illustrated. In Sec.~\ref{sec:collider}, a detailed collider analysis is performed at a 100~TeV $pp$ collider and the sensitivity at the LHC is revisited. Sec.~\ref{sec:smmary} summarizes our results. 

\section{A triply charge Higgs boson model}
\label{sec:model}
We now provide some information about the triply charged Higgs boson in a complex Higgs quadruplet $\Delta\sim (1,4,3/2)$ into the SM to be studied, which is expressed as $\Delta=(\Delta^{+++},\Delta^{++},\Delta^{+},\Delta^{0})^T$, the scalar kinetic Lagrangian is shown in Eq.~\eqref{eq:kinetic} with $\Delta =H_n$.
The covariant derivatives 
\begin{align}
D_{\mu}\Delta&=(\partial_{\mu}-igT^a W_{\mu}^a-ig^\prime Y_{\Delta}B_{\mu})\Delta,\\
D_{\mu}H&=(\partial_{\mu}-ig\tau^a W_{\mu}^a-ig^\prime Y_{H}B_{\mu})H,
\end{align}
and the Higgs doublet $H\sim (1,2,1/2)$ is given by $H=(H^+,H^0)^T$. The hypercharges $Y_\Delta=3/2$, $Y_H=1/2$, and the matrices $\tau^a$ and $T^a$ denote the $SU(2)$ generators in the doublet and quadruplet representations, respectively. 

The Higgs potential is expressed as~\cite{Babu:2009aq}
\begin{align}
V(H,\Delta)&=-\mu_H^2 H^\dagger H+\mu_\Delta^2 \Delta^\dagger \Delta +\lambda_1(H^\dagger H)^2+\lambda_2(\Delta^\dagger\Delta)^2\nn\\
&\quad +\lambda_3(H^\dagger H)(\Delta^\dagger\Delta)+\lambda_4(H^\dagger \tau^a H)(\Delta T^a \Delta)+(\lambda_5H^3\Delta^{*}+\text{H.C.})\;.
\end{align}
The last term is explicitly written as $\lambda_5 H_a H_b H_c \Delta^{*}_{abc}+\text{H.C.}$ with the totally symmetric tensors 
\begin{align}
H_1&=H^+,\quad H_2=H^0,\\
\Delta_{111}&=\Delta^{+++},\quad \Delta_{112}=\dfrac{1}{\sqrt{3}}\Delta^{++},\quad
\Delta_{122}=\dfrac{1}{\sqrt{3}}\Delta^{+},\quad \Delta_{222}=\Delta^{0}\;.
\end{align}

After the electroweak symmetry breaking, $H^0\to (v_H+h^0)/{\sqrt{2}}$ and $\Delta^0\to(v_\Delta+h_\Delta^0)/{\sqrt{2}}$. One obtains that VEVs of the fields $H$ and $\Delta$,
\begin{align}
v_H=\sqrt{\dfrac{\mu_H^2}{\lambda_1}},\quad v_{\Delta}=-\dfrac{v_H^3\lambda_5}{2m_{\Delta}^2}\;.
\end{align} 
Here $m_{\Delta}$ denotes the mass of the neutral field $h_{\Delta}^0$ of the quadruplet $\Delta$, 
\begin{align}
m_{\Delta}^2=\mu_{\Delta}^2+\dfrac{1}{8}v_H^2(4\lambda_3+3\lambda_4)\;.
\end{align}
We can see from the Higgs potential that the $\lambda_4$ term induces the mass splitting between the nearby states of the Higgs quadruplet. To be more concrete, the mass of the field\footnote{Here, $\Delta^{1+}=\Delta^{+}$, $\Delta^{2+}=\Delta^{++}$, and $\Delta^{3+}=\Delta^{+++}$.} $\Delta^{n+}$, is given by
\begin{align}
\label{eq:mass splitting}
m_{\Delta^{n+}}^2=m_{\Delta}^2-n\dfrac{\lambda_4}{4}v_H^2\;.
\end{align}

The singly charged states $H^\pm$ and $\Delta^\pm$ can mix with each other, thus it is necessary to define the normalized and orthogonal states via
\begin{align}
\begin{pmatrix}
G^\pm_W\\
\phi^\pm
\end{pmatrix}
=1/v\begin{pmatrix}
v_H& \sqrt{3}v_\Delta\\
-\sqrt{3}v_{\Delta}&v_H
\end{pmatrix}
\begin{pmatrix}
H^\pm\\
\Delta^\pm
\end{pmatrix}
\end{align}
with $v\equiv\sqrt{v_H^2+3v_\Delta^2}\simeq 246\gev$, where $G_W^\pm$ and $\phi^\pm$ are the would-be Goldstone boson and physical singly charged Higgs boson, respectively. The electroweak $\rho$ parameter is equal to $(v_H^2+3v_\Delta^2)/(v_H^2+9v_{\Delta}^2)$ in this model. After removing the Goldstone mode, one obtains interactions of the physical singly charged Higgs boson $\phi^\pm$ to SM fermions and gauge bosons. With the experimental measurement of $\rho$~\cite{Tanabashi:2018oca}, one obtains $v_{\Delta}\lesssim 1.3\gev$ at $3\sigma$ level. Since the mixing effects are highly suppressed by $v_\Delta/v$, we will not consider them but keep in mind that singly charged Higgs boson can couple to SM leptons even if there are no other fields being introduced. Similarly, neutral Higgs bosons from the doublet and quadruplet can also mix, depending on the parameter $\lambda_5$ in the Higgs potential. For $v_\Delta\ll v$, the mass eigenstates of neutral Higgs bosons are
\begin{align}
\begin{pmatrix}
h_1^0\\
h_2^0
\end{pmatrix}
=1/\sqrt{v_H^2+9v_\Delta^2}\begin{pmatrix}
v_H& 3v_\Delta\\
3v_{\Delta}&-v_H
\end{pmatrix}
\begin{pmatrix}
h^0\\
h_\Delta^0
\end{pmatrix}\;,
\end{align}
where $h_1^0$ is identified as the discovered Higgs boson with mass of about 125~GeV.

Motivated by the non-zero neutrino masses, we consider the scenario, in which a pair of vector-like triplet leptons $\Sigma_{L,R}\sim (1,3,1)$ with $\Sigma=(\Sigma^{++},\Sigma^{+},\Sigma^0)^T$ are introduced into the SM~\cite{Babu:2009aq}. This enables the quadruplet Higgs boson to couple to SM leptons after integrating out the heavy $\Sigma_{L,R}$. The Yukawa Lagrangian is described as
\begin{align}
\mathcal{L}_{\text{Yuk}}=Y_i\overline{L_{ia}^c} \epsilon^{aa^\prime}\Sigma_{La^\prime b}H^{*}_b +\overline{Y}_i\overline{\Sigma_{Rab}}\Delta_{abc}L_{ic^\prime}\epsilon^{cc^\prime}\;,
\end{align}
where $L$ is the left-handed lepton doublet, $Y_i$ and $\bar{Y}_i$ are the Yukawa couplings with $i$ being the generation index. The total symmetric tensors
\begin{align}
\Sigma_{11}=\Sigma^{++},\quad \Sigma_{12}=\dfrac{1}{\sqrt{2}}\Sigma^+,\quad \Sigma_{22}=\Sigma^0\;.
\end{align}
Integrating out $\Sigma_{L,R}$, we obtain the dimension-5 effective operator
\begin{align}
\mathcal{L}_{\text{Yuk}}^{\text{eff}}=-\dfrac{Y_i\overline{Y}_j+Y_j\overline{Y}_i}{m_\Sigma}\overline{L_{ia}^c}L_{ja^\prime}H^*_b\Delta_{bcd}
\epsilon^{ac^\prime}\epsilon^{a^\prime d}+\text{H.C.}\;,
\end{align}
where $m_\Sigma$ is the mass of $\Sigma$ fields. Assuming that the neutrino mass is generated by the above interaction, we obtain 
\begin{align}
\mathcal{L}_{\text{Yuk}}^{\text{eff}}\supset\dfrac{(m_{\nu})_{ij}}{v_\Delta}\big(
\overline{\nu_{Li}^c}\nu_{Lj}\dfrac{v_\Delta}{2}
-\overline{\nu_{Li}^c}\ell_{Lj}\dfrac{\Delta^+}{\sqrt{6}}
-\overline{\ell_{Li}^c}\nu_{Lj}\dfrac{\Delta^+}{\sqrt{6}}
+\overline{\ell_{Li}^c}\ell_{Lj}\dfrac{\Delta^{++}}{\sqrt{6}}
\big)+\text{H.C.}\;.
\end{align}
The first term gives rise to neutrino masses in the flavor basis, the second and third terms contribute to the singly charged Higgs boson decaying into leptons, and the fourth term induces the leptonic decay of the doubly charged Higgs boson as we will discussed in detail in Sec.~\ref{subsec:decay}. 

\section{Constraints}
\label{sec:constraint}
In this section, we will discuss indirect constraints on the model with an extended Higgs quadruplet proposed in Sec.~\ref{sec:model} from the decay of Higgs boson into $\gamma\gamma$, the electroweak precision tests (EWPTs), perturbativity, and low-energy rare process $\mu\to e\gamma$. It is well known that charged Higgs bosons can contribute at 1-loop level to the decay of $h_1^0\to \gamma\gamma$, which has been measured by the ATLAS and CMS Collaboration and combined in terms of signal strengths $\mu_{\gamma\gamma}^{\text{ATLAS}}=1.06\pm 0.12$~\cite{ATLAS:2019slw} and  $\mu_{\gamma\gamma}^{\text{CMS}}=1.20^{+0.17}_{-0.14}$~\cite{CMS:2018lkl} with the integrated luminosities of $80\fbi$ and $35.9\fbi$, respectively. Due to the larger integrated luminosity, we will take the combined signal strength $\mu_{\gamma\gamma}^{\text{ATLAS}}$ to constrain the model parameters.

The couplings between $h_1^0$ and charged Higgs bosons $\Delta^{n\pm}$, i.e,, $h_1^0\Delta^{n+}\Delta^{n-}$, are
\begin{align}
\tilde{\lambda}_n= v_H(\lambda_3+\dfrac{3-2n}{4}\lambda_4),\quad n=1,2,3\;.
\end{align}
From Eq.~\eqref{eq:mass splitting}, one sees that $\lambda_4$ is fixed by the mass splitting $\Delta$, which is given by
\begin{align}
\lambda_4=\dfrac{8m_\Delta}{v_H^2}\Delta m.
\end{align}
The partial width of $h_1^0\to \gamma\gamma$ is thus modified as~\cite{Djouadi:2005gj},
\begin{align}
\label{eq:h2aa}
\dfrac{\Gamma(h_1^0\to \gamma\gamma)}{\Gamma(h_1^0\to \gamma\gamma)_{\text{SM}}}=
\dfrac{|N_c Q_t^2A_{1/2}(\tau_t)+A_1(\tau_W)+\sum\limits_{n=1}^{3}\dfrac{v_H \tilde{\lambda}_n Q_n^2}{2m_{\Delta^{n\pm}}^2} A_0(\tau_{\Delta^{n\pm}})|^2}{|N_c Q_t^2A_{1/2}(\tau_t)+A_1(\tau_W)|^2}\;,
\end{align}
where $\tau_t=m_{h_1^0}^2/(4m_t^2)$, $\tau_W=m_{h_1^0}^2/(4m_W^2)$ and $\tau_{\Delta^{n\pm}}=m_{h_1^0}^2/(4m_{\Delta^{n\pm}}^2)$,
\begin{align}
A_{1/2}(\tau_i)&=2\big[\tau_i+(\tau_i-1)f(\tau_i)\big]\tau_i^{-2}\;,\\
A_{1}(\tau_i)&=-[2\tau_i^2+3\tau_i+3(2\tau_i-1)f(\tau_i)]\tau_i^{-2}\;,\\
A_{0}(\tau_i)&=-\big[\tau_i-f(\tau_i)\big]\tau_i^{-2}\;,
\end{align}
and the function $f(\tau_i)=\arcsin^2\sqrt{\tau_i}$ for $\tau_i<1$. In Eq.~\eqref{eq:h2aa}, we have neglected the terms proportional to $v_\Delta/v_H$. 

\begin{figure}[!htb] 
\centering
\includegraphics[width=0.4\textwidth]{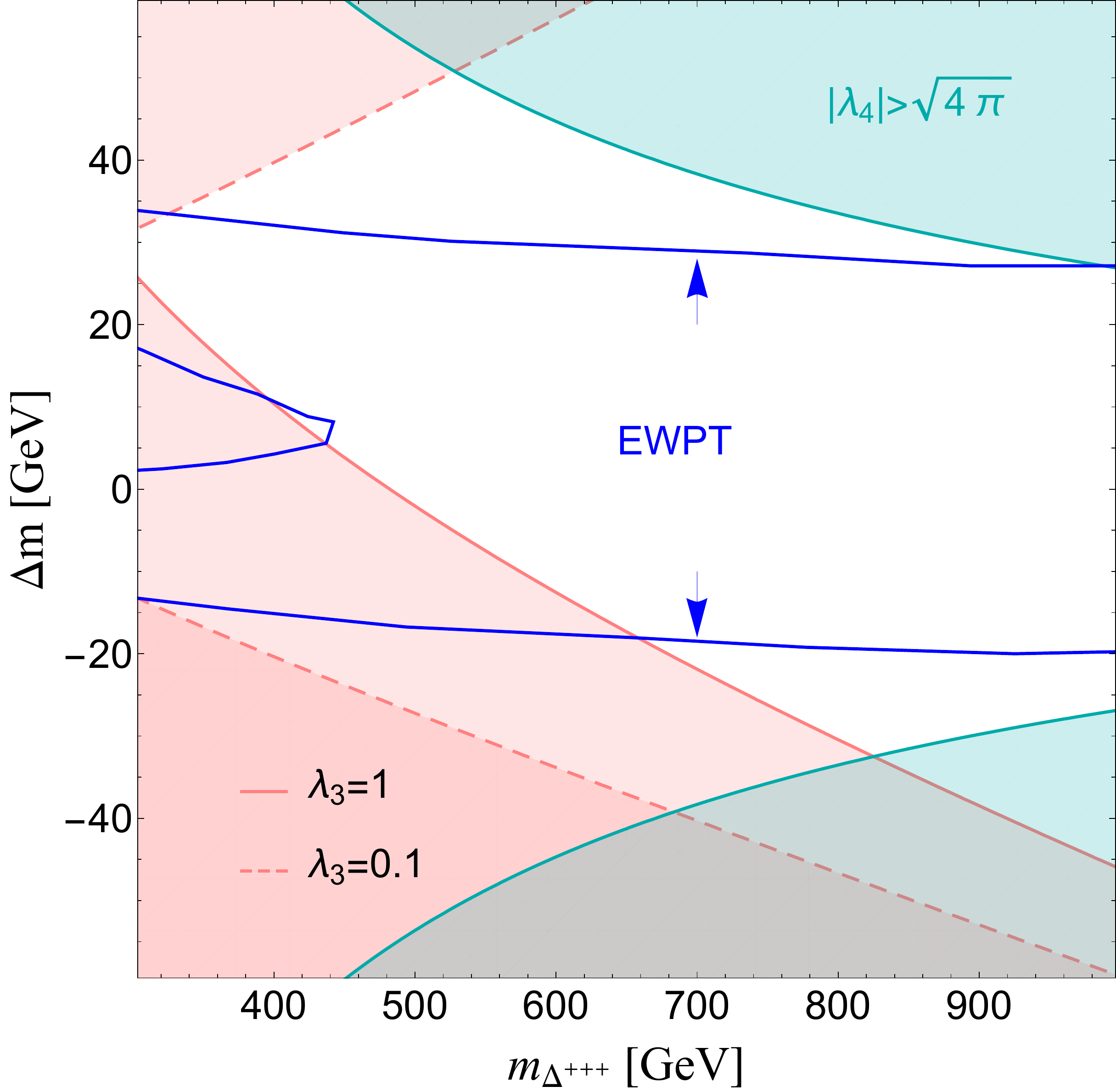}	
\caption{Indirect constraints from $\mu_{\gamma\gamma}^{\text{ATLAS}}=1.06\pm 0.12$, EWPT (taken from Ref.~\cite{Ghosh:2018drw}) and perturbativity in the plane of $m_{\Delta^{\pm\pm\pm}}$ and $\Delta m$. Pink and cyan regions excluded at $2\sigma$ C.L., while the region between blue curves are allowed at $2\sigma$ C.L.. The pink regions with boundaries depicted in solid and dashed curves correspond to $\lambda_3=1$ and 0.1, respectively.}
\label{fig:h2aa}
\end{figure}

Following Ref.~\cite{Ghosh:2018drw}, we also consider the indirect constraints from the EWPTs~\cite{Baak:2012kk} by considering modification to the oblique parameters~\cite{Peskin:1990zt,Peskin:1991sw} and perturbativity, $\lambda_4\leq \sqrt{4\pi}$. In Fig.~\ref{fig:h2aa}, we show the indirect constraints, which are almost independent of $v_\Delta$, in the plane of $m_{\Delta^{\pm\pm\pm}}$ and $\Delta m$. For the $h_{1}^0\to\gamma\gamma$ measurements, we consider the combined signal strength by the ATLAS Collaboration. The pink regions are excluded at $2\sigma$ confidence level (C.L.), where two benchmark values of the coupling $\lambda_3=1,0.1$ are depicted. The cyan regions are excluded at $2\sigma$ C.L. by the perturbativity requirement. The regions between blue curves are however allowed at $2\sigma$ C.L. by the EWPTs. Thus there is still large room in the range of $|\Delta m|\lesssim 30\gev$ satisfying indirect constraints.

\begin{figure}[!htb]
\centering
\includegraphics[width=0.4\textwidth]{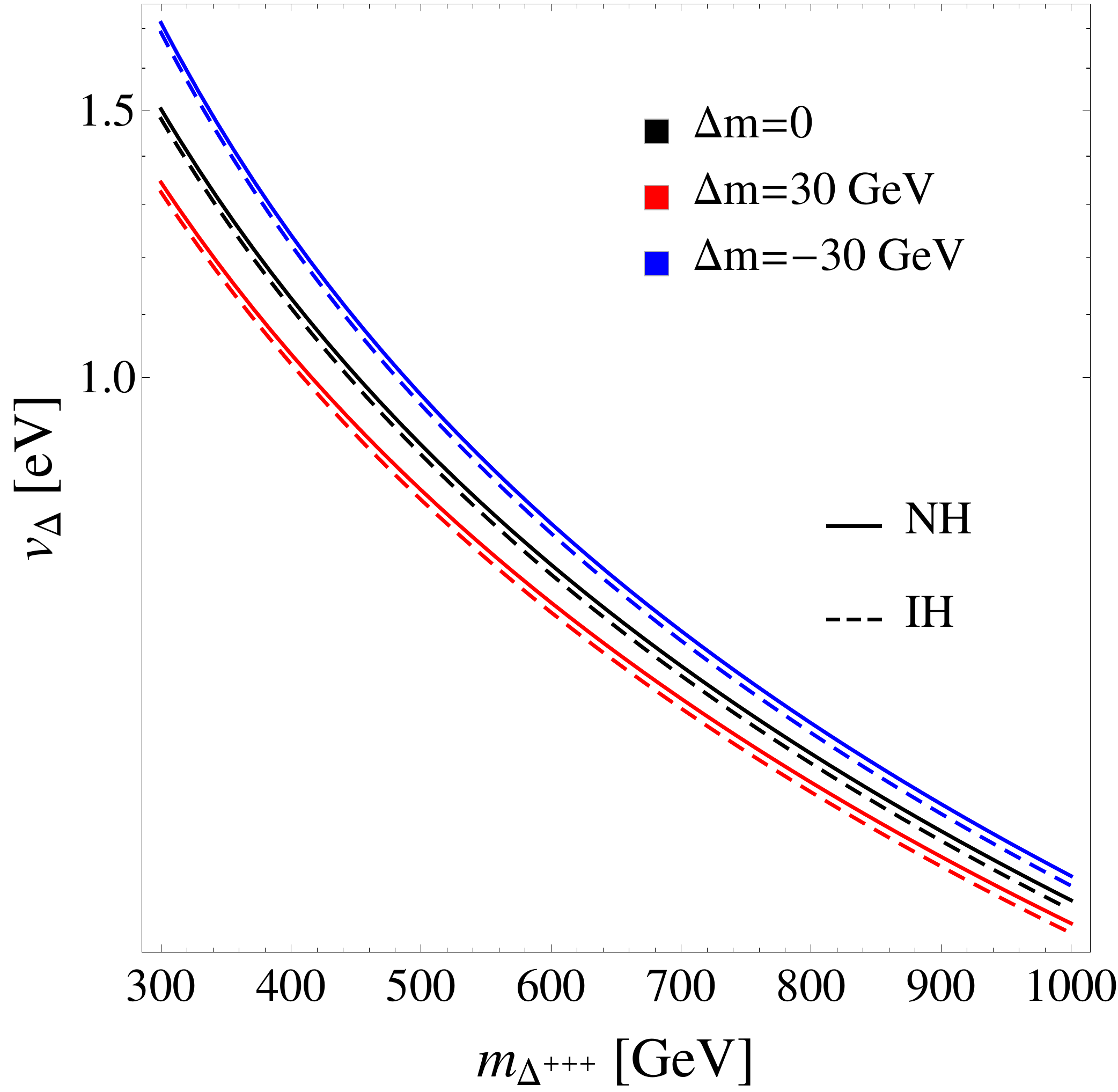}
\caption{Indirect constraint from $\mu\to e\gamma$ measurement in the plane of $m_{\Delta^{\pm\pm\pm}}$ and $v_{\Delta}$. Regions below the curves are excluded at 90\% C.L..  Solid and dashed curves correspond to the NH and IH, respectively. Black, red and blue curves are obtained with $\Delta m=0$, $30\gev$ and $-30\gev$, respectively. }
\label{fig:mu2ea}
\end{figure}

Charged Higgs bosons can contribute to other processes or observables at one-loop level~\cite{Nomura:2019btk}, such as muon anomalous magnetic moment, lepton flavor-violating processes, $Z\to \ell^+\ell^-,\nu\bar{\nu}$, etc. The most stringent constraint comes from the decay branching ratio of $\mu\to e\gamma$, which is~\cite{TheMEG:2016wtm}
\begin{align}
\text{Br}(\mu^+\to e^+\gamma)\leq 4.2\times 10^{-13}
\end{align}
at $90\%$ C.L.. The decay branching ratio of $\mu\to e\gamma$ is given  
by~\cite{Akeroyd:2009nu}
\begin{align}
\text{Br}(\mu^+\to e^+\gamma)=\dfrac{\alpha\large[(m_{\nu}^\dagger m_{\nu})_{12}\large]^2}{108\pi G_F^2v_{\Delta}^4}\Large( \dfrac{1}{m_{\Delta^{\pm\pm}}^2}+\dfrac{1}{4m_{\Delta^{\pm}}^2}\Large)^2\;,
\end{align}
where $\alpha$ and $G_F$ are the fine-structure constant and Fermion coupling constant, respectively. In Fig.~\ref{fig:mu2ea}, the constraint from $\mu\to e\gamma$ measurement is shown. We obtain that $v_{\Delta}\gtrsim 1.5\times 10^{-9}~\text{GeV}$ is allowed by the measurement of $\mu^+\to e^+\gamma$ branching ratio~\cite{TheMEG:2016wtm}. The upper bound on $v_\Delta$ is given by the $\rho$ parameter constraint~\cite{Tanabashi:2018oca}, which is $v_\Delta\lesssim 1.3\gev$.

The triply charged Higgs boson mass and the mass splitting can also be bounded from direct searches for doubly charged Higgs bosons at the LHC. Doubly charged Higgs bosons in these searches are assumed to decay into a pair of same-sign leptons~\cite{Aaboud:2017qph,CMS:2017pet} or $W$ bosons~\cite{Aaboud:2018qcu}. In our work, we emphasize on the discovery prospects of searching for triply charged Higgs bosons as a function of the quadruplet VEV with all decays being included, and on the comparison between the sensitivities at a 100~TeV $pp$ collider and at the LHC. To this end, we will not consider the constraint from the doubly charged Higgs boson direct searches, which however has been discussed in Refs.~\cite{Ghosh:2017jbw,Ghosh:2018drw}.

\section{Production and decay of triply charged Higgs boson}
\label{sec:prod_decay}

\subsection{Production cross sections}
\label{sec:prod}
As mentioned in Sec.~\ref{sec:intro}, triply charged Higgs bosons can be pair produced or associated produced with a doubly charged Higgs boson. In the $s$-channel, they correspond to the Drell-Yan processes through an off-shell photon or $Z$ boson\footnote{We have verified that the contributions from an off-shell photon and a $Z$ boson in the DY process have comparable magnitudes.} and through a $W$ boson, which are termed ``DYZ'' and ``DYW'' processes in this work, respectively. In the $t$-channel, charged Higgs bosons are produced in conjunction with two additional forward jets at leading order~\cite{Han:2007bk} by exchange of $\gamma$, $Z$ and/or $W$ boson. It was found in Refs.~\cite{Han:2007bk,Babu:2016rcr} that the photon fusion (PF) process with collinear initial photons dominates over other contributions involving off-shell photon, $Z$ boson and/or $W$ boson, named as vector boson fusion (VBF) process at the LHC. Following Refs.~\cite{Han:2007bk,Drees:1994zx,Babu:2016rcr}, we use an effective photon approximation~\cite{Budnev:1974de} to describe the PF process, which includes elastic, semi-elastic, and inelastic sub-processes\footnote{We have checked each contribution to the production cross section of a spin-0 resonance in the PF process~\cite{Csaki:2015vek} for validation.} but loses potential tagging forward jets. Since the cross section of PF process is proportional to $Q_\Delta^4$ with $Q_\Delta$ being the electric charge of $\Delta^{n\pm}$, it can even surpass the cross sections of DY processes for the production of triply charged Higgs boson. 
On the other hand, the VBF process, the cross section of which is expected to increase fairly with the collider energy, can be separated by tagging the forward jets. In the following, we will concentrate on the production of triply charged Higgs boson in DYW, DYZ and PF processes. 

\begin{figure}[!htb]
\centering
\includegraphics[width=0.35\textwidth]{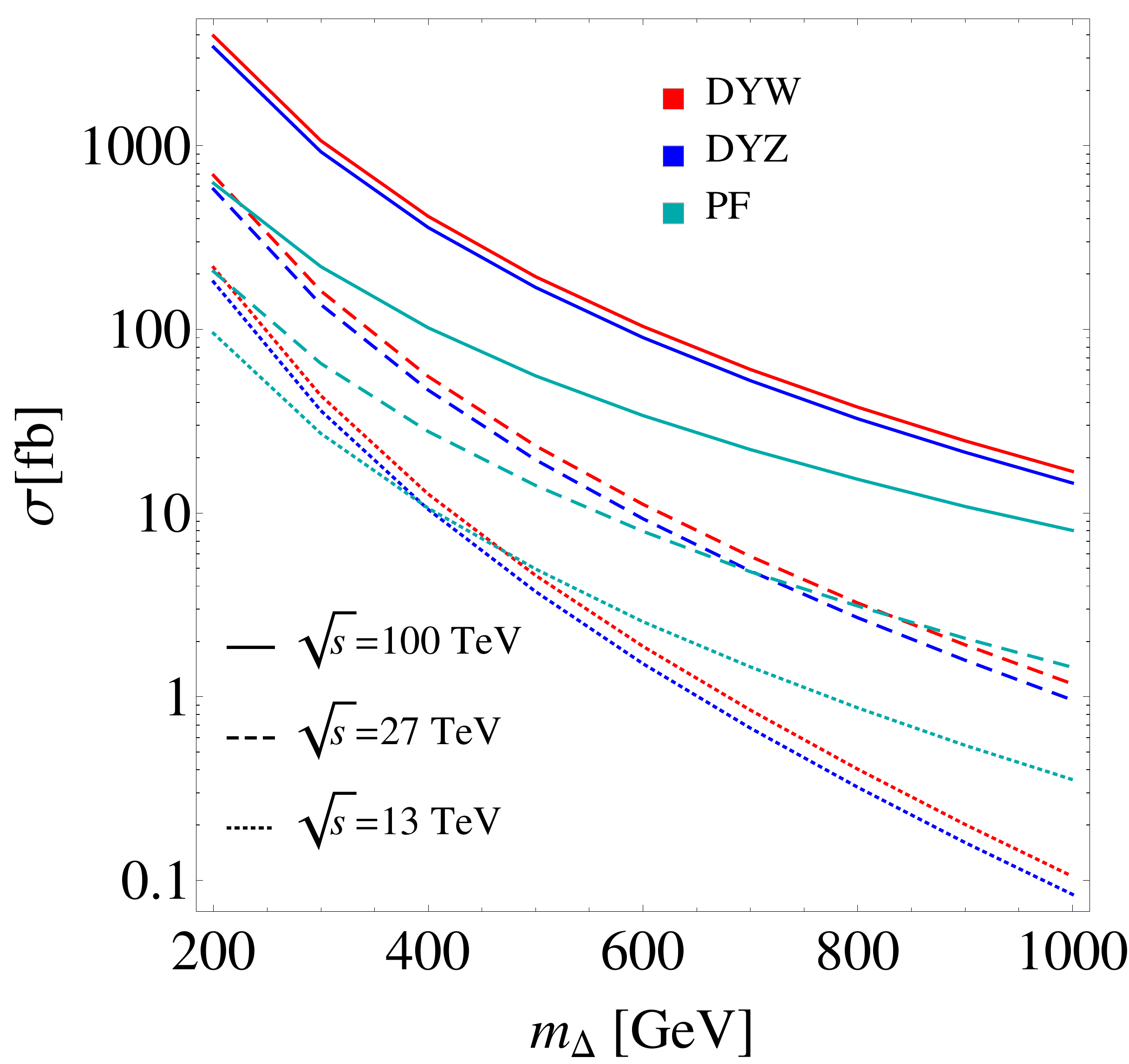}
\includegraphics[width=0.35\textwidth]{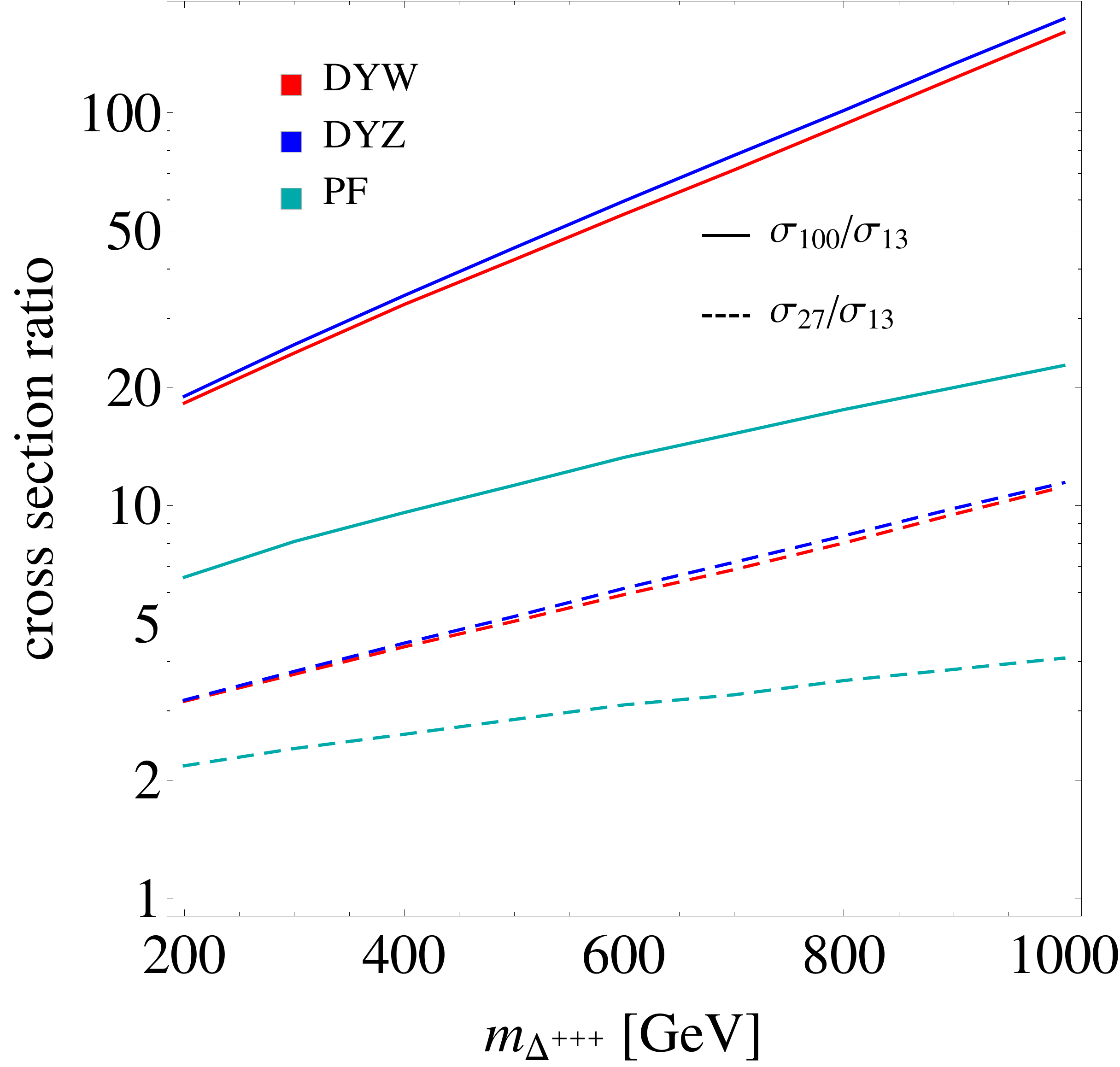}	
\caption{Left: Production cross sections of triply charged Higgs bosons via the Drell-Yan processes (DYW, DYZ) and photon fusion process (PF) at $\sqrt{s}=13$, 27 and 100~TeV as a function of $m_{\Delta^{\pm\pm\pm}}$ with the mass splitting neglected. Right: ratios of cross sections at $\sqrt{s}=27$ and $100\tev$ to that at $\sqrt{s}=13\tev$, denoted as $\sigma_{27}/\sigma_{13}$ and $\sigma_{100}/\sigma_{13}$, respectively.}
\label{fig:cross_section}
\end{figure}

The studies of triply charged Higgs boson at the LHC with the center-of-mass energy $\sqrt{s}=13\sim 14\tev$ can be found in Refs.~\cite{Bambhaniya:2013yca,Ghosh:2017jbw,Ghosh:2018drw}. In the potential era of LHC update, named as High Energy LHC (HE-LHC)~\cite{Zimmermann:2651305,CidVidal:2018eel}, the collider energy can reach 27~TeV, which increases the LHC mass reach of triply charged Higgs boson. A 100~TeV $pp$ collider such as proton-proton Future Circular Collider (FCC-hh)~\cite{Benedikt:2018csr,Abada:2019lih} or Super Proton-Proton Collider (SPPC)~\cite{CEPCStudyGroup:2018rmc,CEPCStudyGroup:2018ghi} is also designed, which provides new possibilities of discovering triply charged Higgs bosons at $pp$ colliders. 

In the left panel of Fig.~\ref{fig:cross_section} we show the cross sections of triply charged Higgs production at $\sqrt{s}=13$, 27 and 100~TeV obtained with \texttt{MG5\_aMC@NLO v2.6.5}~\cite{Alwall:2014hca} and \texttt{NNPDF23\_lo\_as\_0130\_qed} PDF set~\cite{Ball:2013hta} for the charged Higgs boson mass range between 300~GeV and 1000~GeV. It is notable that the DY cross sections have been multiplied by a next-to leading order (NLO) $K$-factor of 1.25~\cite{Muhlleitner:2003me,ATLAS-CONF-2016-051}, while higher order corrections to the PF process are small and neglected~\cite{Han:2007bk}. We can see that the cross section increases with the center-of-mass energy $\sqrt{s}$; the PF cross section dominates over the DY cross sections for $m_{\Delta^{\pm\pm\pm}}\lesssim 400$ and $500\gev$ at $\sqrt{s}=13$ and 27~TeV, respectively, while at $\sqrt{s}=100\tev$ the PF cross section is always smaller than the DY cross sections for $m_{\Delta^{\pm\pm\pm}}\leq 1000\gev$~\footnote{It is worthy to note that there is a large uncertainty of photon PDF in \texttt{NNPDF23\_lo\_as\_0130\_qed} PDF set, which could overestimate the photon-fusion production cross section at the LHC~\cite{Cai:2017mow,Bertone:2017bme}, but the impact is small since the dominant contribution comes from the DY processes.}. The ratios of cross sections at $\sqrt{s}=27\tev$ and $\sqrt{s}=100\tev$ to that at $\sqrt{s}=13\tev$, denoted as $\sigma_{100}/\sigma_{13}$ and $\sigma_{27}/\sigma_{13}$, are depicted in the right panel, which highlights the improvement of mass reach at $\sqrt{s}=27$ and $100\tev$. We will postpone a detailed analysis at the HE-LHC to a future work.

In Fig.~\ref{fig:cross_section}, we have set the masses of all the charged Higgs bosons to be the same, namely the mass splitting $\Delta m=0$. For $\Delta m \neq 0$, the production cross section of the DYW process is altered. We have checked that for $300\gev\leq m_{\Delta^{\pm\pm\pm}}\leq 1000\gev$, the DYW production cross section is reduced by at most 5\% for $\Delta m=10\gev$ and 15\% for $\Delta m=30\gev$.

\subsection{Decays of charged Higgs bosons}
\label{subsec:decay}

To evaluate the significance of the production processes, it is essential to investigate the decays of charged Higgs bosons. Triply charged Higgs boson can decay in cascade into doubly charged Higgs boson or in three-body through an off-shell doubly charged Higgs boson. Therefore, we will first discuss the decay of doubly charged Higgs boson. 

One can easily obtain the decay widths of doubly charged Higgs boson into $W^+W^+$ and $\ell^+\ell^+$ by rescaling those in the Type-II seesaw model~\cite{Perez:2008ha,Aoki:2011pz,Yagyu:2012qp,kang:2014jia}, which are
\begin{align}
\label{eq:with_ww_ll}
\Gamma(\Delta^{\pm\pm}\to \ell_{i}^{\pm}\ell_{j}^{\pm})&=\dfrac{m_{\Delta^{\pm\pm}}}{12\pi(1+\delta_{ij})}|h_{ij}|^2,\nn\\
\Gamma(\Delta^{\pm\pm}\to W^{\pm} W^{\pm}) &=\dfrac{3g^4v_{\Delta}^2m_{\Delta^{\pm\pm}}^3}{64\pi m_W^4}\sqrt{1-4\xi_W}(1-4\xi_W+12\xi_W^2)
\end{align}
with $\xi_W\equiv m_W^2/m_{\Delta^{\pm\pm}}^2$, $\ell_{1,2,3}=e,\mu,\tau$, where we have defined~\cite{Aoki:2011pz}
\begin{align}
h_{ij}=m_{\nu}^{ij}/(\sqrt{2}v_{\Delta})
\end{align}
with $m_{\nu}^{ij}$ denoting the neutrino mass matrix in the flavor basis and assumed $m_{\Delta^{\pm\pm}}>2m_W$. Here, we only consider the contribution of a Higgs quadruplet to neutrino mass at tree level; for 1-loop level contribution, one could refer to Refs.~\cite{Bambhaniya:2013yca,Ghosh:2017jbw,Ghosh:2018drw}. In the neutrino mass basis, the diagonal neutrino mass matrix is
\begin{align}
m_{\nu}^{\text{diag}}=U^T m_{\nu}U\;,
\end{align}
where $U$ is the Pontecorvo-Maki-Nakagawa-Sakata (PMNS) mixing matrix. Assuming that the CPV phases in the PMNS matrix are zero, we can determine the explicit form of $m_{\nu}$ using the central values of recent data~\cite{Tanabashi:2018oca}~\footnote{There is an update of the mixing $\sin^2\theta_{23}$~\cite{Esteban:2018azc}, which slightly changes the neutrino mass matrix $m_\nu$. } on the mixing angles and neutrino mass squared differences for both normal hierarchy (NH) and inverted hierarchy (IH) mass spectra. 

Doubly charged Higgs boson $\Delta^{\pm\pm}$ can also decay into singly or triply charged Higgs boson, depending on the mass spectrum of the Higgs quadruplet. There are two cases for the mass spectrum, which is determined by the parameter $\lambda_4$ in the Higgs potential, see Eq.~\eqref{eq:mass splitting}. Defining the mass splitting between the nearby states of the Higgs quadruplet, $\Delta m \equiv m_{\Delta^{\pm\pm}}-m _{\Delta^{\pm\pm\pm}}= m_{\Delta^{\pm}}-m _{\Delta^{\pm\pm}}=m_{\Delta}-m_{\Delta^{\pm}}$, we obtain
\begin{itemize}
\item Case $\Delta m >0$: $m_{\Delta^{\pm\pm\pm}}<m_{\Delta^{\pm\pm}}<m_{\Delta^{\pm}}<m_{\Delta}$;
\item Case $\Delta m<0$: $m_{\Delta^{\pm\pm\pm}}>m_{\Delta^{\pm\pm}}>m_{\Delta^{\pm}}>m_{\Delta}$.
\end{itemize}

For case $\Delta m >0$, $\Delta^{\pm\pm}$ can decay into $\Delta^{\pm*}W^\pm$, $\Delta^{\pm\pm\pm*}W^\mp$ and $\Delta^{\pm\pm\pm}W^{\mp*}$, while for case $\Delta m <0$, $\Delta^{\pm\pm}$ can decay into $\Delta^{\pm\pm\pm*}W^\mp$, $\Delta^{\pm*}W^\pm$ and $\Delta^{\pm}W^{\pm*}$. Here, $\Delta^{\pm\pm\pm*}$, $\Delta^{\pm*}$ and $W^{\pm*}$ denote off-shell particles.  The decay $\Delta^{\pm\pm}\to \Delta^{\pm*}W^\pm$ depends on the couplings of $\Delta^{\pm}W^{\mp}Z$ and $\Delta^\pm \ell^\mp \nu$, which are proportional to $v_\Delta$ or $1/v_{\Delta}$ similar to $\Delta^{\pm\pm}\to W^{\pm}W^\pm$ or $\Delta^{\pm\pm} \to \ell^\pm \ell^\pm$. Therefore, we can neglect the contribution of $\Delta^{\pm\pm}\to \Delta^{\pm*}W^\pm$ in the total width of $\Delta^{\pm\pm}$. The decay $\Delta^{\pm\pm}\to \Delta^{\pm\pm\pm*}W^\mp$ depends on the interaction of $\Delta^{\pm\pm\pm}$ to SM particles through an off-shell $\Delta^{\pm\pm}$ and can also be neglected.

The cascade decays $\Delta^{\pm\pm}\to \Delta^{\pm} W^{\pm*}$ and $\Delta^{\pm\pm}\to \Delta^{\pm\pm\pm} W^{\mp*}$ only depend on the mass splitting $\Delta m$ approximately with the widths being given by~\cite{Perez:2008ha,Aoki:2011pz}
\begin{align}
\Gamma(\Delta^{\pm\pm}\to \Delta^{\pm} W^{\pm*})&=-\dfrac{3g^4 \Delta m^5}{40\pi^3m_W^4}\;,\\
\Gamma(\Delta^{\pm\pm}\to \Delta^{\pm\pm\pm} W^{\mp*})&=\dfrac{9g^4 \Delta m^5}{160\pi^3m_W^4}\;.
\end{align}
From the constraints by the EWPTs, $|\Delta m|\lesssim 30\gev$ as shown in Fig.~\ref{fig:h2aa}. We will thus choose the benchmark values $\Delta m=0,\pm 1\gev,\pm 10\gev$ for simplicity.

For $\Delta m <0$ $(\Delta m >0)$, $\Delta^{\pm\pm}$ can also decay into $\Delta^\pm \pi^\pm$ ($\Delta^{\pm\pm\pm} \pi^\mp$) with the decay widths~\cite{Perez:2008ha}
\begin{align}
\Gamma(\Delta^{\pm\pm}\to \Delta^{\pm} \pi^{\pm})&=-\dfrac{g^4 \Delta m^3 f_{\pi}^2}{8\pi m_W^4}\;,\\
\Gamma(\Delta^{\pm\pm}\to \Delta^{\pm\pm\pm} \pi^{\mp})&=\dfrac{3g^4 \Delta m^3 f_{\pi}^2}{32\pi m_W^4}\;,
\end{align}
where the decay constant of $\pi$ meson $f_{\pi}=131~\text{MeV}$. It is easy to check that the cascade decay width of $\Delta^{\pm\pm}$ into off-shell $W$ boson is much larger than that into $\pi$ meson for $|\Delta m|\gtrsim 1\gev$.

The total width of $\Delta^{\pm\pm\pm}$ can thus be expressed as
\begin{align}
\label{eq:total_width}
\Gamma_{\Delta^{\pm\pm}}&=\Gamma(\Delta^{\pm\pm}\to \ell_i^\pm\ell_j^\pm)+
\Gamma(\Delta^{\pm\pm}\to W^\pm W^\pm)\nn\\
&\quad +\theta(-\Delta m) \Big[
\Gamma(\Delta^{\pm\pm}\to \Delta^{\pm} W^{\pm*})+
\Gamma(\Delta^{\pm\pm}\to \Delta^{\pm} \pi^{\pm})\Big]\nn\\
&\quad +\theta(\Delta m) \Big[
\Gamma(\Delta^{\pm\pm}\to \Delta^{\pm\pm\pm} W^{\mp*})+
\Gamma(\Delta^{\pm\pm}\to \Delta^{\pm\pm\pm} \pi^{\mp})\Big]\;,
\end{align}
where the Heviside function $\theta(x)=1$ for $x>0$ and 0 for $x<0$. For $\Delta m=0$, only the first two terms contribute.

\begin{figure}[!htb]
\centering  	
\includegraphics[width=0.38\textwidth]{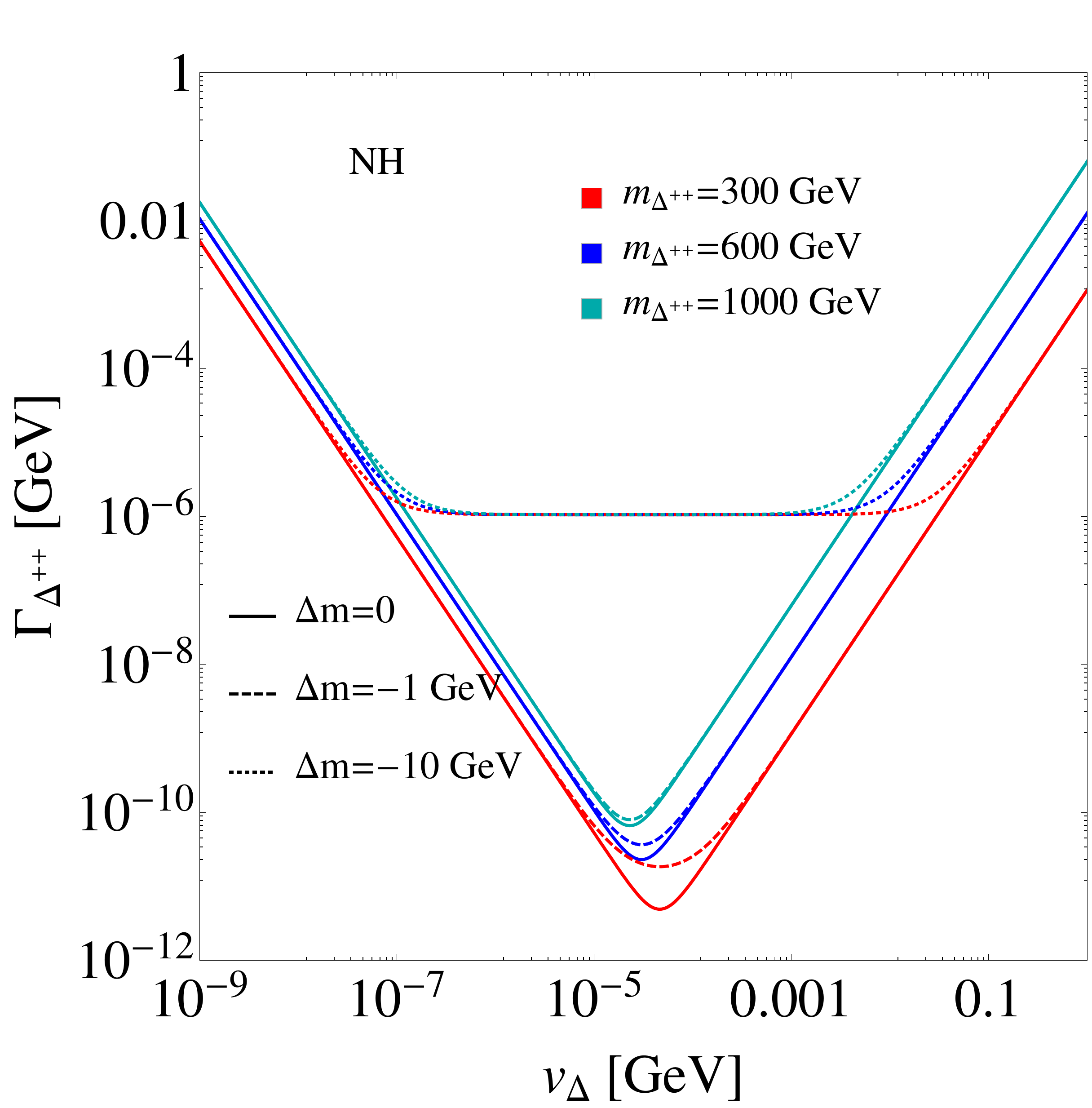}
\includegraphics[width=0.4\textwidth]{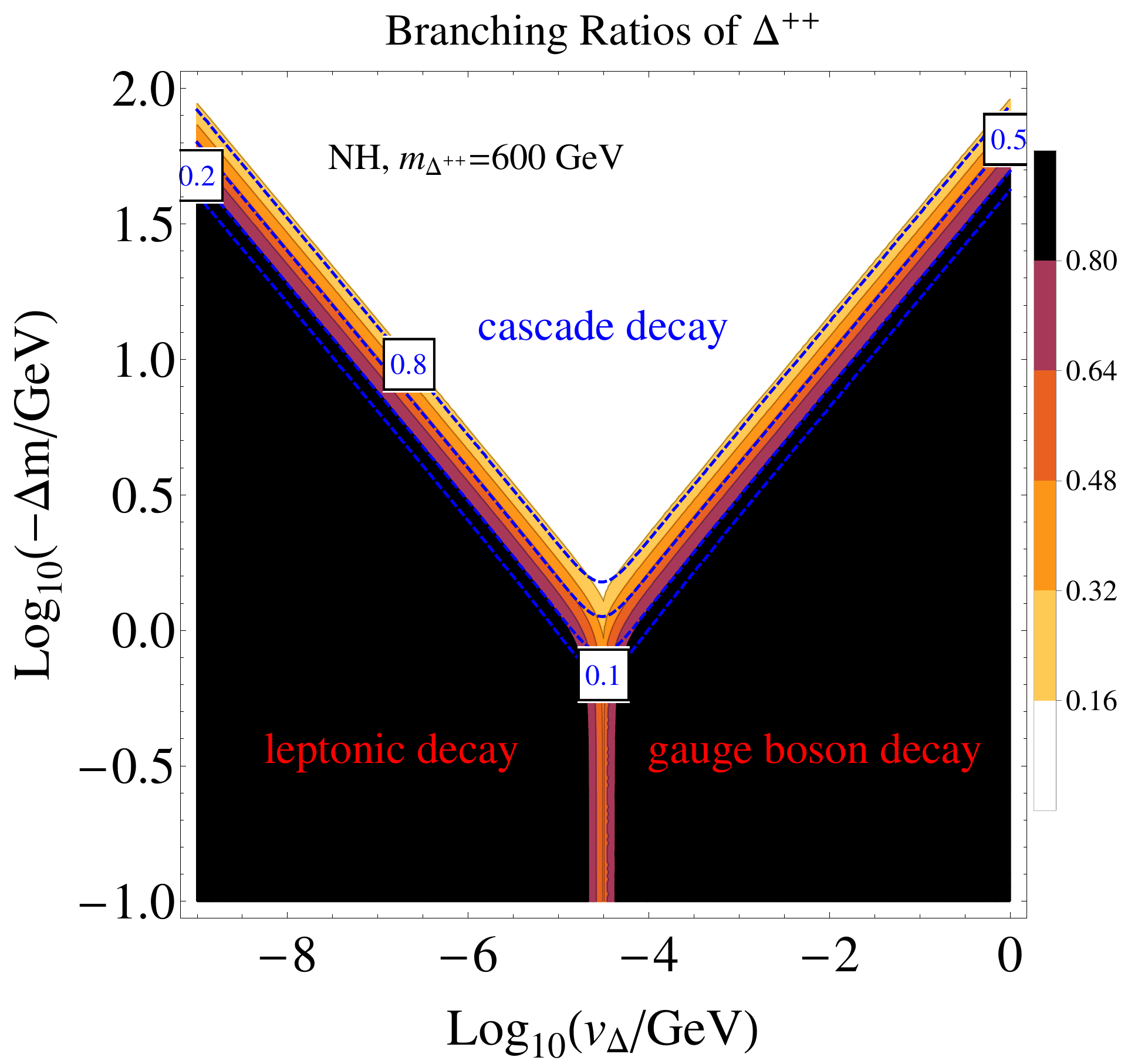}	
\caption{Left: total decay width of $\Delta^{\pm\pm}$ defined in Eq.~\eqref{eq:total_width} as a function of the quadruplet VEV $v_{\Delta}$ for $m_{\Delta^{\pm\pm}}=300$, $600,900\gev$ for the NH. The solid, dashed, dotted curves correspond to $\Delta m=0$, $-1,-10\gev$, respectively. Right: decay branching ratios of $\Delta^{\pm\pm}$ in the plane of $\log_{10}(v_{\Delta}/\text{GeV})$ and $\log_{10}(-\Delta m/\text{GeV})$ for $m_{\Delta^{\pm\pm}}=600\gev$ and the NH. The shaded regions represent the branching ratios in the leptonic decay and gauge boson decay channels for $v_{\Delta}\lesssim 10^{-4.5}\gev$ and $v_{\Delta}\gtrsim 10^{-4.5}\gev$, respectively. The blue curves denote the branching ratio in the cascade decay channel.}
\label{fig:br_dm}
\end{figure}

These three decay modes of $\Delta^{\pm\pm}$ compete with each other controlled by the quadruplet VEV $v_\Delta$ and the mass splitting $\Delta m$.
To evaluate the fraction of the cascade decays, we depict the total width and decay branching ratios of $\Delta^{\pm\pm}$ for $\Delta m\leq 0$ and the NH in Fig.~\ref{fig:br_dm}. For $\Delta m\geq 0$ and/or the IH, we can get similar results. The branching ratio of cascade decays increases with $|\Delta m|$. For $\Delta m=-1\gev~(-10\gev)$, it is larger than 0.1 (0.8) in the range $10^{-5}\gev \lesssim v_{\Delta}\lesssim 10^{-4}\gev$ $(10^{-6.5}\gev \lesssim v_{\Delta}\lesssim 10^{-2.5}\gev)$, as shown in the right panel. Given the total width in the left panel, the proper decay length $c\tau_{\Delta^{\pm\pm}}=\hbar c/\Gamma_{\Delta^{\pm\pm}}$ can be easily obtained and is smaller than $0.1~\text{mm}$ for $300\gev\leq m_{\Delta^{\pm\pm}}\leq 1000\gev$, which ensures the validity of prompt search of $\Delta^{\pm\pm}$ at $pp$ colliders. 

Triply charged Higgs boson $\Delta^{\pm\pm\pm}$ can decay into $W\ell_i^\pm \ell_j^\pm$ and $W^\pm W^\pm W^\pm$ if kinetically allowed. The partial widths of three-body decays through an off-shell $\Delta^{\pm\pm}$ are
\begin{align}
\label{eq:width_Delta+++}
\Gamma(\Delta^{\pm\pm\pm} \to W^\pm \ell_i^\pm \ell_j^\pm)&=\frac{g^2 m^3_{\Delta^{\pm\pm\pm}}|h_{ij}|^2}{768\pi^3 m^2_{W}(1+\delta_{ij})}\int _{0}^{(m_{\Delta^{\pm\pm\pm}}-m_W)^2}ds F(s)\;,\nn\\
\Gamma(\Delta^{\pm\pm\pm} \to W^\pm W^\pm W^\pm)&=\frac{3g^6 v^2_\Delta m^5_{\Delta^{\pm\pm\pm}}}{4096\pi^3 m^6_W}\int _{4m_W^2}^{(m_{\Delta^{\pm\pm\pm}}-m_W)^2}\int_{t_{\text{min}}}^{t_{\text{max}}}dt ds G(s,t)
\end{align}
with
\begin{align}
F(s)&=\dfrac{m_W^2}{m_{\Delta^{\pm\pm\pm}}^4}\big[6(-2-2r_s+r_W+1/r_W(1-r_s)^2)\big]D(s)r_s\lambda(1,r_s,r_W)^{1/2},\\
G(s, t)&=\dfrac{1}{m_{\Delta^{\pm\pm\pm}}^4}\Big[\big[24r_W(-2-2r_s+r_W+1/r_W(1-r_s)^2)\big] D(s)\nn\\
&\quad \times(2r_W^2+1/4(r_s-2r_W)^2)\nn\\
&\quad +48 [(1-r_s)(1-r_t)-(1/2r_W r_s+1/2r_W r_t+5/2 r_W-3/2r_W^2)] E(s,t)\nn\\
&\quad \times [3r_W^2+1/4(r_s-4r_W)(r_t-4r_W)]\Big]\\
t_{\text{max}}&=\frac{1}{4s}[(m^2_{\Delta^{+++}}-m^2_W)^2-(\lambda(s, m^2_W, m^2_W)^{\frac{1}{2}}-\lambda(m^2_{\Delta^{+++}}, s, m^2_W)^{\frac{1}{2}})^2]\\
t_{\text{min}}&=\frac{1}{4s}[(m^2_{\Delta^{+++}}-m^2_W)^2-(\lambda(s, m^2_W, m^2_W)^{\frac{1}{2}}+\lambda(m^2_{\Delta^{+++}}, s, m^2_W)^{\frac{1}{2}})^2].
\end{align}
and
\begin{align}
&D(s)=\dfrac{1}{(r_s-(1+\Delta m/m_{\Delta^{\pm\pm\pm}})^2)^2+(1+\Delta m/m_{\Delta^{\pm\pm\pm}})^2\Gamma_{\Delta^{\pm\pm}}^2/m_{\Delta^{\pm\pm\pm}}^2}\;,\\
&E(s, t)=\dfrac{1}{(r_s-(1+\Delta m/m_{\Delta^{\pm\pm\pm}})^2)(r_t-(1+\Delta m/m_{\Delta^{\pm\pm\pm}})^2)+(1+\Delta m/m_{\Delta^{\pm\pm\pm}})^2\Gamma_{\Delta^{\pm\pm}}^2/
m_{\Delta^{\pm\pm\pm}}^2}\;.
\end{align}
Here, $s, t$ denote the invariant mass of the $W$ boson pair from the decay of $\Delta^{\pm\pm}$, $r_s\equiv s/m_{\Delta^{\pm\pm\pm}}^2$, $r_t\equiv t/m_{\Delta^{\pm\pm\pm}}^2$, $r_W\equiv m_W^2/m_{\Delta^{\pm\pm\pm}}^2$, and $\lambda(x,y,z)\equiv (x-y-z)^2-4yz$. In the limit of $m_W/m_{\Delta^{\pm\pm\pm}}\to 0$, the above integrations over $F(s)$ and $G(s, t)$ are equal to 1. It is noted that the total width $\Gamma_{\Delta^{\pm\pm}}\lesssim 0.01\gev$ for $10^{-9}\gev\leq v_{\Delta}\leq 1\gev$ (see Fig.~\ref{fig:br_dm}), which has negligible effect on the three-body decay widths.

\begin{figure}[!htb]
\centering
\includegraphics[width=0.4\textwidth]{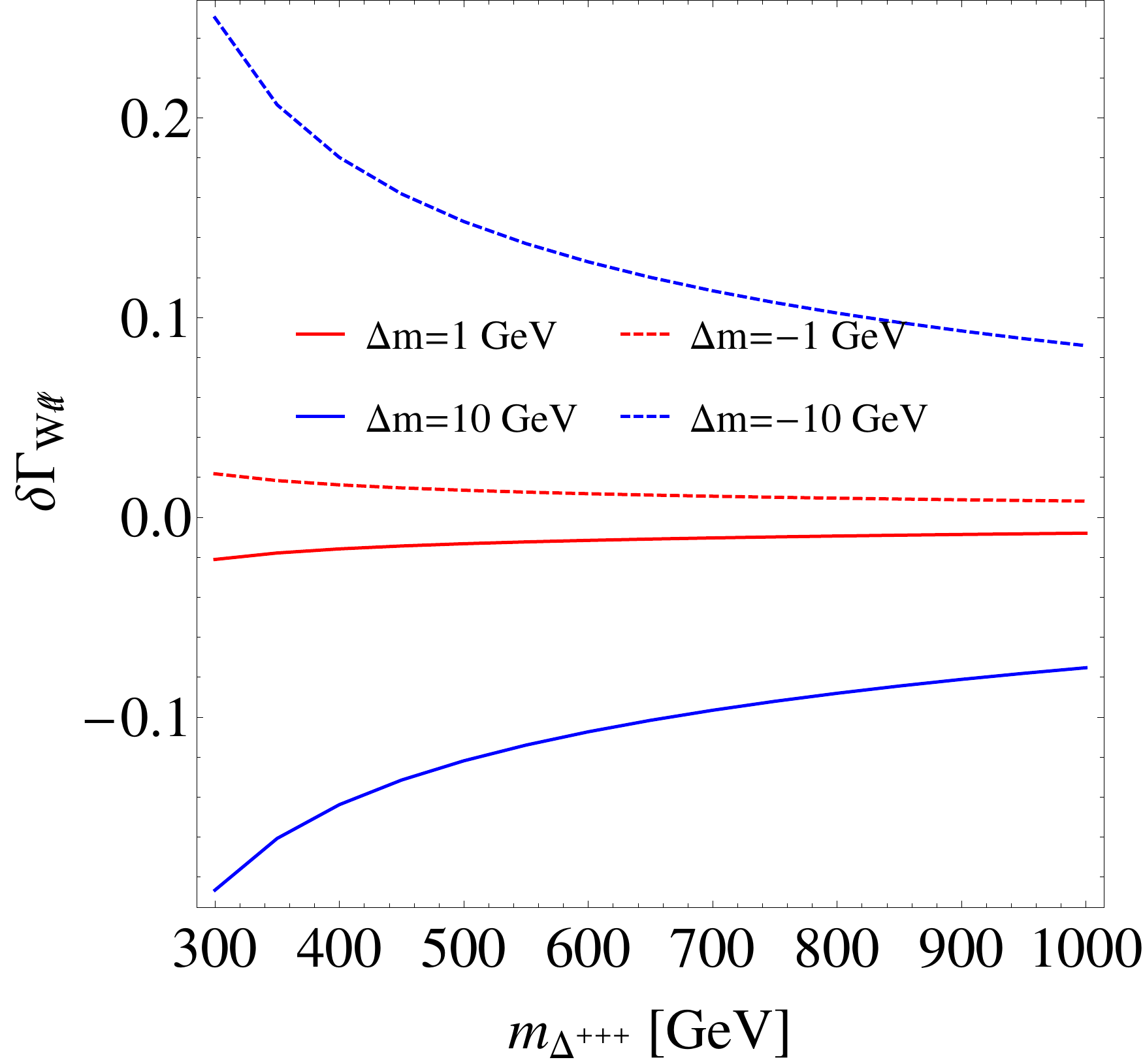}
\includegraphics[width=0.4\textwidth]{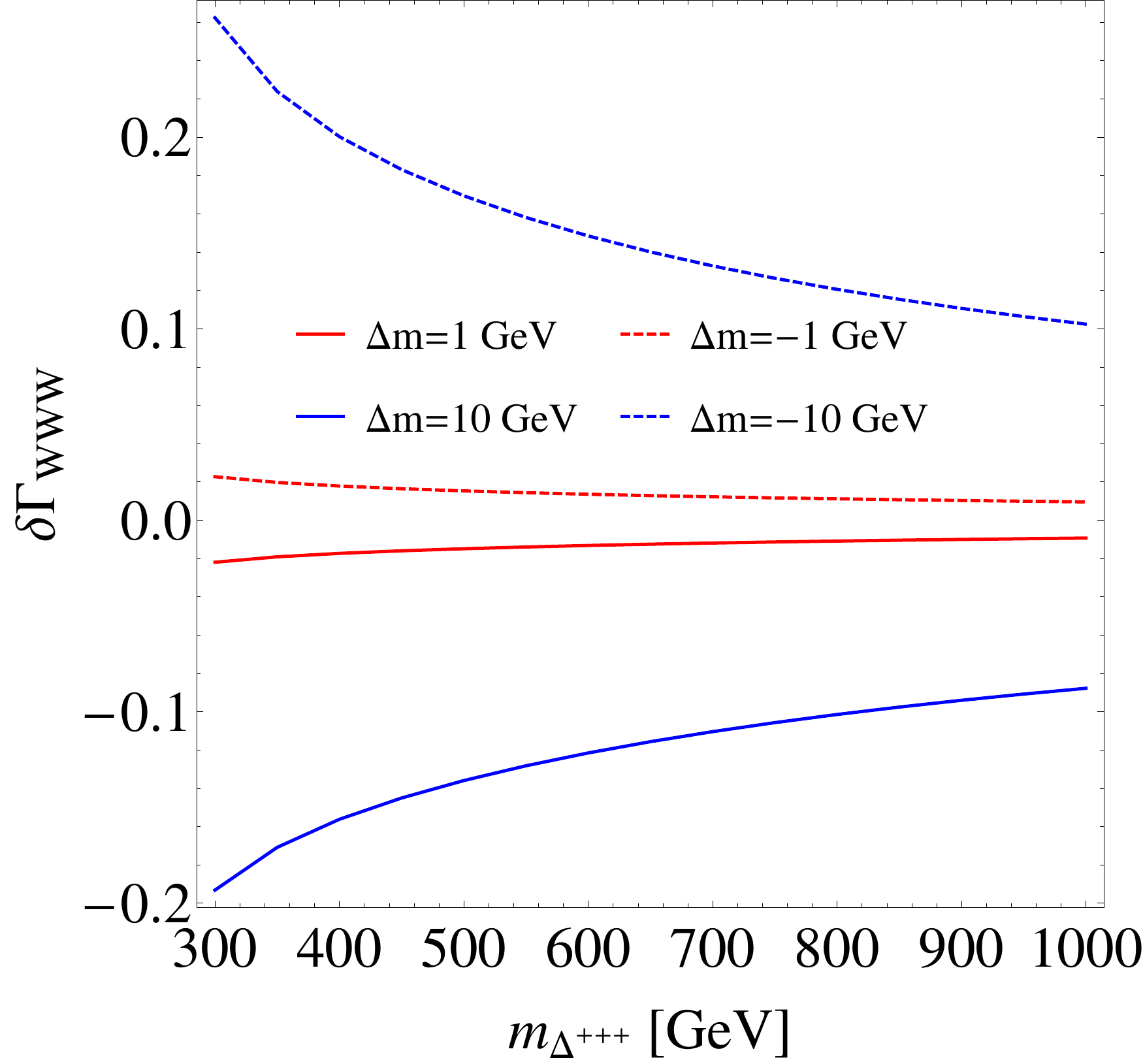}		
\caption{Impact of $\Delta m$ on the partial decay widths of $\Delta^{\pm\pm\pm}$ into $W^{\pm}\ell^{\pm}_i\ell^{\pm}_j$ and $W^{\pm}W^{\pm}W^{\pm}$ in the left and right panels, respectively. Benchmark values of $|\Delta m|=1$, $10\gev$ are considered.}
\label{fig:delta_gamma_dm}
\end{figure}

Different from the decays of $\Delta^{\pm\pm}\to \ell_i^\pm \ell_j^\pm$, $W^\pm W^\pm$, the three-body decays of $\Delta^{\pm\pm\pm}$ in Eq.~\eqref{eq:width_Delta+++} depend on the mass splitting $\Delta m$. To estimate its impact, we introduce 
\begin{align}
\delta\Gamma_{W\ell\ell} =(\Gamma_{W\ell\ell}-\Gamma_{W\ell\ell}^{0})/\Gamma_{W\ell\ell}^{0},\quad \delta\Gamma_{WWW} =(\Gamma_{WWW}-\Gamma_{WWW}^{0})/\Gamma_{WWW}^{0},
\end{align}
where $\Gamma_{W\ell\ell}=\sum_{i,j} \Gamma(\Delta^{\pm\pm\pm} \to W^\pm \ell_i^\pm \ell_j^\pm)$ and $\Gamma_{WWW}=\Gamma(\Delta^{\pm\pm\pm} \to W^\pm W^\pm W^\pm)$ and $\Gamma_{W\ell\ell}^0$ and $\Gamma_{WWW}^0$ are the corresponding values with $\Delta m=0$. In Fig.~\ref{fig:delta_gamma_dm}, the values of $\delta\Gamma_{W\ell\ell}$ and $\delta\Gamma_{WWW}$ are shown. We find that both $\delta\Gamma_{\ell\ell W}$ and $\delta\Gamma_{WWW}$ are negligible for $|\Delta m|=1\gev$ and increase to $10\%-25\%$ for $|\Delta m|=10\gev$ in the mass range $300\gev\leq m_{\Delta^{\pm\pm\pm}}\leq 1000\gev$.

The interplay between the decays $\Delta^{\pm\pm\pm}\to W^\pm \ell_i^\pm\ell_j^\pm $ and $\Delta^{\pm\pm\pm}\to W^\pm W^\pm W^\pm$ is the similar to that for $\Delta^{\pm\pm}$ in two-body decays. Therefore, we need to include the cascade decays of $\Delta^{\pm\pm\pm}$ with $\Delta^{\pm\pm}$ being on shell in the medium $v_{\Delta}$ region for $\Delta m<0$ with the widths being approximately given by
\begin{align}
\Gamma(\Delta^{\pm\pm\pm}\to \Delta^{\pm\pm} W^{\pm*})&=-\dfrac{9g^4 \Delta m^5}{160\pi^3m_W^4}\;,\\
\Gamma(\Delta^{\pm\pm\pm}\to \Delta^{\pm\pm} \pi^{\pm})&=-\dfrac{3g^4 \Delta m^3f_{\pi}^2}{32\pi m_W^4}.
\end{align}

For $\Delta m>0$, the cascade decay of $\Delta^{\pm\pm\pm}$ is kinetically forbidden. Hence, the total width of $\Delta^{\pm\pm\pm}$ is expressed as
\begin{align}
\label{eq:total_width_triply}
\Gamma_{\Delta^{\pm\pm\pm}}&=\Gamma(\Delta^{\pm\pm\pm} \to W^+ \ell_i^\pm \ell_j^\pm)+
\Gamma(\Delta^{\pm\pm\pm} \to W^\pm W^\pm W^\pm)\nn\\
&\quad +\theta(-\Delta m)\Big[
\Gamma(\Delta^{\pm\pm\pm}\to \Delta^{\pm\pm} W^{\pm*})+
\Gamma(\Delta^{\pm\pm\pm}\to \Delta^{\pm\pm} \pi^{\pm})\Big].
\end{align}

\begin{figure}[!htb]
\centering  
\includegraphics[width=0.38\textwidth]{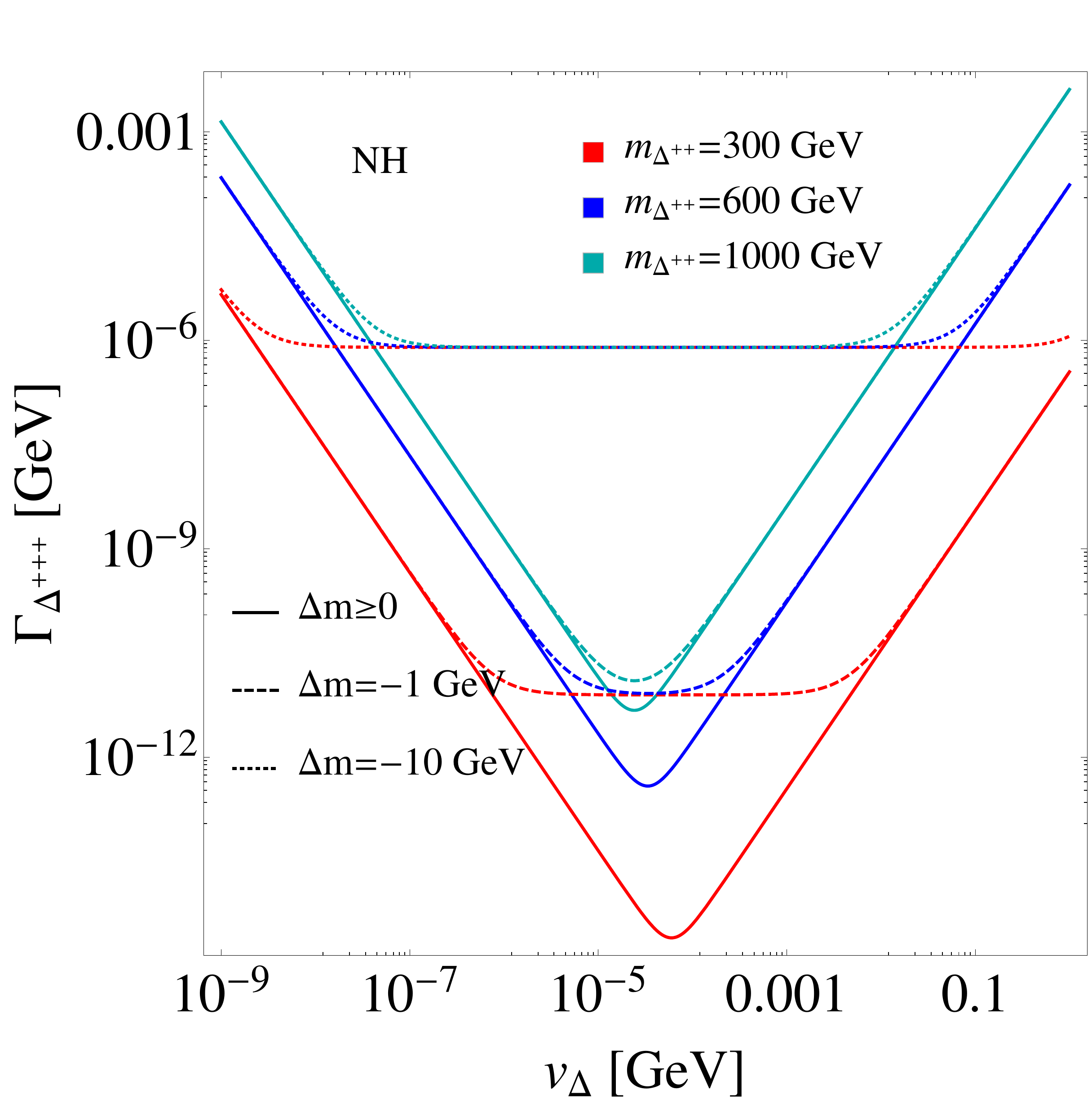}		
\includegraphics[width=0.38\textwidth]{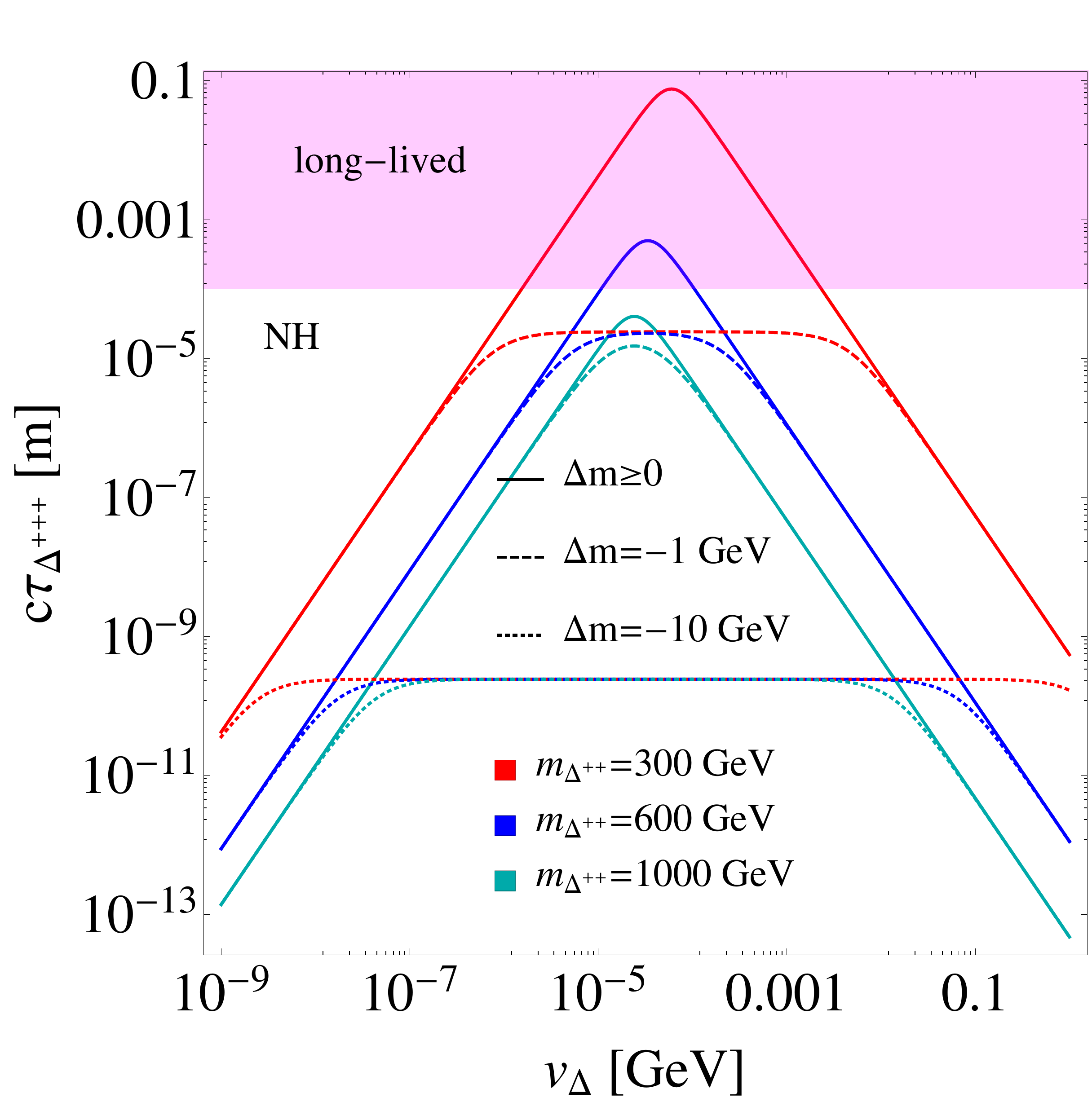}		
\caption{Total decay width (let panel) and proper decay length (right panel) of $\Delta^{\pm\pm}$ defined in Eq.~\eqref{eq:total_width} as a function of the quadruplet VEV $v_{\Delta}$ for $m_{\Delta^{\pm\pm}}=300$, $600,1000\gev$ with the NH. The solid, dashed, dotted curves correspond to $\Delta m=0$, $-1,-10\gev$, respectively. In the magneta shaded region, $c\tau_{\Delta^{\pm\pm\pm}}\geq 0.1~\text{mm}$, $\Delta^{\pm\pm\pm}$ is long-lived. }
\label{fig:length_Deltappp}
\end{figure}

The total width and proper decay width of $\Delta^{\pm\pm\pm}$ are depicted in Fig.~\ref{fig:length_Deltappp}. It is interesting to observe that since the three-body decay widths of $\Delta^{\pm\pm}$ are much smaller than the two-body decay widths of $\Delta^{\pm\pm}$, the cascade decay dominates in the medium $v_\Delta$ region for $\Delta m<0$ even with $\Delta m=-1\gev$. For $\Delta m>0$, the cascade decays are not allowed and the dependence of three-body decays on $\Delta m$ is not shown explicitly for simplicity. In the left panel of Fig.~\ref{fig:length_Deltappp}, the total width can be as small as $10^{-15}\gev$ so that the proper decay length can reach $0.1~\text{mm}\sim 0.1~\text{m}$ -- the region that is inappropriate for prompt search~\cite{ATLAS:2016jza}, as shown in the right panel. For $\Delta m<0$, however, the proper decay length is large enough for the prompt search with the contribution of cascade decays added.

Although the partial widths of three-body decays and cascade decays depend on the mass splitting $\Delta m$, the decay branching ratios of $\Delta^{\pm\pm\pm}$ are almost independent of $\Delta m$. For $\Delta m>0$, the $\Delta m$ dependence of the partial widths cancels in the branching ratios of $\Delta^{\pm\pm\pm}$, resulting in a modification smaller than 1.7\% for $\Delta m =10\gev$. For $\Delta m <0$, the cancellation is similar in the low and high $v_\Delta$ regions as that for $\Delta m>0$. In the medium $v_\Delta$ region, the cascade decay dominates over the three-body decays, which ensures that the branching ratio is independent of $\Delta m$.

\section{Collider analysis}
\label{sec:collider}

In this section, we will perform collider studies of triply charged Higgs bosons at $pp$ colliders. Given the couplings of $\Delta^{\pm\pm}$ to charged leptons and $W$ bosons, we have the decay channels: for the DYZ and PF processes $\Delta^{+++}\Delta^{---}\to$ $(\ell^+\ell^+ W^+)(\ell^-\ell^- W^-)$, $(\ell^\pm\ell^\pm W^\pm)(W^\mp W^\mp W^\mp)$, and $(W^+W^+W^+)(W^-W^-W^-)$; for the DYW process $\Delta^{\pm\pm\pm}\Delta^{\mp\mp}\to$ $(\ell^\pm\ell^\pm W^\pm)(\ell^\mp\ell^\mp)$, $(\ell^\pm\ell^\pm W^\pm)(W^\mp W^\mp)$, $(W^\pm W^\pm W^\pm) (\ell^\mp \ell^\mp)$, and $(W^\pm W^\pm W^\pm)(W^\mp W^\mp )$, where $\ell\equiv e,\mu,\tau$. As shown in Sec.~\ref{sec:prod_decay}, $\Delta m$ has negligible impact on the decay branching ratios of $\Delta^{\pm\pm\pm}$. Therefore, we could simulate the above processes with $m_{\Delta^{\pm\pm}}=m_{\Delta^{\pm\pm\pm}}$ in the production but keep $m_{\Delta^{\pm\pm}}=m_{\Delta^{\pm\pm\pm}}+\Delta m$ in the decays of $\Delta^{\pm\pm}$. Besides, we need further to include the cascade decays, which are important in the medium $v_\Delta$ region. For $\Delta m<0$, $\Delta^{\pm\pm\pm}\to\Delta^{\pm\pm}W^{\pm*}$ and $\Delta^{\pm\pm}\to \Delta^{\pm}W^{\pm*}$. For $\Delta m>0$, $\Delta^{\pm\pm}\to \Delta^{\pm\pm\pm}W^{\pm*}$. Here we do not consider the signals with cascade decays into $\pi$ mesons since their decay branching ratios are negligible as compared to those into off-shell $W$ bosons for $\Delta m\gtrsim 1\gev$.
 
For an inclusive final state, we can always achieve at least three same-sign leptons in case of $\Delta m>0$ if on-shell $W$ bosons decay into leptons. In case of $\Delta m<0$, however, the leptons or jets from off-shell $W$ bosons are soft and are unlikely to be detected without a delicate study\footnote{The experimental preselection cut on the transverse momentum on the lepton is $p_{T,e/\mu}>20\gev$ at the LHC~\cite{Aaboud:2017qph,CMS:2017pet,Aaboud:2018qcu}. While the momentum of charged lepton from off-shell $W$ boson is limited by the mass splitting $|\Delta m|\lesssim 30\gev$. Therefore, it is hard to isolate such soft leptons~\cite{Ghosh:2018drw}.}. Henceforth, we will concentrate on sensitivities in the case of $\Delta m\geq 0$. 

The SM backgrounds are those with at least three same-sign charged leptons in the final states. In previous studies, the backgrounds $t\bar{t}W$~\cite{Mukhopadhyaya:2010qf,Bambhaniya:2013yca,
Ghosh:2017jbw,Ghosh:2018drw,Agarwalla:2018xpc}, $t\bar{t}Z$~\cite{Bambhaniya:2013yca,Agarwalla:2018xpc}, $t\bar{t}t\bar{t}$~\cite{Mukhopadhyaya:2010qf,Bambhaniya:2013yca,Agarwalla:2018xpc}, $t\bar{t}b\bar{b}$~\cite{Mukhopadhyaya:2010qf,Bambhaniya:2013yca}, $t\bar{t}h$~\cite{Agarwalla:2018xpc}, $WWZ$~\cite{Agarwalla:2018xpc}, $WZZ$~\cite{Agarwalla:2018xpc} and $ZZZ$~\cite{Agarwalla:2018xpc} were considered. In Ref.~\cite{Ghosh:2018drw}, the backgrounds $WZ$ and $ZZ$ were discussed with charge misidentification of leptons taken into account.

In our study, we consider the backgrounds with at least two same-sign leptons at parton level and the third same-sign lepton could come from heavy-flavor hadron decays or charge misidentification. Besides, the $t\bar{t}$ background is also taken into account, since its cross section is huge. The set of backgrounds can be read off from the experimental searches for final states with same-sign leptons or multiple leptons~\cite{CMS:2019nig,CMS:2019see,Aaboud:2017dmy}, which are classified into $t\bar{t}$ production in association with a boson $(t\bar{t}W$, $t\bar{t}Z/\gamma^*$, $t\bar{t}h$ with $h$ being the SM Higgs boson), multi-top production $(t\bar{t}$, $t\bar{t}t/\bar{t}$, $t\bar{t}t\bar{t})$, multi-boson production ($WZ$, $Z/\gamma^*$ $WWW$, $WWZ$, $WZZ$, $ZZZ$, $WW\gamma^*$, $WZ\gamma^*$) and rare processes $(t\bar{t}b\bar{b}$, $tWZ$, $t/\bar{t}Zq)$ with $q$ denoting one of quarks except $t/\bar{t}$. 

The comments on the backgrounds are made as follows. Backgrounds with an off-shell photon, such as $t\bar{t}\gamma^*$, are not generated since their contributions are expected to be reduced significantly after imposing the lower cuts on the invariant mass of opposite-sign same-flavor leptons in Cut-3 (see the definition below) as compared to the corresponding backgrounds with an on-shell $Z$ boson. The backgrounds $t\bar{t}h,h\to b\bar{b},WW^*$ are not considered since their cross sections are much smaller than those of $t\bar{t}b\bar{b},t\bar{t}W$. For the background $t\bar{t}Z$, we only consider the decay $ Z\to \ell^+\ell^-$ and neglect $Z\to q\bar{q}$ since the latter cross section is much smaller as compared to $t\bar{t}jj$ and $t\bar{t}b\bar{b}$. The tri-top production $t\bar{t}t/\bar{t}$~\cite{Barger:2010uw} with a much smaller cross section than that of $t\bar{t}W$~\cite{Aaboud:2019njj} can be neglected. 

The charge misidentification probability is about $10^{-5}\sim 10^{-3}$ for electrons $(\epsilon_e)$ due to bremsstrahlung interactions with the inner detector material and negligible for muons at the 13~TeV LHC~\cite{CMS:2019nig,CMS:2019see,Khachatryan:2016kod,Sirunyan:2017uyt}. At a 100~TeV $pp$ collider, we assume a conservative and uniform rate $\epsilon_e=10^{-3}$~\cite{Alva:2014gxa}. The charge-misidentified backgrounds are obtained from reweighting the background by the charge misidentification probabilities~\cite{Khachatryan:2016kod}, see Tab.~\ref{tbl:probability}. Backgrounds with a
non-prompt lepton may fake the signal, which originates from hadron decays or in photon
conversions as well as hadrons misidentified as leptons. It is shown at the 13~TeV LHC that the non-prompt leptons mainly come from heavy-flavor hadron decays in events containing top quark, $W$ boson or $Z$ boson~\cite{Aaboud:2017dmy}. Besides, the probability of jet faking lepton can also be reduced with the cut on missing energy~\cite{CMS:2019see,Sirunyan:2017uyt,Mukhopadhyaya:2010qf,Mukhopadhyay:2011xs,Agarwalla:2018xpc}, i.e., Cut-5 below\footnote{Non-prompt leptons from jet faking can be distinguished from the prompt leptons in $W/Z$ decays with delicated isolation variables~\cite{Khachatryan:2016kod}.}. Therefore, we will only consider non-prompt leptons from heavy-flavor hadron decays at $pp$ colliders in this study.  
 
We generate parton-level signal and background events at $\sqrt{s}=100\tev$ using \texttt{MG5\_aMC@NLO v2.6.5}~\cite{Alwall:2014hca}, which are passed to \texttt{Pythia8}~\cite{Sjostrand:2014zea} for possible sequential decays, parton shower and hadronization. The default factorization and renormalization scales are used. The backgrounds $WZ$ and $t\bar{t}$ are matched upto two additional jets~\cite{Du:2018eaw}, $t\bar{t}t\bar{t}$, $t\bar{t}b\bar{b}$ and $t/\bar{t}Zq$ are generated without additional partons for simplicity, while the other backgrounds are matched to additional one jet.

The next-leading-order QCD overall $K$-factors of the background processes are available at the LHC colliding energy $\sqrt{s}=14\tev$ ranging from 1.2 to 2.0~\cite{Binoth:2009wk,Cascioli:2014yka,Grazzini:2016swo,Binoth:2008kt,
Nhung:2013jta,Campanario:2008yg,Ahrens:2011px,Campbell:2012dh,Garzelli:2012bn,
Lazopoulos:2008de,Bevilacqua:2012em,Bredenstein:2009aj}. As an estimate, we apply these $K$-factors to the corresponding processes at $\sqrt{s}=100\tev$~\cite{Alva:2014gxa}.
The detector response is simulated using \texttt{Delphes}~\cite{deFavereau:2013fsa} with the built-in baseline FCC-hh detector configuration. The probability of one $b$ quark to be identified as $b$-jet is $[1-p_T/(20~\text{TeV})]\cdot 85\%$ and the mis-taggig efficiencies for light-flavor quarks and $c$-quark wrongly identified as $b$-jets are $[1-p_T/(20~\text{TeV})]\cdot 1\%$ and $[1-p_T/(20~\text{TeV})]\cdot 5\%$ in the central region ($|\eta|<2.5$)~\cite{Jamin:2019mqx}.

In order to identify objects, we impose the following criteria~\cite{Kling:2018xud,Alva:2014gxa}
\begin{align}
\label{eq:criteria}
p_{T,e/\mu}>20\gev,\quad p_{T,j/b}>30\gev,\quad |\eta_{e/\mu/j/b}|<6,
\end{align}
where $j$ and $b$ denote the light-flavor jets and $b$-tagged jet, respectively. The lepton candidates are isolated within a cone of radius of 0.3, and the jet candidates are clustered with the anti-$k_t$ algorithm~\cite{Cacciari:2008gp} and a radius parameter of 0.4 implemented in the \texttt{FastJet} package~\cite{Cacciari:2011ma}.

\begin{table}[!htb]
\tabcolsep=12pt
\caption{The charge misidentification probabilities of backgrounds with $e^+e^+/e^+\mu^+/\mu^+\mu^+$ and one electron $e^-$ or two electrons $e^-e^-$. The same probabilities can be obtained for the charge-conjugated combinations. }
\begin{tabular}{|l|c|c|}
\hline
\hline
 & $e^-$ & $e^-e^-$ \\
\hline
$e^+e^+$& $\epsilon_e$ & 4$\epsilon_e$ \\ \hline
$e^+\mu^+$& $\epsilon_e$ & 3$\epsilon_e$ \\ \hline
$\mu^+\mu^+$& $\epsilon_e$ & 2$\epsilon_e$ \\ \hline
\hline
\end{tabular}
\label{tbl:probability}
\end{table}

Events are then selected with a series of cuts. It is demanded the angular separation between any two reconstructed objects satisfies\footnote{$\Delta\eta$ and $\Delta \phi$ denote the pseudo-rapidity and azimuthal angle difference between any two reconstructed objects.} $\Delta R\equiv\sqrt{(\Delta\eta)^2+(\Delta\phi)^2}>0.3$~\cite{Kling:2018xud} (Cut-1), which can help to reject leptons from the decay of a $b$-hadron or $c$-hadron~\cite{Aaboud:2017rzf}. Three or more charged leptons are required with the $p_T$ of the leading, sub-leading and sub-sub-leading leptons larger than 50~GeV, 35~GeV and 25~GeV, respectively and at least two of them have the same charge (Cut-2), where $\ell_1$. 
To reduce backgrounds from Drell-Yan processes and $Z$ boson decays, events with opposite-sign same-flavor lepton pairs or same-sign electron pairs with the invariant mass below 12~GeV or within the mass window of $15\gev$ around the $Z$ boson mass are rejected~\cite{CMS:2019nig,CMS:2019see} (Cut-3). For the signal processes, the final states can be $\ell^+\ell^+\ell^+\ell^-\ell^-(\ell^-/jj)E_T^{\text{miss}}$, $\ell^+\ell^+\ell^+\ell^-jj(\ell^-/jj)E_T^{\text{miss}}$, $\ell^+\ell^+\ell^+jjjj(\ell^-/jj)E_T^{\text{miss}}$, and the charge-conjugated ones. Therefore, we further impose the following selection cuts:
\begin{itemize}
\item exactly three same-sign leptons are required (Cut-4);
\item missing transverse momentum ${E}_T^{\text{miss}}>50\gev$ (Cut-5);
\item $b$-tagged jets are vetoed (Cut-6);
\end{itemize}
It is noted that experimental search for signals in final state with SS3L signature and at least one $b$-tagged jet has be performed~\cite{Aaboud:2017dmy}, which is typically different from our context.
Cuts on objects other than the three same-sign leptons can also be imposed. For example, one can require the sum of the residual lepton number and jet number to be larger than 2. In this paper, we however only consider Cut-1 to Cut-6 for an easier comparison with previous studies.

\begin{figure}[!htb]
\centering
\includegraphics[width=0.45\textwidth]{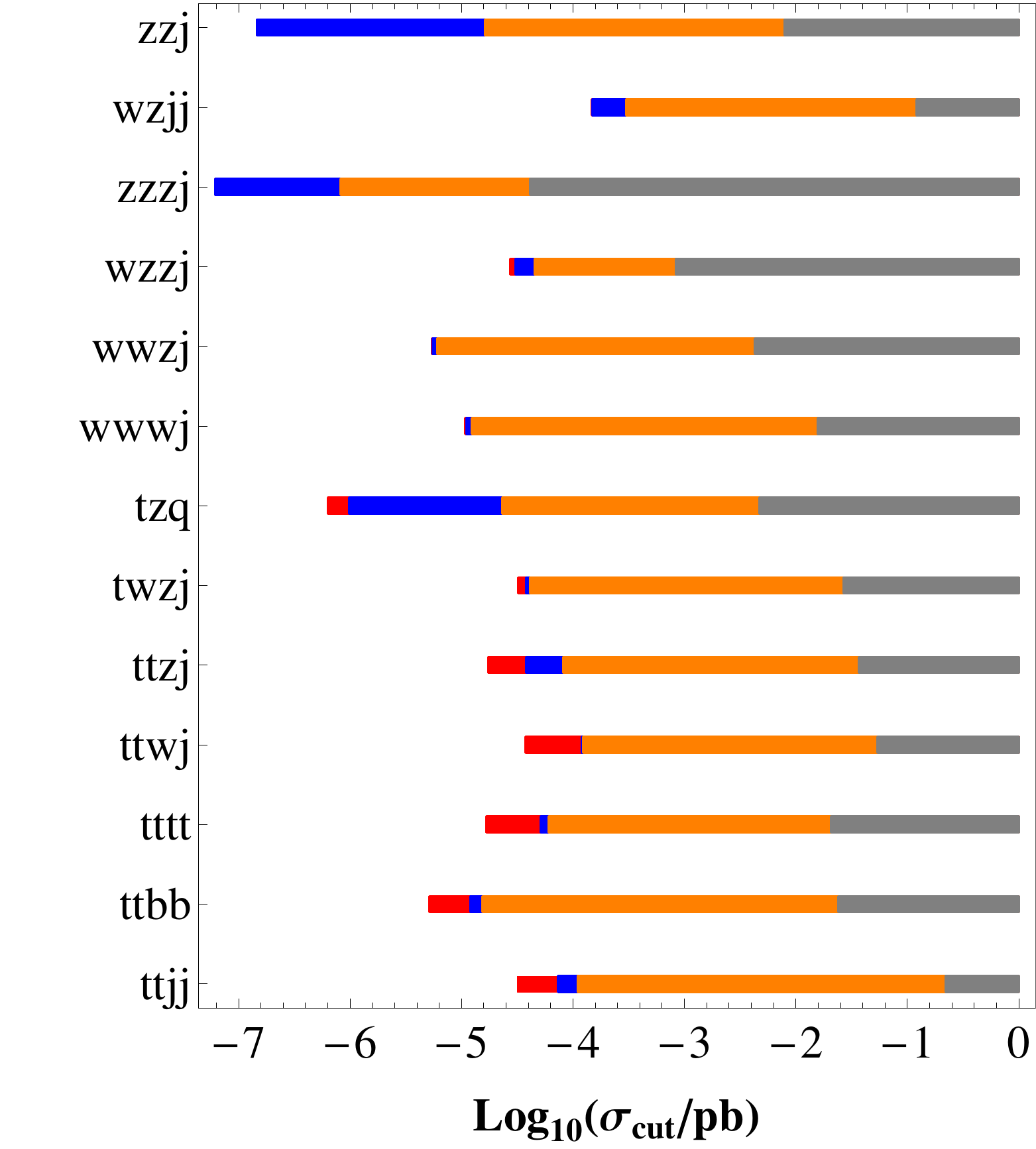}	
\caption{Cut flow of the background processes. The left endpoints of the gray, orange, blue and red bands correspond to the cross sections after Cut-3 to Cut-6, respectively.}
\label{fig:cuteff_bkg}
\end{figure}

It is straightforward to obtain the cut flow of cross sections after the selection cuts. In Fig.~\ref{fig:cuteff_bkg}, cross sections of the background processes after Cut-3 to Cut-6 are depicted, where the left endpoints of the gray, orange, blue and red bands correspond to the cross sections after Cut-3 to Cut-6, respectively. Assuming that the cross section after Cut-$i$ is $\sigma_{\text{cut}}^i$ and the corresponding cut efficiency is $\epsilon_i=\sigma_{\text{cut}}^{i}/\sigma_0$ with $\sigma_0$ being the background cross section before any cut, one obtains the relation
\begin{align}
\log_{10}\sigma_{\text{cut}}^i-\log_{10}\sigma_{\text{cut}}^{i-1}
&=\log_{10}\dfrac{\epsilon_i}{\epsilon_{i-1}}\;.
\end{align}
Therefore, the length of each colored band characterizes the cut efficiency of an individual cut. The total cross section of backgrounds $\sim 0.34\fb$ after selection cuts is dominated by $WZ$, $t\bar{t}W$, $t\bar{t}$, $t\bar{t}Z$, while the backgrounds $t\bar{t}t\bar{t}$ and $t\bar{t}b\bar{b}$ are less important~\cite{Mukhopadhyaya:2010qf}. We can see that the background $ZZ$ becomes negligible after imposing selection cut on the missing transverse momentum.

\begin{figure}[!htb]
\centering
\includegraphics[width=0.3\textwidth]{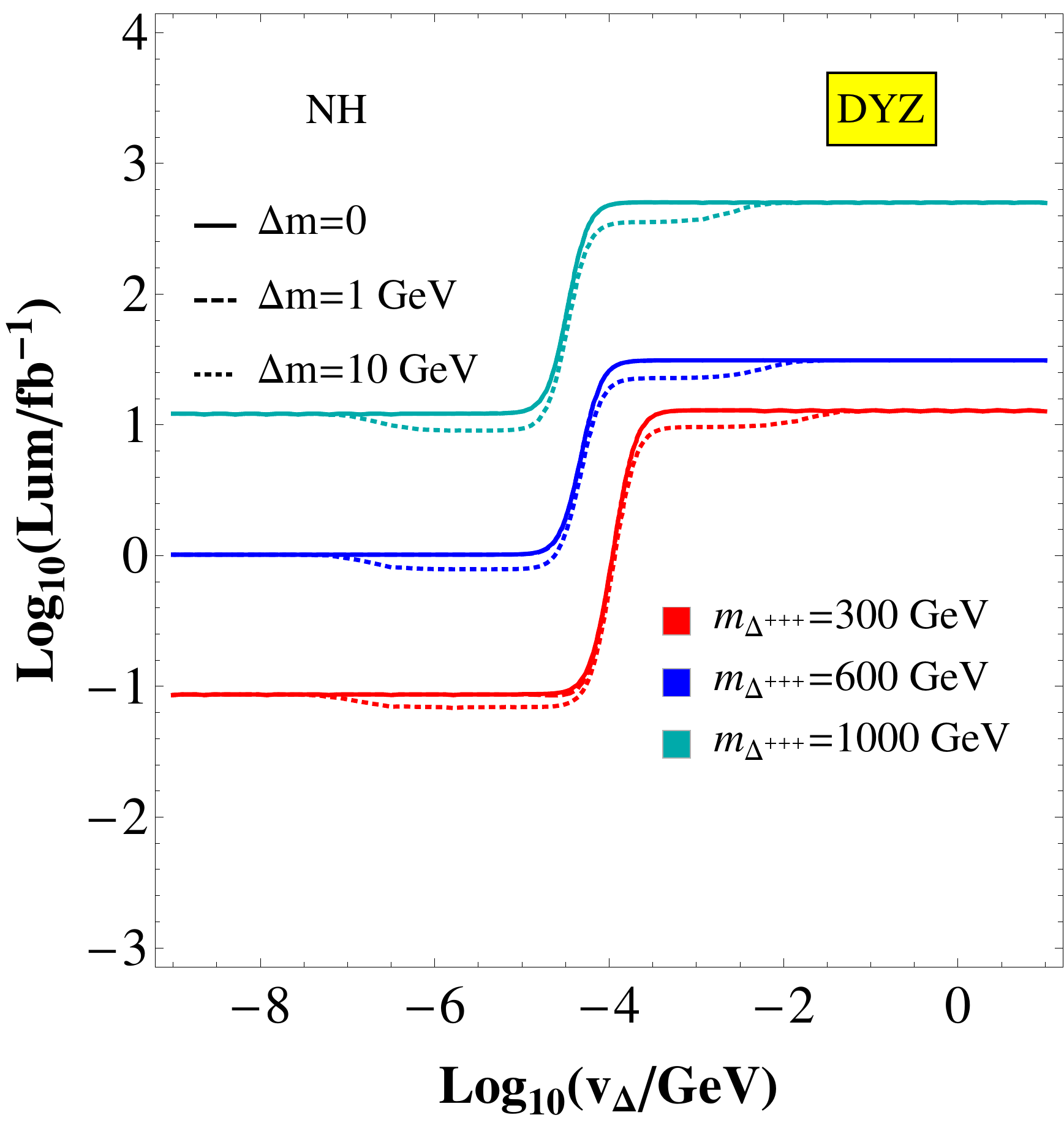}	
\includegraphics[width=0.3\textwidth]{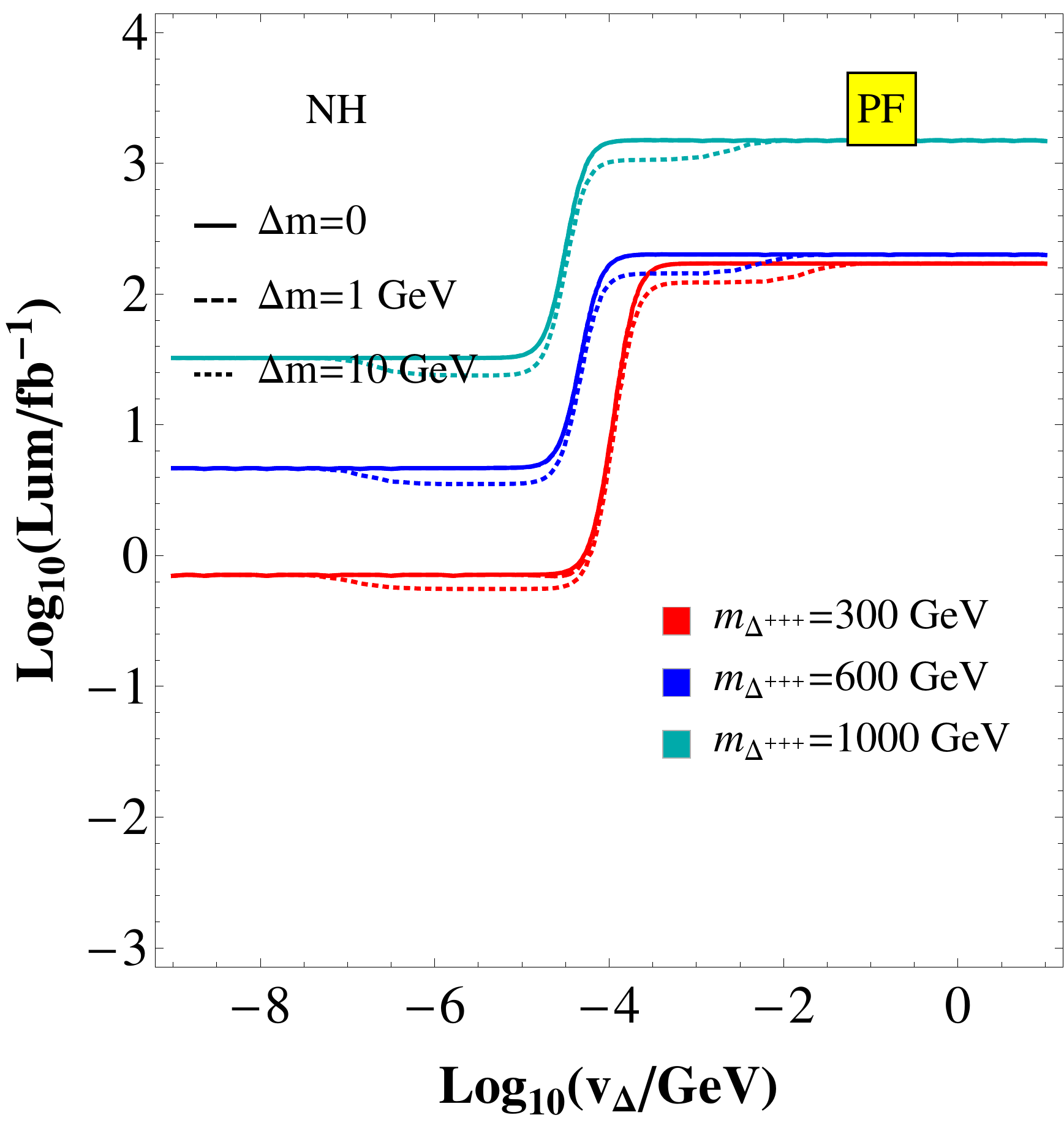}	
\includegraphics[width=0.3\textwidth]{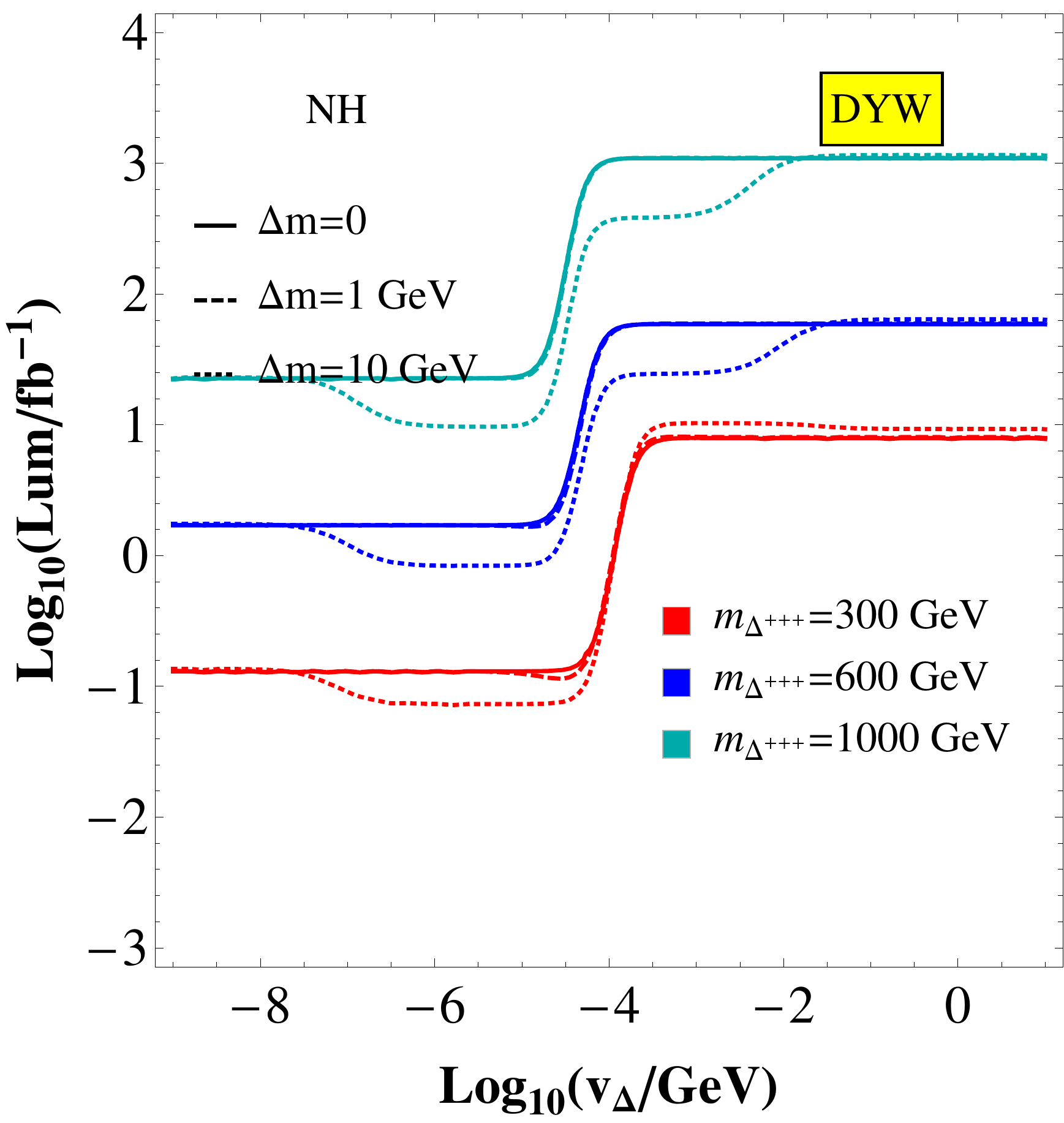}	
\caption{$5\sigma$ discovery prospects of searching for triply charged Higgs boson at a 100~TeV $pp$ collider with SS3L signature in the DYZ, PF and DYW processes. The benchmark scenarios with $\Delta m=0,1,10\gev$ and $m_{\Delta^{\pm\pm\pm}}=300,600,1000\gev$ for the NH are depicted in the plane of $\log_{10}(v_\Delta/\text{GeV})$ and $\log_{10}(\mathcal{L}/\text{fb}^{-1})$. }
\label{fig:lum-vD1}
\end{figure}

To evaluate the signal significance, we use
~\cite{Cowan:2010js}
\begin{align}
\label{eq:discovery}
\mathcal{Z}&=\sqrt{2\bigg [(n_s+n_b)\log\dfrac{n_s+n_b}{n_b}-n_s\bigg ]}\;,
\end{align}
where $n_s$ and $n_b$ denote the numbers of signal and background events after selection cuts. This formula is valid even for $n_b\ll n_s$~\cite{Cowan:2010js}. The discovery prospects of individual signal process are depicted in Fig.~\ref{fig:lum-vD1}. The solid, dashed and dotted curves correspond to the required integrated luminosities to reach $5\sigma$ discovery for $\Delta m=0$, 1~GeV and 10~GeV, respectively. Doubly and triply charged Higgs bosons decay into on-shell leptons and gauge bosons in the low and high $v_\Delta$ regions, respectively, alongside with a smooth transition in the medium $v_\Delta$ region due to the cascade decay of $\Delta^{\pm\pm}$. It is apparent that a larger integrated luminosity is required for $v_\Delta\gtrsim 10^{-3}\gev$ than for $v_\Delta\lesssim 10^{-5}\gev$ since in the latter case $\Delta^{\pm\pm(\pm)}$ mainly decays into $W$ bosons and the signal cross section is dissipated by the decays of $W$ bosons. We can see that with the integrated luminosity of about $0.1~(10)$, $1~(25)$, $10\fbi~(400\fbi)$, the triply charged Higgs boson with mass being 300, 600, 1000~GeV can be discovered in the DYZ processe for $v_{\Delta}\lesssim 10^{-5}\gev$ $(v_{\Delta}\gtrsim 10^{-3}\gev)$. Although the production cross section for the DYW process is slightly larger than the DYZ production cross section (cf. Fig.~\ref{fig:cross_section}), there are more combinations of decays in the DYZ process so that the integrated luminosities required to reach $5\sigma$ discovery in the DYW process are larger except for $m_{\Delta^{\pm\pm\pm}}=300\gev$ and $v_\Delta\gtrsim 10^{-3}\gev$ as a result of more dramatic phase suppression from the decay $\Delta^{\pm\pm\pm}\to W^\pm W^\pm W^\pm$ in the DYZ process.
The sensitivity in the PF process is the lowest, which is limited by its small production cross section at $\sqrt{s}=100\tev$ as shown in Fig.~\ref{fig:cross_section}. However, since the PF process is composed of $t$-channel sub-processes~\cite{Ghosh:2017jbw}, the production cross section is less suppressed with the increase of the triply charged Higgs boson mass as compared to the DY processes. Similarly, the PF cross section for $m_{\Delta^{\pm\pm\pm}}=300\gev$ and $v_\Delta\gtrsim 10^{-3}\gev$  is suppressed due to the phase suppression. Consequently, the sensitivity required to reach $5\sigma$ discovery for $m_{\Delta^{\pm\pm\pm}}=300\gev$ is close to that for $m_{\Delta^{\pm\pm\pm}}=600\gev$. 

From the right panel of Fig.~\ref{fig:lum-vD1}, we can also find that the DYW process is more sensitive to the cascade decay $\Delta^{\pm\pm}\to\Delta^{\pm\pm\pm}W^{\mp*}$ as compared to the DYZ and PF processes.
For the DYZ and PF processes, the production of both $\Delta^{+++}\Delta^{---}$ and $\Delta^{++}\Delta^{--}$ with the decays $\Delta^{\pm\pm}\to \Delta^{\pm\pm\pm}W^{\mp*}$ and $\Delta^{\pm\pm\pm}\to \ell^\pm \ell^\pm W^\pm$, $W^\pm W^\pm W^\pm$ are considered. Since the production cross section of $\Delta^{++}\Delta^{--}$ are about $20\%$ of $\Delta^{+++}\Delta^{---}$, the required luminosities to reach $5\sigma$ discovery  are lowered slightly for $\Delta m=10\gev$ as compared to that for $\Delta m=0$ in the medium $v_\Delta$ region. For the DYW process, we consider the production of $\Delta^{\pm\pm\pm}\Delta^{\mp\mp}$ with the decays $\Delta^{\pm\pm}\to \Delta^{\pm\pm\pm}W^{\mp*},\ell^\pm\ell^\pm, W^\pm W^\pm$ and $\Delta^{\pm\pm\pm}\to \ell^\pm \ell^\pm W^\pm$, $W^\pm W^\pm W^\pm$. In the medium $v_\Delta$ region the sensitivity is remarkably improved due to more combinations of decays, except for $m_{\Delta^{\pm\pm\pm}}=300\gev$ and $v_{\Delta}\gtrsim 10^{-3}\gev$ when the decay $\Delta^{\pm\pm\pm}\to W^\pm W^\pm W^\pm$ is suppressed kinematically.

\begin{figure}[!htb]
\centering
\includegraphics[width=0.35\textwidth]{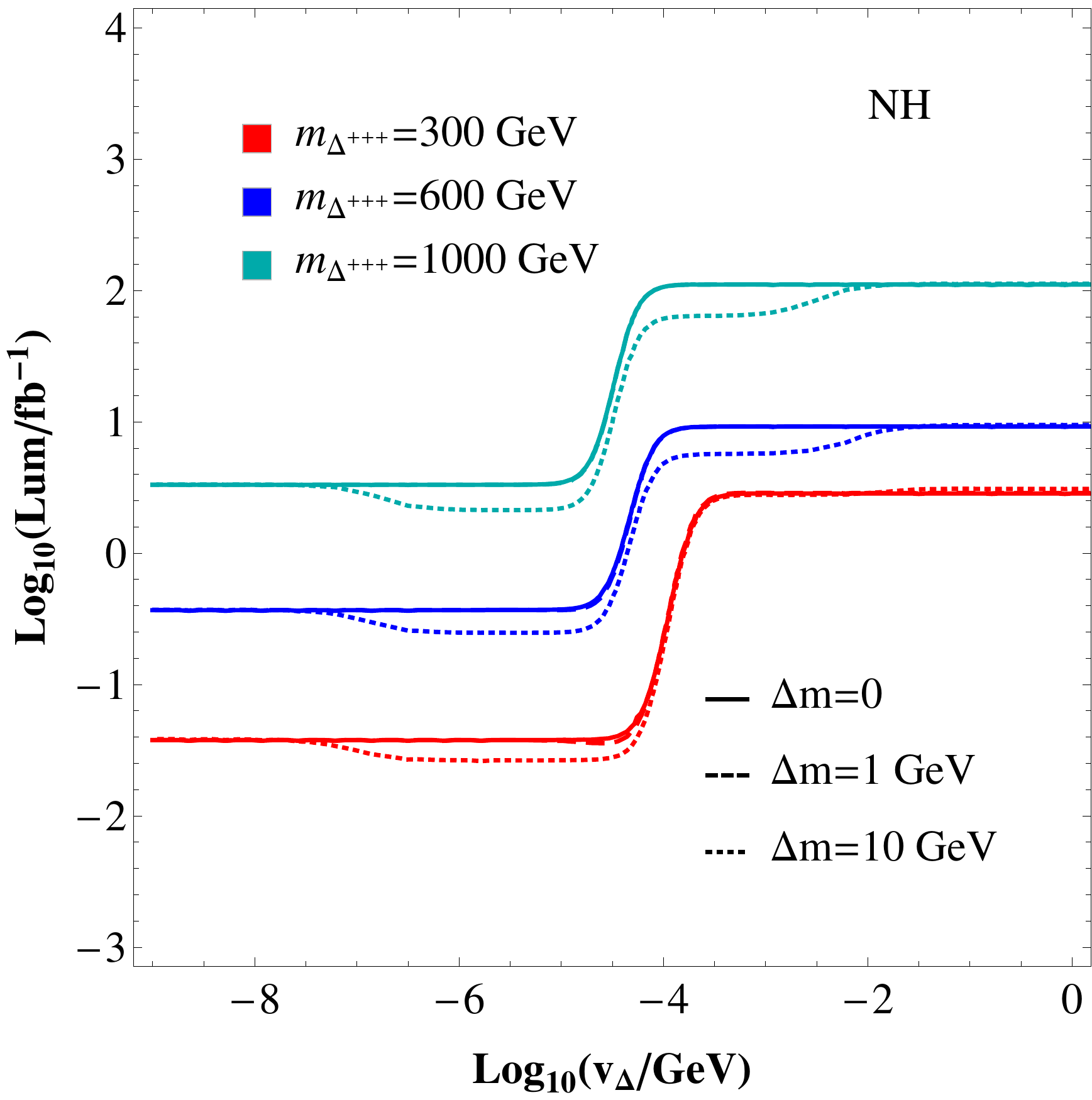}	
\includegraphics[width=0.35\textwidth]{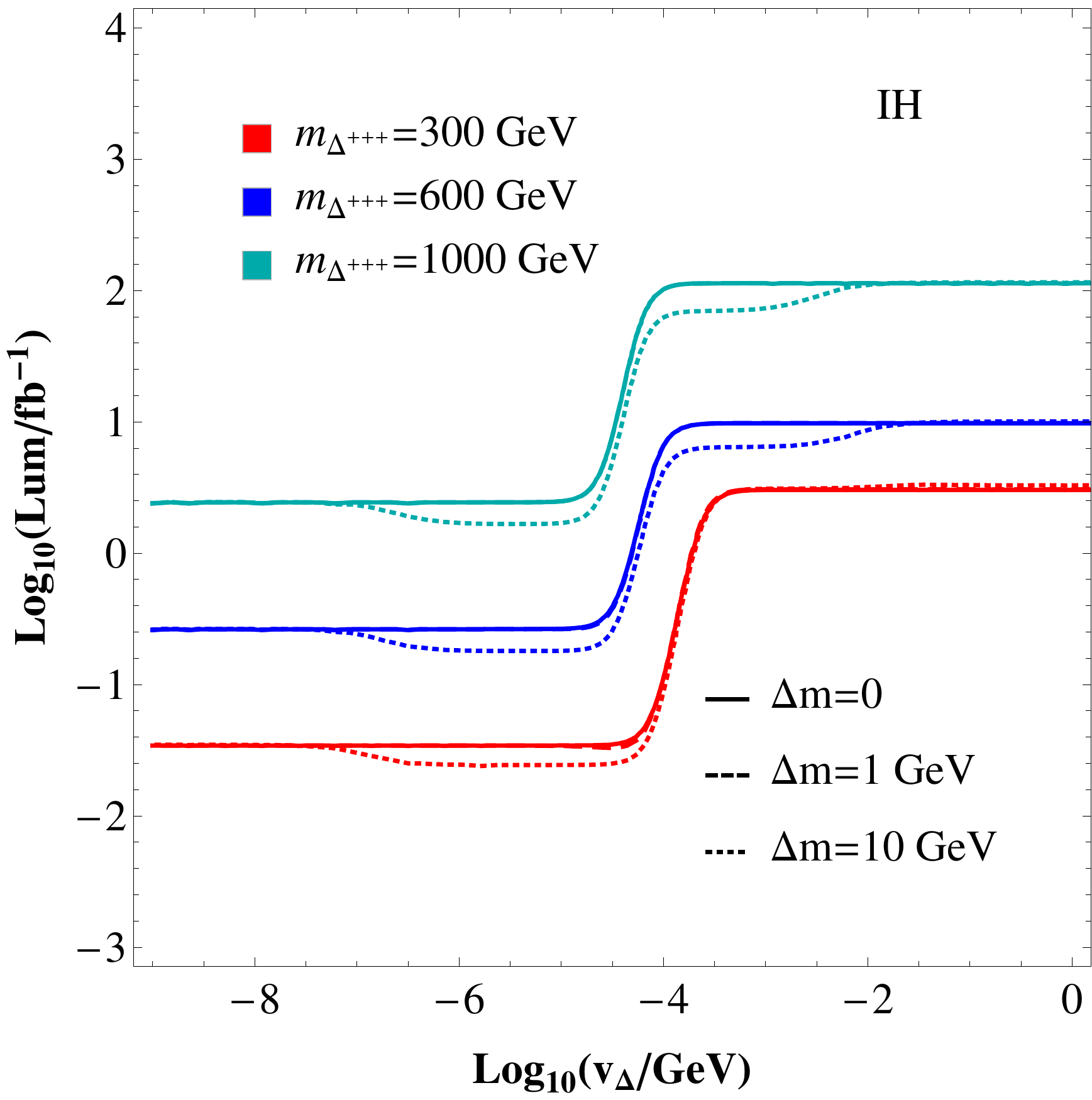}	
\caption{$5\sigma$ discovery prospects of searching for triply charged Higgs boson at a 100~TeV $pp$ collider with SS3L signature. The benchmark scenarios with $\Delta m=0,1,10\gev$ and $m_{\Delta^{\pm\pm\pm}}=300,600,1000\gev$ for the NH and IH are depicted in the plane of $\log_{10}(v_\Delta/\text{GeV})$ and $\log_{10}(\mathcal{L}/\text{fb}^{-1})$.}
\label{fig:lum-vD2}
\end{figure}

The discovery prospects after combining the signals in the DYW, DYZ and PF processes are shown in Fig.~\ref{fig:lum-vD2}. With the integrated luminosity of about $100\fbi$, the triply charged Higgs boson with mass below 1000~GeV can be discovered. Besides, the required integrated luminosity to reach $5\sigma$ in the region of $v_{\Delta}\lesssim 10^{-5}\gev$ for the IH is smaller than that for the NH, since the coupling of $\Delta^{\pm\pm}$ to the electron pair for the IH is larger. 
For $v_\Delta \gtrsim 10^{-3}\gev$, the sensitivities for the NH and IH are are the same since $\Delta^{\pm\pm(\pm)}$ mainly decays into $W^\pm W^\pm (W^\pm)$, which is independent of the neutrino mass hierarchy. Below, we will concentrate on the sensitivities for the NH.

\begin{figure}[!htb]
\centering
\includegraphics[width=0.35\textwidth]{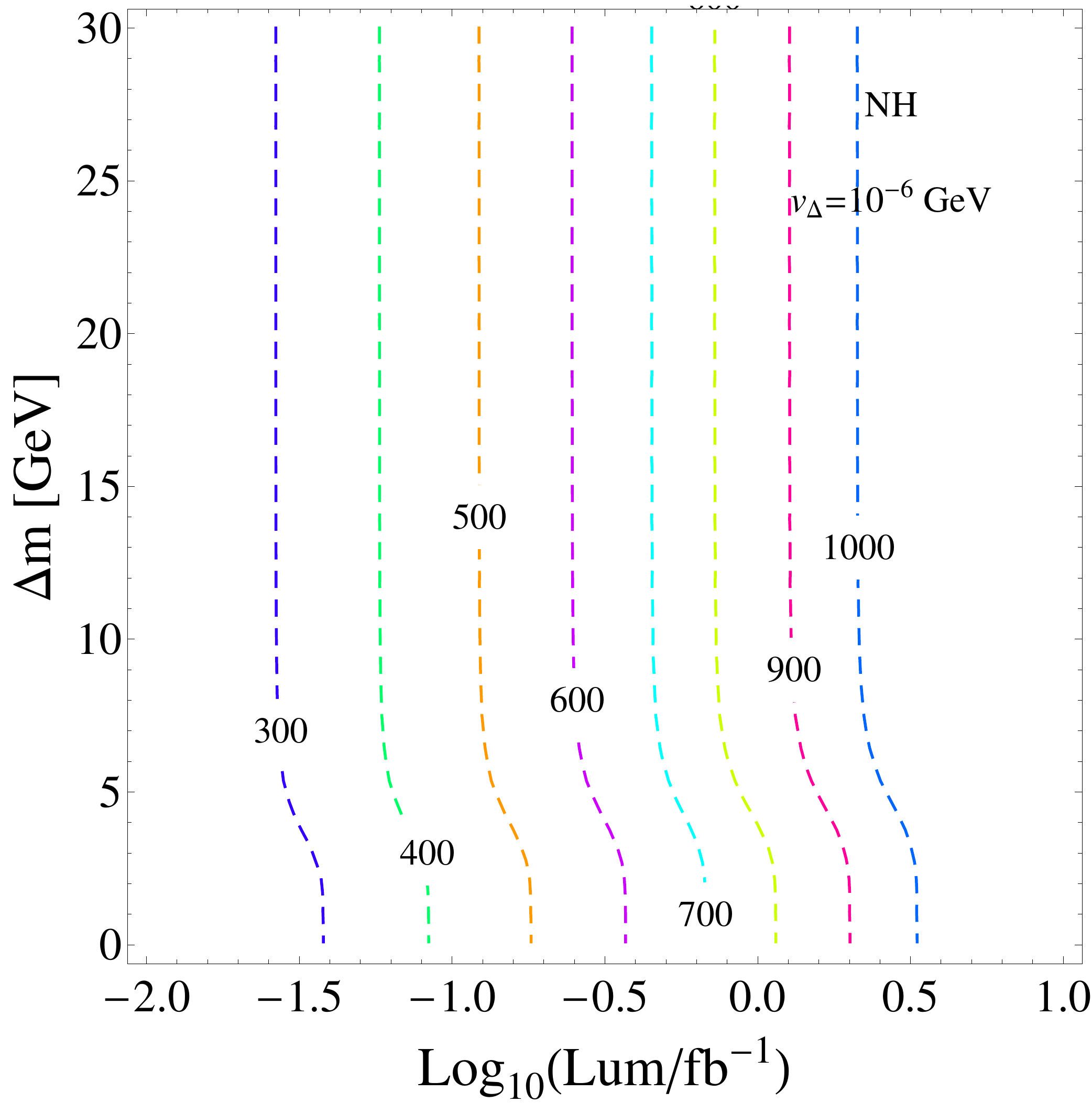}	
\includegraphics[width=0.35\textwidth]{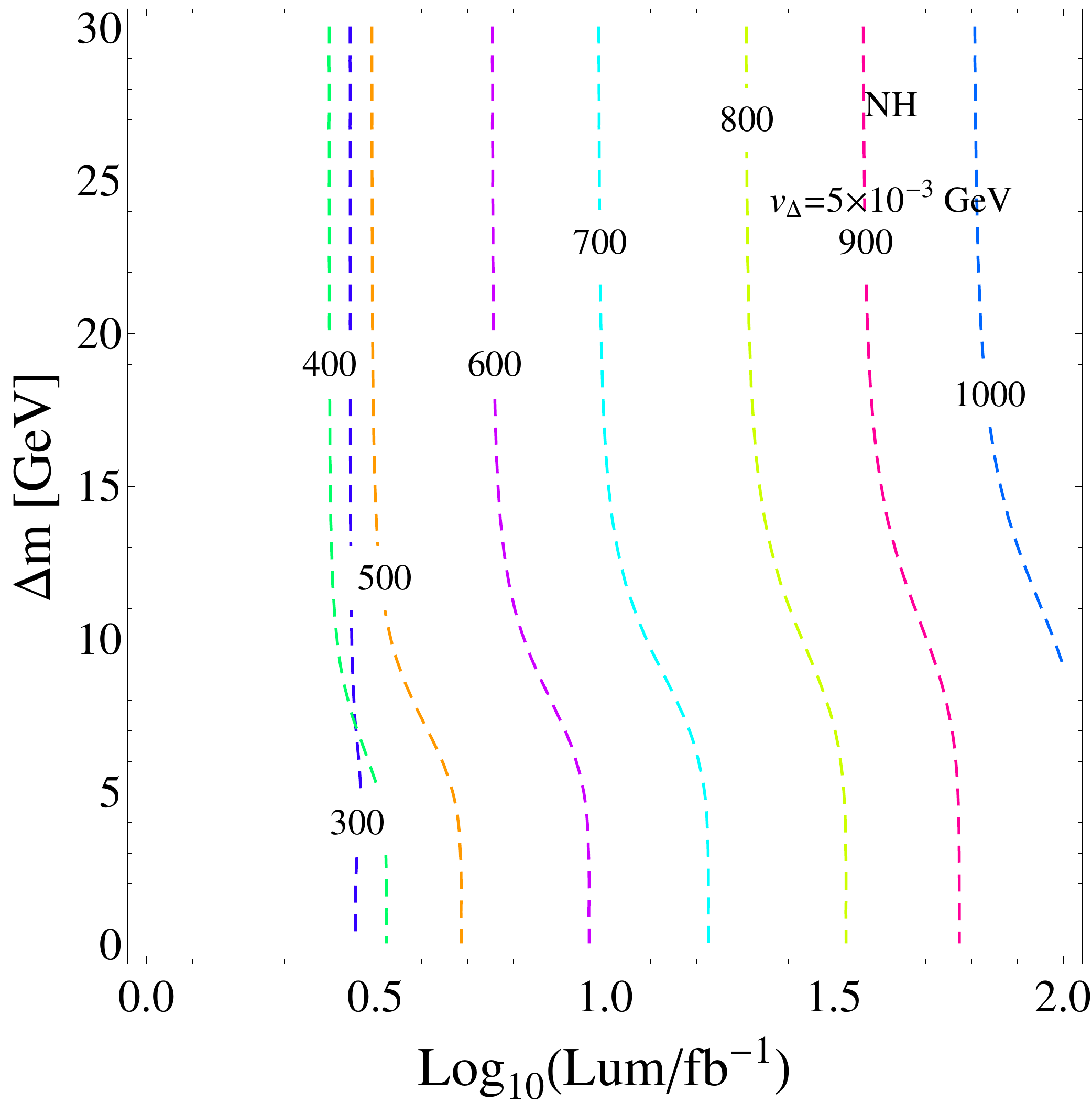}	
\caption{$5\sigma$ discovery reach in the plane of $\log_{10}(\mathcal{L}/\text{fb}^{-1})$ and $\Delta m$ with two benchmark values of $v_{\Delta}=10^{-6}\gev$ and $5\times 10^{-3}\gev$ in the left and right panels, respectively. The $5\sigma$ contours are labelled by the mass $m_{\Delta^{\pm\pm\pm}}$ in units of GeV.}
\label{fig:lum-dm}
\end{figure}

\begin{table}[!htb]
\tabcolsep=5pt
\caption{Cross sections (in units of pb) of the backgrounds at 100/13 TeV before cuts and after cuts, which are denoted as $\sigma_0({100/13\tev})$ and $\sigma_{\text{cut}}({100/13\tev})$, respectively. The $K$-factors for each process are listed in the second column. The notation of ``aE$\pm$0b'' stands for $a\times 10^{\pm b}$.}
\begin{tabular}{c|c|c|c|c|c}
\hline
\hline
background & $K$-factor & $\sigma_0({100\tev})$& $\sigma_{\text{cut}}({100\tev})$ & $\sigma_0({13\tev})$ & $\sigma_{\text{cut}}({13\tev})$ \\ \hline
ttjj	&	1.5	&	1.07E+03	&	3.15E-05	&	6.58E+01	&	1.94E-06	\\
ttbb	&	1.77	&	1.07E+02	&	4.86E-06	&	1.21E+00	&	5.49E-08	\\
tttt	&	1.21	&	2.39E-01	&	1.36E-05	&	7.20E-04	&	4.09E-08	\\
ttwj	&	1.28	&	8.41E-01	&	9.04E-05	&	2.23E-02	&	2.40E-06	\\
ttzj	&	1.35	&	1.05E+00	&	1.22E-05	&	1.41E-02	&	1.65E-07	\\
twzj	&	1.45	&	7.06E-01	&	1.55E-05	&	1.44E-02	&	3.16E-07	\\
tzq	&	1.1	&	4.88E-01	&	6.15E-07	&	1.13E-02	&	1.43E-08	\\
wwwj	&	1.74	&	1.55E-01	&	1.20E-05	&	7.15E-03	&	5.53E-07	\\
wwzj	&	1.98	&	7.27E-02	&	2.87E-06	&	2.88E-03	&	1.14E-07	\\
wzzj	&	1.96	&	3.28E-02	&	1.73E-05	&	1.17E-03	&	6.20E-07	\\
zzzj	&	1.58	&	2.02E-03	&	3.93E-08	&	1.18E-04	&	2.30E-09	\\
wzjj	&	1.83	&	7.23E+00	&	1.34E-04	&	4.94E-01	&	9.14E-06	\\
zzj	&	1.47	&	5.19E-01	&	1.23E-07	&	4.60E-02	&	1.09E-08	\\
\hline
\hline
\end{tabular}
\label{tbl:xsection}
\end{table}

To illustrate the dependence of $5\sigma$ contours on the mass splitting $\Delta m$, we show the $5\sigma$ discovery reach in the plane of $\log_{10}(\mathcal{L}/\text{fb}^{-1})$ and $\Delta m$ in Fig.~\ref{fig:lum-dm} with two benchmark values $v_{\Delta}=10^{-6}\gev$ and $5\times 10^{-3}\gev$, which ensure $\text{Br}(\Delta^{\pm\pm\pm}\to \ell^\pm \ell^\pm W^\pm)=1$ and $\text{Br}(\Delta^{\pm\pm\pm}\to W^\pm W^\pm W^\pm)=1$ for $\Delta m\geq 0$, respectively. Moreover, from the right panel of Fig.~\ref{fig:length_Deltappp}, the proper decay lengths for $v_{\Delta}=10^{-6}\gev$ and $5\times 10^{-3}\gev$ are both larger than 0.1~mm, which ensures the validity of prompt search. 
The integrated luminosities to reach $5\sigma$ discovery decreases with $\Delta m$ for $0<\Delta m\lesssim 10\gev ~(15\gev)$ for $v_{\Delta}=10^{-6}\gev~(5\times 10^{-3}\gev)$ as shown in Fig.~\ref{fig:lum-dm}. It is notable that for the production cross sections of charged Higgs bosons (cf. Fig.~\ref{fig:cross_section}) we have always set $\Delta m=0$ for simplicity. This can increase the DYW production cross section by 5\%-15\% at most depending on the value of $\Delta m$ and $m_{\Delta^{\pm\pm\pm}}$, and overestimates the signal cross section by several percent for $10\gev\lesssim \Delta m\lesssim 30\gev$.

Finally, the sensitivities at the FCC-hh and the LHC are compared. The latter one has been investigated in Refs.~\cite{Ghosh:2018drw,Ghosh:2017jbw} with the above benchmark values of $v_\Delta$ being chosen. Different from the significance formula in Eq.~\eqref{eq:discovery}, they used $n_s/\sqrt{n_s+n_b}$ to quantify the significance and found that at $5\sigma$ level $m_{\Delta^{\pm\pm\pm}}\lesssim 950\gev$ can be reached at the LHC with the integrated luminosity of $3\abi$ for $v_\Delta=10^{-6}\gev$, while it is reduced to $m_{\Delta^{\pm\pm\pm}}\lesssim 600\gev$ for $v_\Delta=5\times 10^{-3}\gev$. However, it is known~\cite{Cowan:2010js} that $n_s/\sqrt{n_s+n_b}$ is a good approximation of the significance $\mathcal{Z}$ in Eq.~\eqref{eq:discovery} if $n_s\ll n_b$. In our case, $n_s$ and $n_b$ can be comparable. As a result, we find that $n_s/\sqrt{n_s+n_b}$ underestimates the significance by several times. It is striking that the integrated luminosities required to reach $\mathcal{Z}=5$ can be smaller than that with $n_s/\sqrt{n_s+n_b}=5$ by one or two orders. To be more concrete, in Tab.~III of Ref.~\cite{Ghosh:2018drw}, the signal cross section with $(m_{\Delta^{\pm\pm\pm}},\Delta m, v_\Delta)=(400\gev,0,10^{-6}\gev)$ for the NH is $1.19\times 10^{-3}\pb$ and the total background cross section is $1.21\times 10^{-3}\pb$. The integrated luminosities required to reach $\mathcal{Z}=5$ and $n_s/\sqrt{n_s+n_b}=5$ are $3.2\fbi$ and $21.2\fbi$, respectively. From the left panel of Fig.~22 of Ref.~\cite{Ghosh:2018drw}, $n_s/\sqrt{n_s+n_b}=5$ is reached for $m_{\Delta^{\pm\pm\pm}}\simeq 600\gev$ and the integrated luminosity $\mathcal{L}=3\abi$, we can thus infer that the signal cross section is $2.66\times 10^{-4}\pb$ and the significance value $\mathcal{Z}=32.5$ with the integrated luminosity of $3\abi$. Conversely, $\mathcal{Z}=5$ can be reached with only $23.7\fbi$ of data.

\begin{figure}[!htb]
\centering
\includegraphics[width=0.3\textwidth]{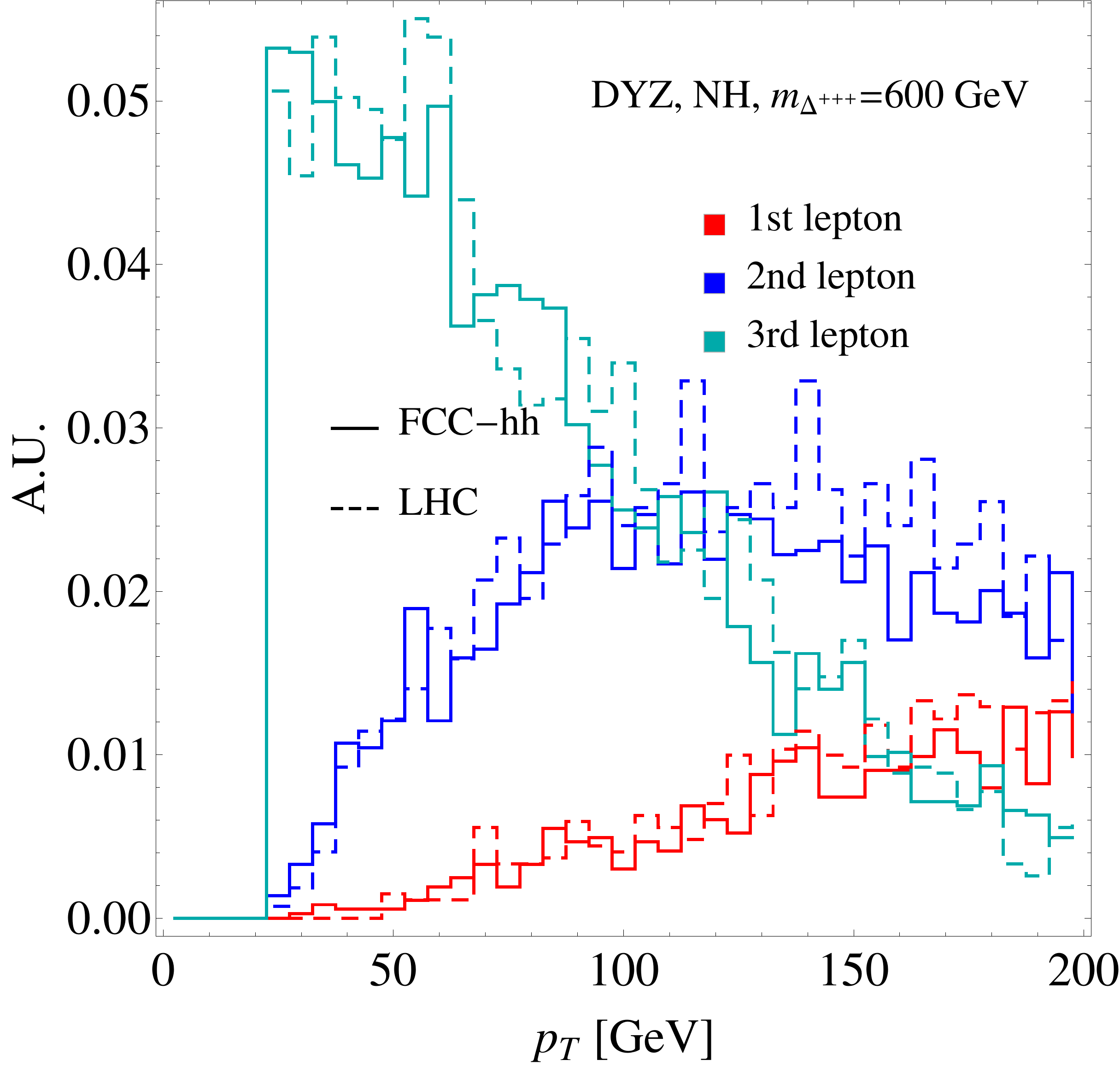}	
\includegraphics[width=0.3\textwidth]{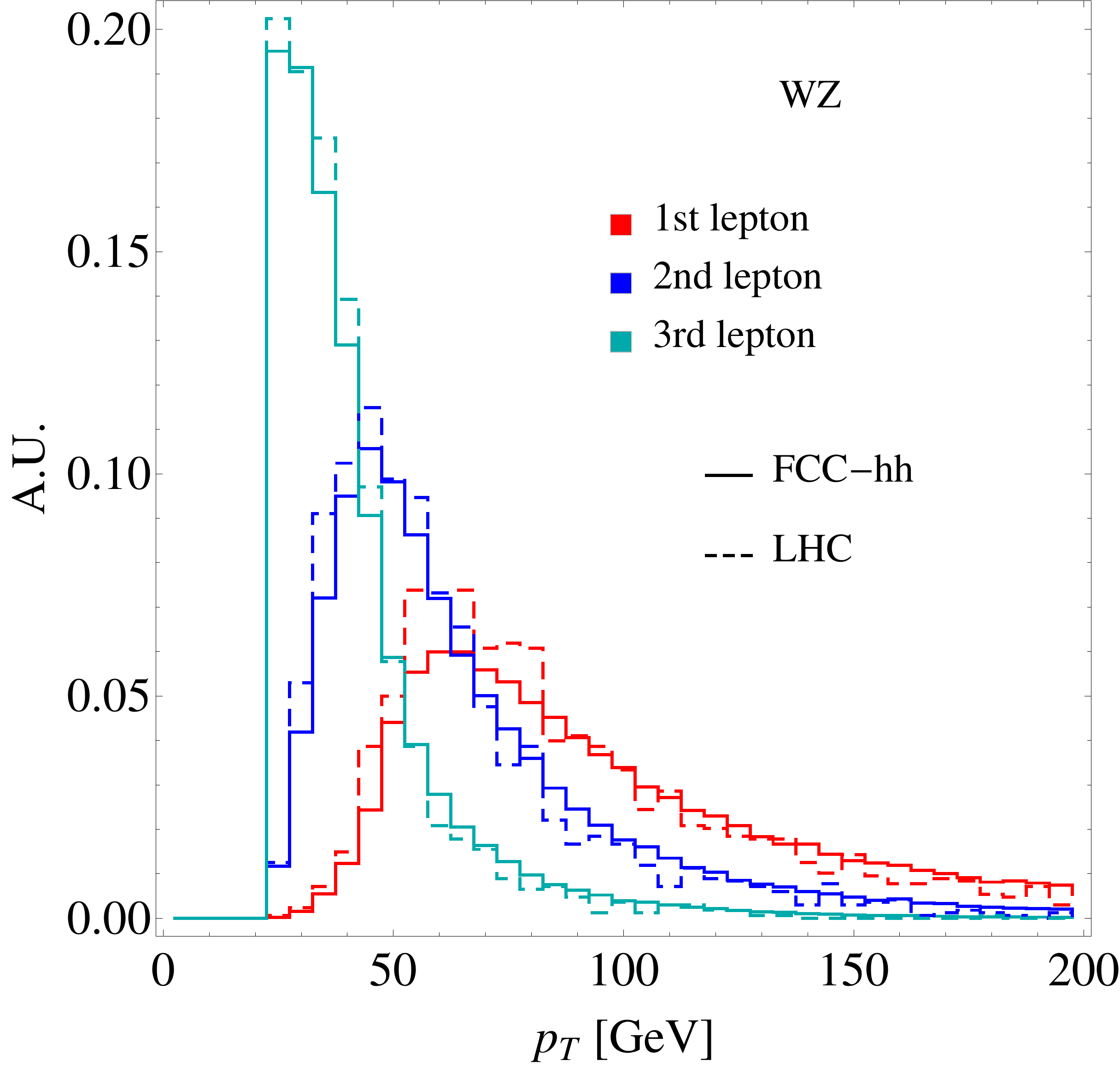}	\\
\includegraphics[width=0.3\textwidth]{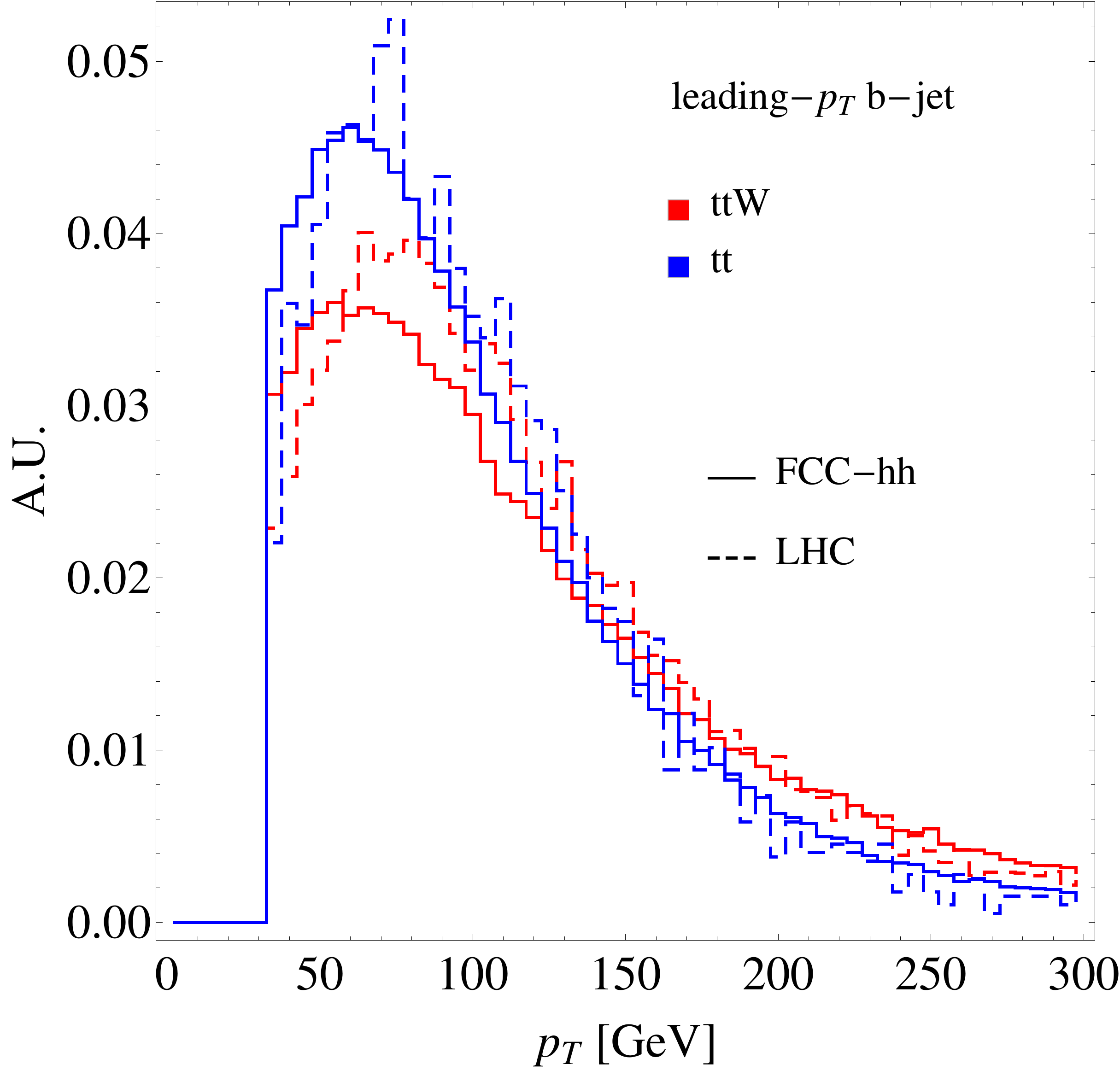}
\includegraphics[width=0.3\textwidth]{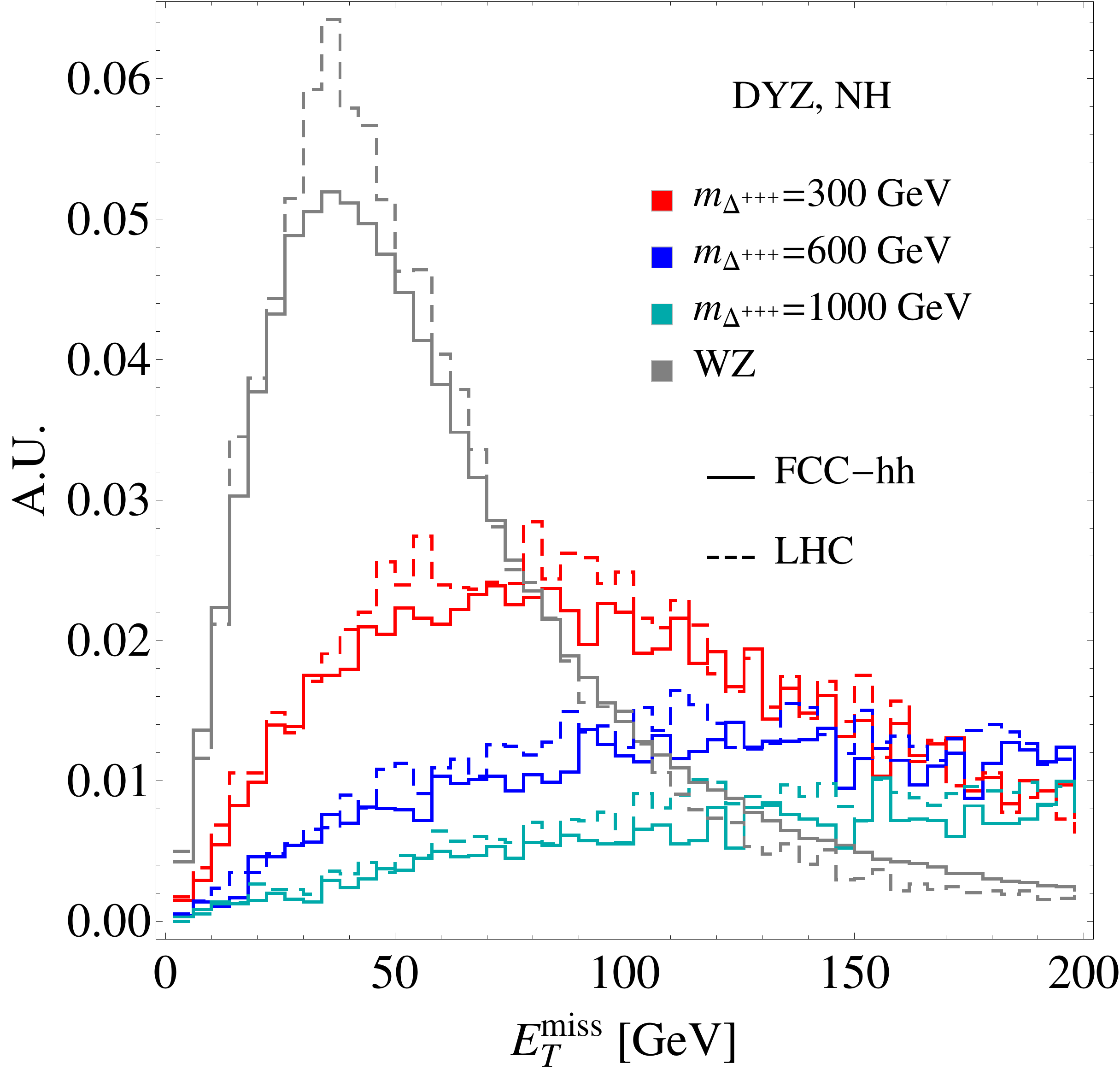}	
\caption{The normalized distributions with an arbitrary unit (A.U.) of $p_T$ and ${E}_T^{\text{miss}}$. Upper left: $p_T$ for leptons in the DYZ process for the NH and $m_{\Delta^{\pm\pm\pm}}=600\gev$; upper right: $p_T$ for leptons in the background $WZ$ process; lower left: $p_T$ for leading-$p_T$ $b$-jet in the backgrounds $ttW$ and $t\bar{t}$ processes; lower right: $E_T^{\text{miss}}$ in the DYZ process for the NH and $m_{\Delta^{\pm\pm\pm}}=600\gev$ and in the background $WZ$ process. Leptons ordered by $p_T$ are denoted by 1st, 2nd and 3rd ones.}
\label{fig:kin_leptons}
\end{figure}

\begin{figure}[!htb]
\centering	
\includegraphics[width=0.35\textwidth]{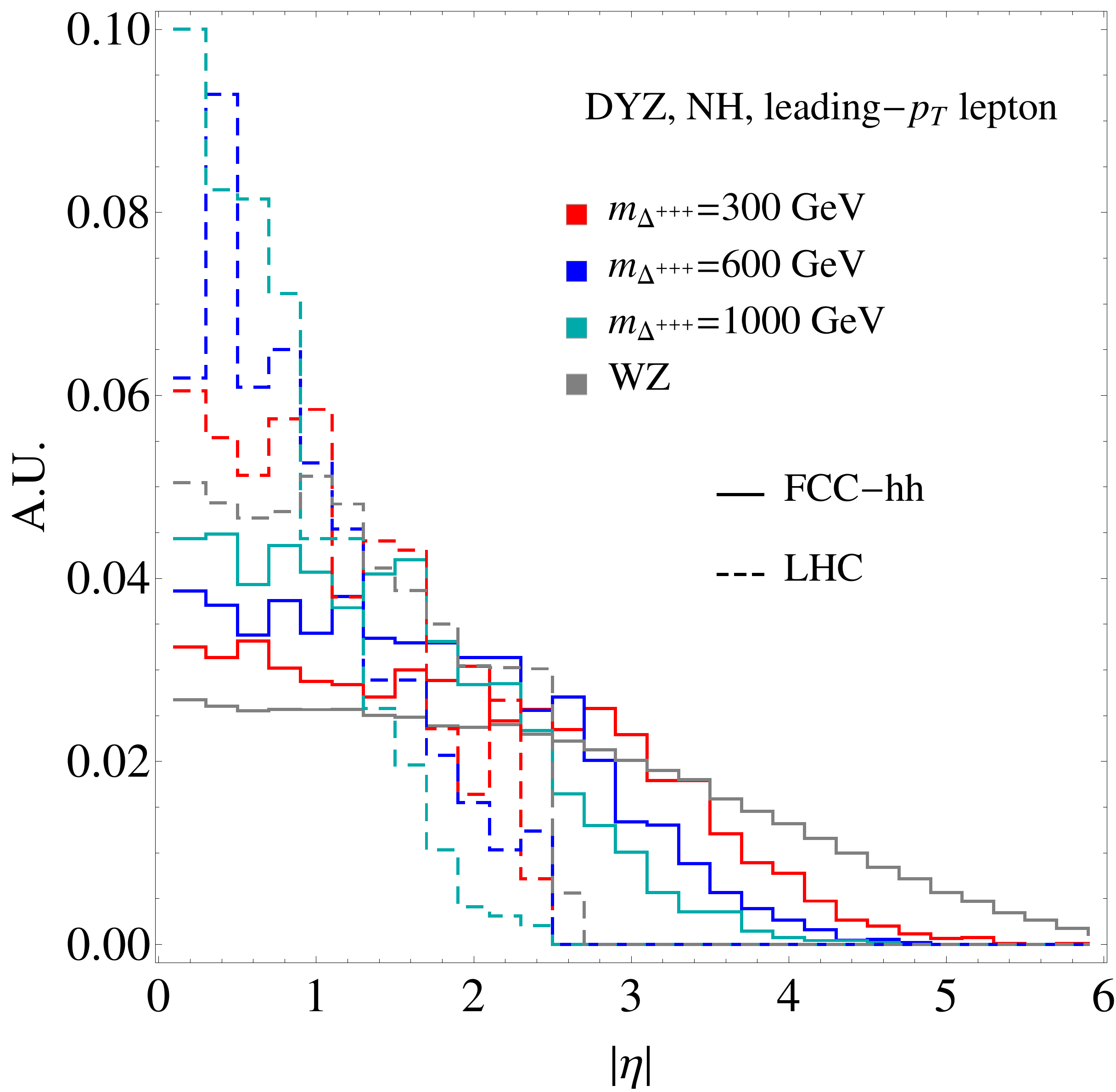}	
\includegraphics[width=0.35\textwidth]{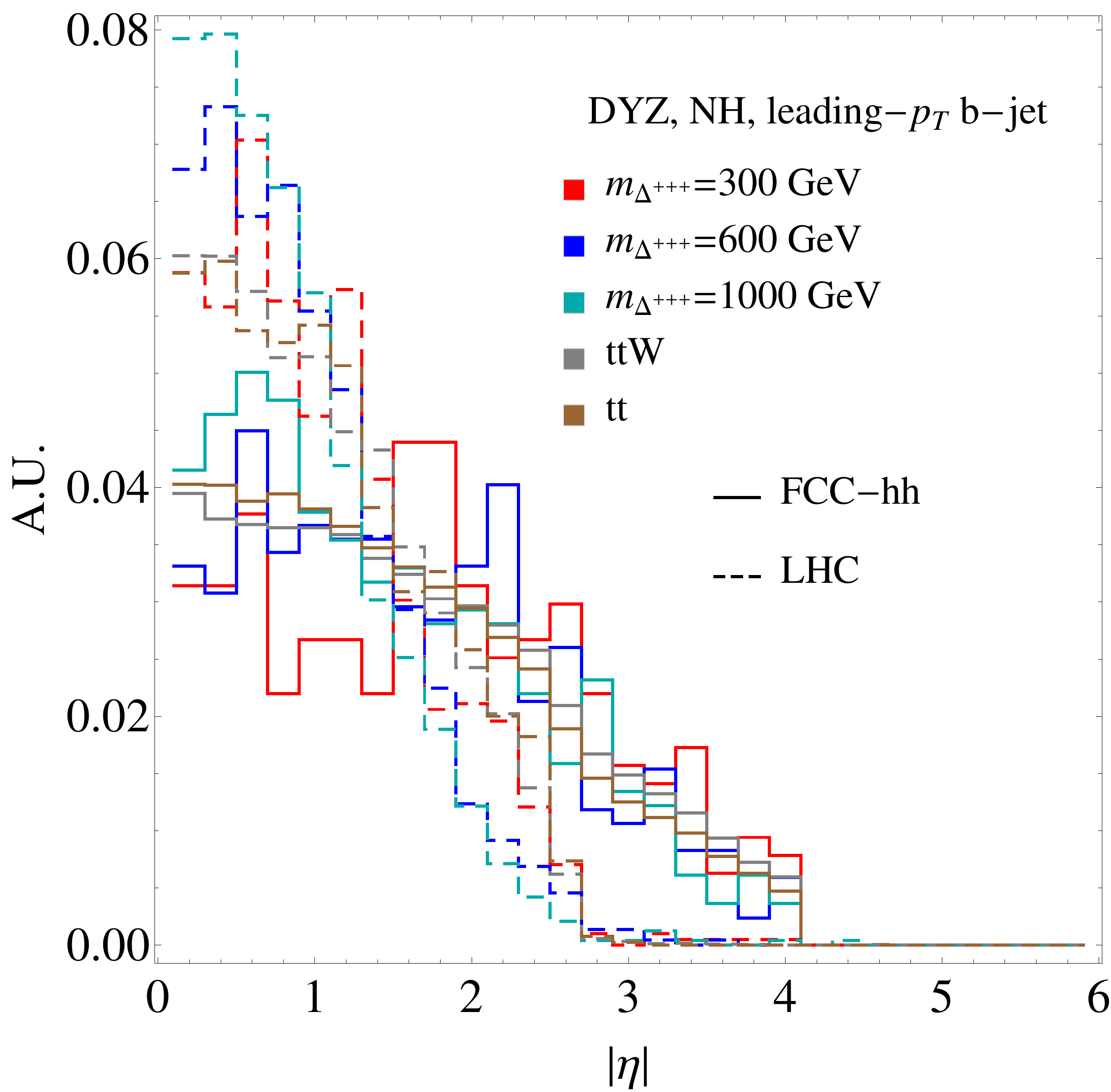}	
\caption{The normalized distributions of the rapidity $|\eta|$ in the DYZ process for the NH and $m_{\Delta^{\pm\pm\pm}}=300$, $600$ and $1000\gev$ and backgrounds. $|\eta|$ for the leading $p_T$ lepton and $b$-jet for the left and right panels, respectively. }
\label{fig:kin_jet}
\end{figure}

Ref.~\cite{Ghosh:2018drw} present the discovery prospects of $n_s/\sqrt{n_s+n_b}$=5 with the integrated luminosities of $100\fbi$ and $3\abi$, which are unable to be converted into the discovery prospects of $\mathcal{Z}=5$ with varying $\Delta m$ and $m_{\Delta^{\pm\pm\pm}}$. Therefore, we will not use their $5\sigma$ curves. Furthermore, we find that it is feasible to obtain the sensitivity at the LHC by projecting the result at the FCC-hh, which is obtained by the delicate detector simulation shown above. To verify the validity, we depict the kinematic distributions of the signals and backgrounds at the 13~TeV LHC and FCC-hh in Figs.~\ref{fig:kin_leptons} and \ref{fig:kin_jet}. Figure~\ref{fig:kin_leptons} displays the distributions of $p_T$ for leptons and leading-$p_T$ $b$-jet and ${E}_T^{\text{miss}}$. One can  see that these distributions at the 13~TeV LHC and FCC-hh are close to each other. The most notable difference at these two colliders comes from the rapidity distributions~\cite{Alva:2014gxa}, which are shown in Fig.~\ref{fig:kin_jet}. The leptons and $b$-jets tend to have a larger rapidity at the FCC-hh than that at the LHC. The cut efficiencies mainly depend on the $p_T$ and $\eta$ of leptons and $E_T^{\text{miss}}$ for Cut-1 to Cut-5. For Cut-6, the veto of $b$-tagged jets depends on the $b$-tagging efficiency. Although the recommended $b$-tagging efficiency at the LHC by the CMS Collaboration~\cite{Chatrchyan:2012jua} is lower than that at the FCC-hh, this does not have large impact since the most dominant background is $WZ$. Therefore, if we impose the same cuts\footnote{From Fig.~\ref{fig:kin_jet}, imposing the cuts $|\eta_{e/\mu/b}|<2.5$ and $|\eta_{e/\mu/b}|<6$ does not make much difference at the LHC.} at the LHC as that at the FCC-hh, the cut efficiencies at these two colliders are expected to be roughly the same.

\begin{figure}[!htb]
\centering
\includegraphics[width=0.35\textwidth]{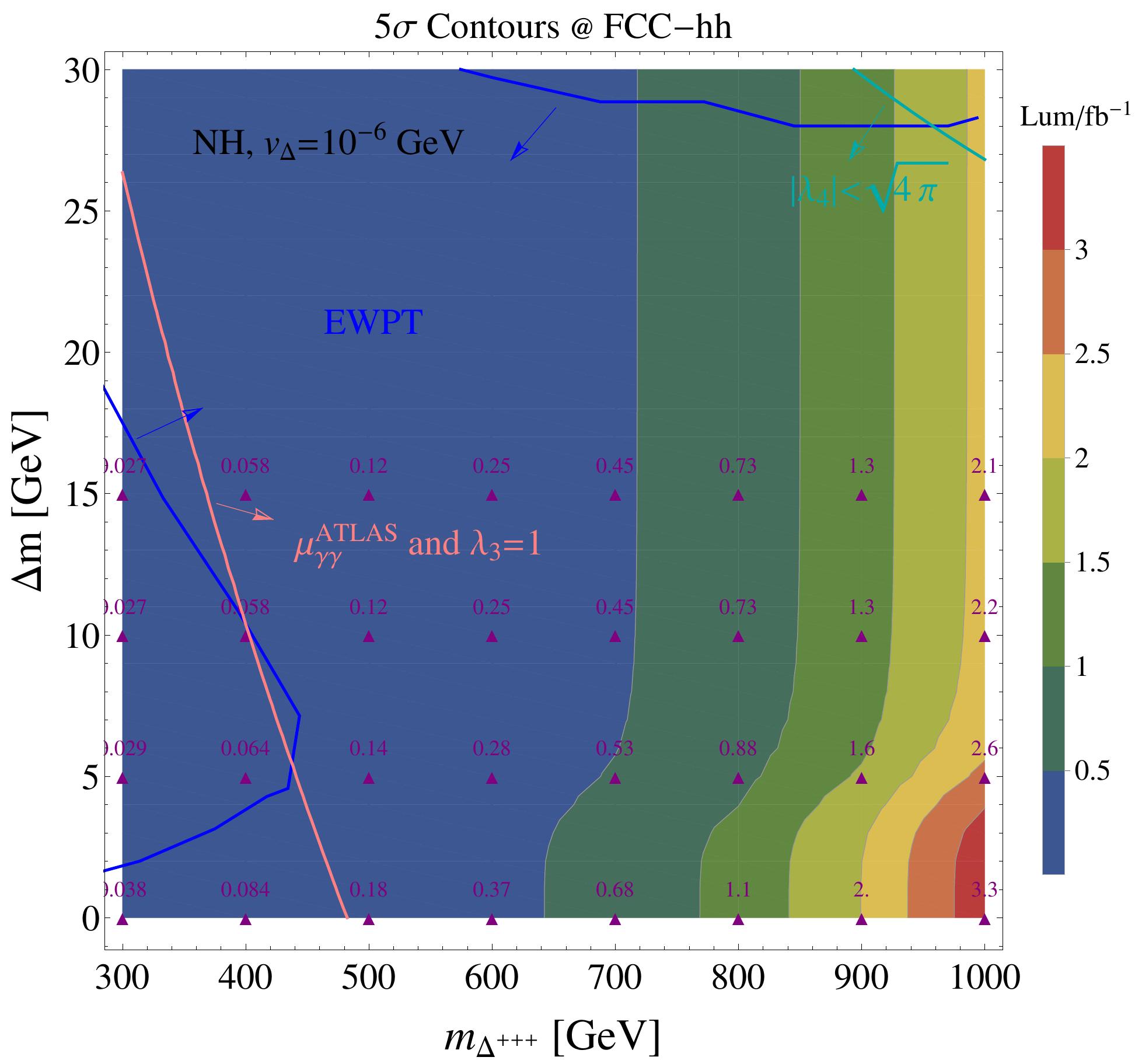}
\includegraphics[width=0.35\textwidth]{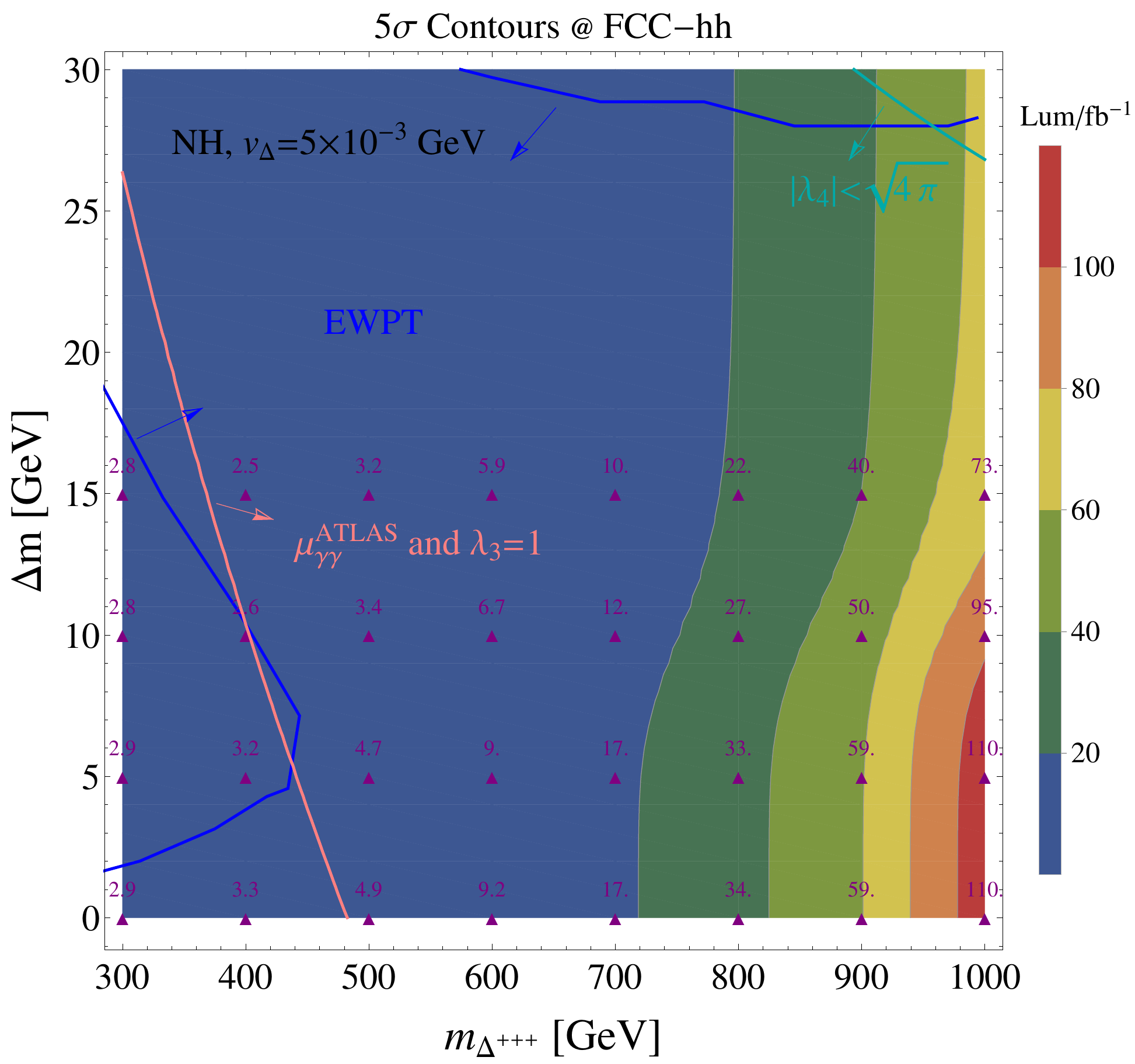}	
\includegraphics[width=0.35\textwidth]{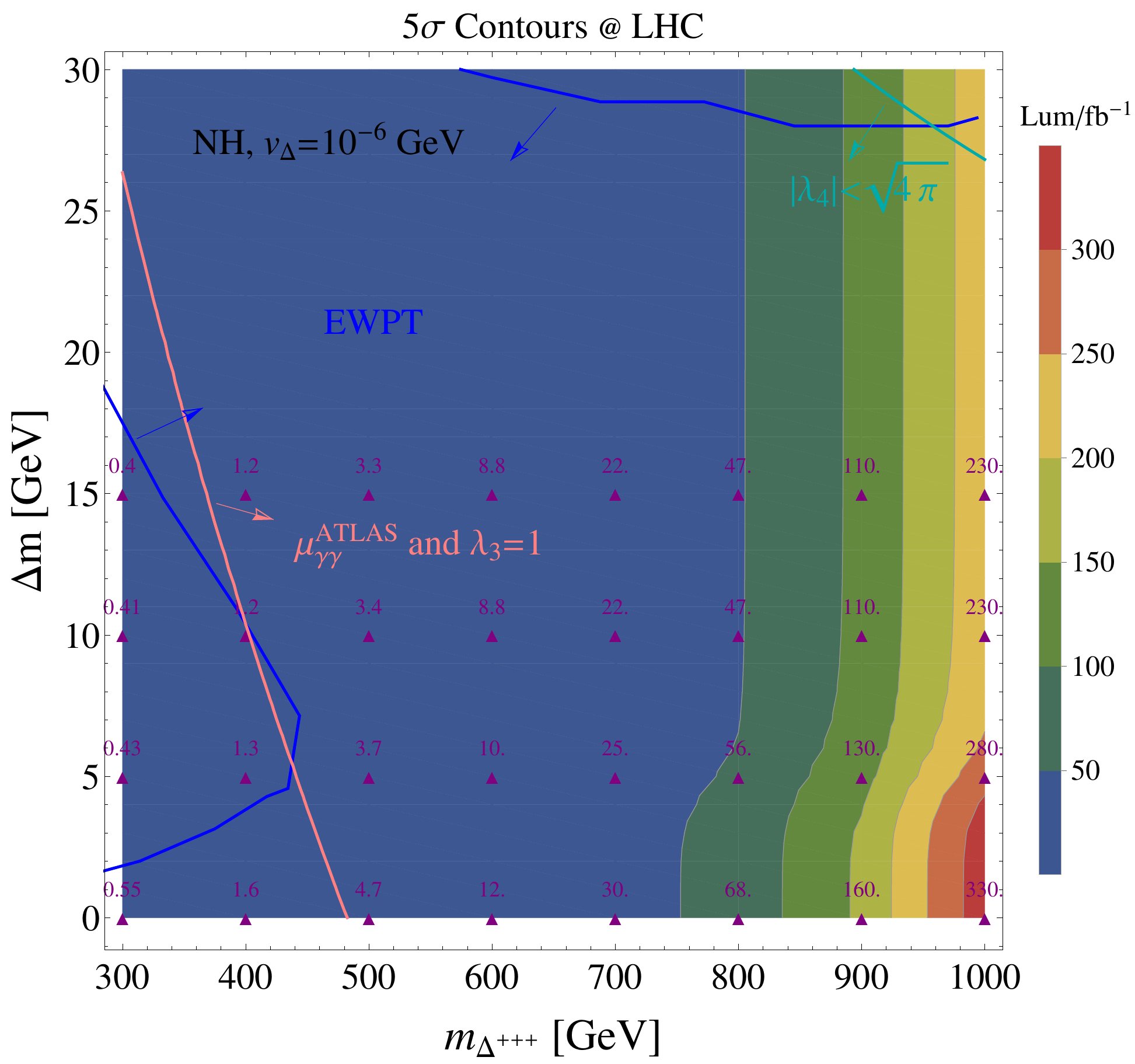}
\includegraphics[width=0.35\textwidth]{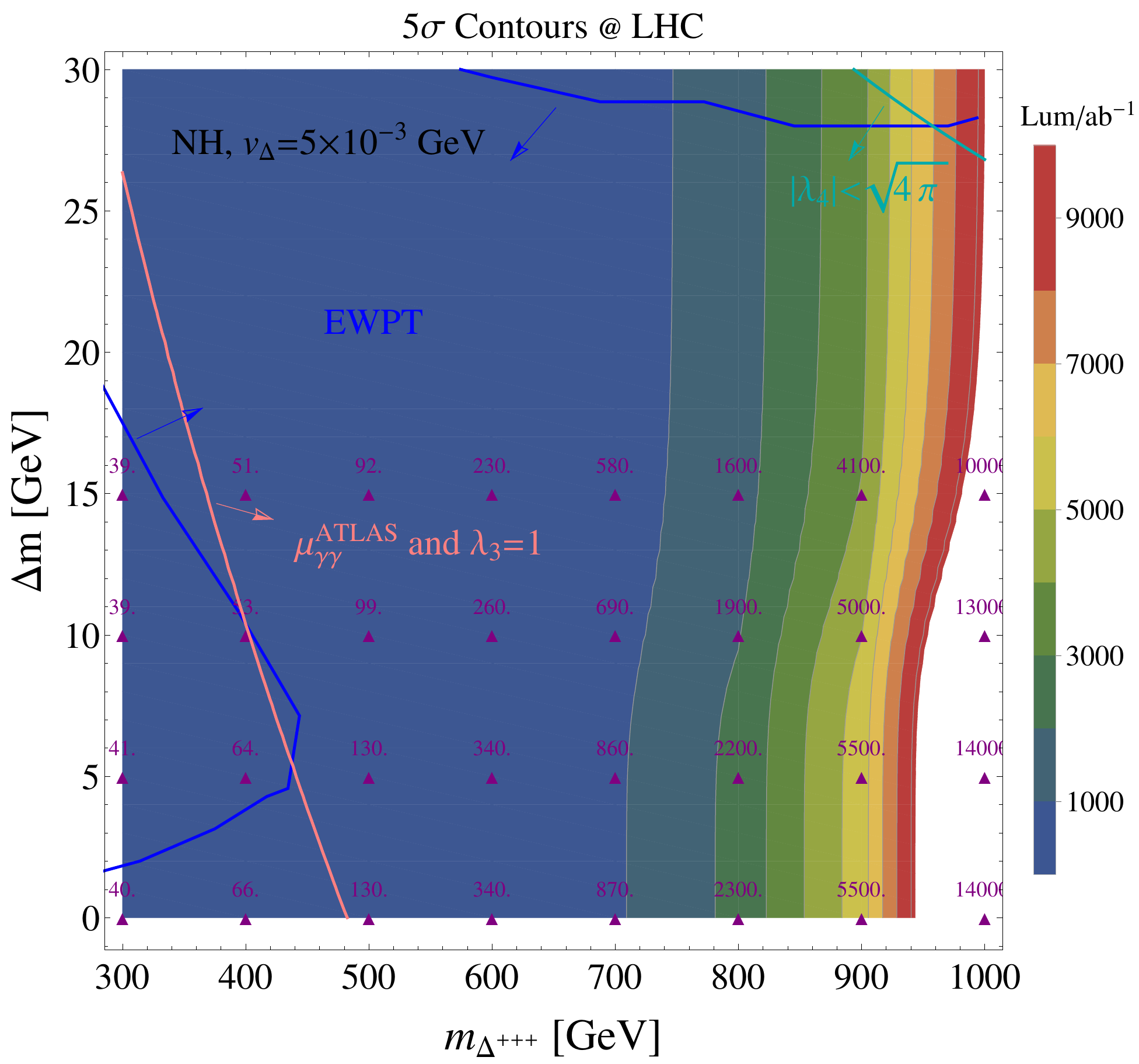}		
\caption{ $5\sigma$ discovery prospects in the plane of $m_{\Delta^{\pm\pm\pm}}$ and $\Delta m$ with the banchmark values of $v_{\Delta}=10^{-6}\gev$ (left panels) and $5\times 10^{-3}\gev$ (right panels) for the NH at the 13~TeV LHC and FCC-hh. The contours correspond to the integrated luminosities (in units of $\fbi$) required to satisfy $\mathcal{Z}=5$. Benchmark points with $\Delta m=1$, $5,10,20\gev$ and $m_{\Delta^{\pm\pm\pm}}=300\gev-1000\gev$ are depicted in purple triangles with the numbers denoting the required integrated luminosities. The allowed regions from indirect constraints are also indicated.}
\label{fig:MD-dm}
\end{figure}

In the last two columns of Tab.~\ref{tbl:xsection}, we show the cross sections of background processes without any cut and after all cuts at the 13~TeV LHC as a recast of cross sections at the FCC-hh assuming the same cut efficiencies, which are denoted as $\sigma_{0}(13\tev)$ and $\sigma_{\text{cut}}(13\tev)$, respectively. We obtain that the dominant backgrounds at the LHC are $WZ$, $t\bar{t}W$ and $t\bar{t}$. The cross section of $WZ$ is $9.14\times 10^{-6}\pb$ after cuts, which is consistent with that in Ref.~\cite{Ghosh:2018drw}. Nevertheless, it is notable that the charge misidentification rate is missing in Ref.~\cite{Ghosh:2018drw}, which should be also order of $10^{-3}$ as we assumed; although the cross section of $WZ$ is slightly smaller than that in Ref.~\cite{Ghosh:2018drw}, this could come from that fact that we impose a larger ${E}_T^{\text{miss}}$ cut. The cross section of background $ZZ$ we obtain is smaller than that in Ref.~\cite{Ghosh:2018drw} since we have further rejected events with lepton invariant mass below 12 GeV as in Cut-2. The cross section of background $t\bar{t}W$ we obtain is larger since we have considered both leptonic and hadronic decays of top quark, while only the leptonic decay was considered in Ref.~\cite{Ghosh:2018drw}. For the hadronic decay of top quark, the third charged lepton comes from the decay of heavy-flavor hadrons. We also obtain that the signal cross section with $(m_{\Delta^{\pm\pm\pm}},\Delta m, v_\Delta)=(400\gev,0,10^{-6}\gev)$ for the NH is $2.2\times 10^{-3}\pb$, which is about 2 times of that in Ref.~\cite{Ghosh:2018drw}. This is because we have multiplied a $K$-factor of 1.25 for the DY cross section. Besides, the PF cross section for $m_{\Delta^{\pm\pm\pm}}=400\gev$ is comparable to that of the DYW or DYZ cross section (see Fig.~\ref{fig:cross_section}),  which has been included in our analysis but not been considered in Ref.~\cite{Ghosh:2018drw}.

Finally, we show the summaries of constraints and discovery prospects in the plane of $m_{\Delta^{\pm\pm\pm}}$ and $\Delta m$ with the banchmark values of $v_{\Delta}=10^{-6}\gev$ and $5\times 10^{-3}\gev$ for the NH in Fig.~\ref{fig:MD-dm}. The $5\sigma$ contours correspond to the integrated luminosities required to satisfy $\mathcal{Z}=5$. At the 13~TeV LHC, the regions of $\Delta m\geq 0$ and $300\gev\leq m_{\Delta^{\pm\pm\pm}}\leq 1000\gev$ can be discovered with the integrated luminosity of $300\fbi$ for $v_{\Delta}=10^{-6}\gev$, while the region of $ m_{\Delta^{\pm\pm\pm}}>800\gev$ for $v_{\Delta}=5\times 10^{-3}\gev$ is unable to be discovered even with the integrated luminosity of $3\abi$. At the FCC-hh, the regions of $\Delta m\geq 0$ and $m_{\Delta^{\pm\pm\pm}}$ for $v_{\Delta}=10^{-6}\gev$ and $v_{\Delta}=5\times 10^{-3}\gev$ can be discovered with the integrated luminosities of $3.3\fbi$ and $110\fbi$, respectively.
It thus clearly indicates that a 100~TeV $pp$ collider, or FCC-hh in our study, is able to extend the kinematic region beyond the LHC significantly in the searches for triply charged Higgs bosons.

\section{Conclusions}
\label{sec:smmary}

In this work, we have studied the potential of searching for triply charged Higgs bosons at the LHC and a 100~TeV $pp$ collider. We first discuss the methodology of producing and detecting a multi-charged Higgs boson at $pp$ colliders. While the singly and doubly charged Higgs bosons have been discussed thoroughly, the triply charged Higgs boson has not been paid much attention. The details of a specifically non-trivial model with a Higgs quadruplet and a pair of vector-like triplet leptons are given. The indirect constraints on this model are subsequently discussed, which indicate that the magnitude of mass splitting $\Delta m$ between the nearby states of the Higgs quadruplet is restricted to be smaller than 30~GeV while the quadruplet VEV larger than $1.5\times 10^{-9}\gev$ is allowed.

We then discuss the production cross section and decay branching ratio of the triply charged Higgs boson. With the increase of collider energy, the production cross section becomes larger significantly. This motivates us to study the sensitivity of searching for a triply charged Higgs bsons at a 100~TeV $pp$ collider. Triply charged Higgs boson can decay into $W^\pm W^\pm W^\pm $ or $\ell^\pm \ell^\pm W$ through an off-shell doubly charged Higgs boson with the decay branching ratios being nearly independent of the mass splitting $\Delta m$. The cascade decays $\Delta^{\pm\pm\pm}\to \Delta^{\pm\pm}W^{\pm*},\Delta^{\pm\pm}\pi^{\pm} $ are open if $\Delta m<0$. The interplay between these decay modes are determined by $v_\Delta$ and $\Delta m$. In case of $\Delta m\geq 0$, however, the proper decay length of the triply charged Higgs boson can be larger than $0.1~\text{mm}$ (long-lived), which makes the conventional prompt search inappropriate.

Thanks to the high charge, three same-sign leptons can be produced in the decays of triply charged Higgs boson. In previous studies with SS3L signature at the LHC, only part of SM backgrounds were considered. We make a complete list of backgrounds, simulate them at a 100~TeV $pp$ collider by taking the FCC-hh as an example and perform a detailed collider analysis with at least three same-sign leptons in the final state being selected, which is inclusive for the signal processes with one or two $\Delta^{\pm\pm\pm}$ and the decays $\Delta^{\pm\pm(\pm)}\to \ell^{\pm}\ell^{\pm\pm}(W^\pm)$ and $W^\pm W^\pm (W^\pm)$. The cascade decays giving rise to the SS3L signature for $\Delta m>0$ are also properly included. Signal events are generated according to their dependence on the mass splitting $\Delta m$ and the quadruplet VEV $v_\Delta$, so that we can obtain the discovery significance as a function of $v_\Delta$, which is allowed in the range $1.5\times 10^{-9}\gev\lesssim v_\Delta \lesssim 1.3\gev$.

For a comparison, we choose two benchmark values of $v_\Delta$, for which prompt searches are valid. We find that previous studies at the LHC underestimated the significance by several times. We revisit the sensitivity at the LHC by projecting that at the FCC-hh since the differential distributions at these two colliders are close except the rapidity distributions. From the comparison, it is clearly shown that at the FCC-hh is powerful for the discovery of triply charged Higgs bosons, which extends the kinematic region of the LHC and improves the sensitivity significantly.
\\
\\

\appendix

{\bf Appendix}

\section{The $Z$, $W^\pm$ masses and the would-be Goldstone modes}
\label{sec:app1}

In this Appendix, we will give more details of the model with a SM Higgs doublet $H$ and a Higgs multiplet $H_n$ in Sec.~\ref{sec:intro} by expanding Eq.~\eqref{eq:kinetic}.

The VEV of $H_n$ will modify the $W$ and $Z$ boson masses compared to the model with just $H$ to have
\begin{eqnarray}
m^2_W = {g^2\over 4}(v_{H}^2 + 2(I_n (I_n+1)-Y^2_n)v^2_n)\;,\;\;\;\;m^2_Z ={g^2\over 4 c^2_W} (v_{H}^2 +4Y^2_nv^2_n)\;.
\end{eqnarray} 
where $I_n$ is the isospin of the $SU(2)_L$ of a $n$-th rank Higgs representation. Therefore, the $\rho$ parameter is expressed as
\begin{eqnarray}
\rho ={v_{H}^2 + 2(I_n (I_n+1)-Y^2_n)v^2_n\over v_{H}^2 +4 Y^2_nv^2_n}\;.
\end{eqnarray}

Experimentally, the $\rho$ parameter is determined to be very close to unity, $\rho = 1.00039\pm 0.00019$~\cite{Tanabashi:2018oca}. If the VEVs satisfy $v_n=v_H$, $\rho=1$ is predicted at tree level for $n=2$ with $I=1/2$, $Y=1/2$ or $n=7$ with $I = 3$, $Y=2$~\cite{Hisano:2013sn,Harris:2017ecz,Chiang:2018irv}. In the usual Higgs representation, the VEV $v_n$ is constrained to be small compared with the doublet VEV $v_{H}$. The new Higgs boson couplings to SM fermions are small proportional to $v_n/v_{H}$. If $n$ is larger than 3, $H_n$ does not couple to SM fermions directly for a Lagrangian that is renormalizable.

 The Goldstone bosons and charged Higgs fields of $H_n$ are
\begin{eqnarray}
\label{goldstones}
&&G_Z ={ v_{H} I^0 + 2 Y_n v_n I^0_n\over \sqrt{v_{H}^2 +4 Y^2_n v^2_n}}\;,\nonumber\\
&&G_{W}^+ = { v_{H} h^+ + v_n\sqrt{2(I_n(I_n +1) - Y^2_n)}\phi^+\over \sqrt{v_{H}^2+v^2_n 2(I_n(I_n+1) -Y^2_n)}}\;,
\end{eqnarray}
where $h^+$ denotes the singly charged field from the doublet representation $H$. 

After removing the Goldstone bosons, one can obtain the physical pseudoscalar $A^0$ and singly charged Higgs bosons $h^+_i$ as given by
\begin{eqnarray}
\label{imag_neutral}
A^0 = {2 Y_n v_n I^0 - v_{H} I^0_n\over \sqrt{v_{H}^2 +4 Y^2_n v^2_n}}\;,
\end{eqnarray}
and
\begin{eqnarray}
\label{sing_charged}
&&h^+_1 = { v_n\sqrt{2(I_n(I_n +1) - Y^2_n)}h^+ - v_{H} \varphi^+\over \sqrt{v_{H}^2+v^2_n 2(I_n(I_n+1) -Y^2_n)}}\;,\nonumber\\
&&h^+_2 = {\sqrt{(I_n + Y_n +1)(I_n-Y_n)}h^+_n +  \sqrt{(I_n - Y_n +1)(I_n + Y_n)}  h^{-\ast}_n \over \sqrt{2(I_n(I_n +1) - Y^2_n)}}\;,
\end{eqnarray}
with $\varphi^+$ given by
\begin{eqnarray}
&&\varphi^+ = {\sqrt{(I_n - Y_n +1)(I_n + Y_n)} h^+_n - \sqrt{(I_n + Y_n +1)(I_n-Y_n)} h^{-\ast}_n \over \sqrt{2(I_n(I_n +1) - Y^2_n)}}\;,
\end{eqnarray}
Note that $h^+_i$ may or may not be mass eigenstates depending on the details of Higgs potential. For simplicity, we will assume  that they are mass eigenstates. 

It is convenient to write the two real neutral components $h^0$ and $h^0_n$ as
\begin{eqnarray}
\label{real_neutral}
&&h^0_1 ={ v_{H} h^0 + 2 Y_n v_n h^0_n\over \sqrt{v_{H}^2 +4 Y^2_n v^2_n}}\;,\;\;\;\;h^0_2 = {2 Y_n v_nh^0 - v_{H} h^0_n\over \sqrt{v_{H}^2 +4 Y^2_n v^2_n}}\;.
\end{eqnarray}

In general, $h^0$, $h^+$, $h_n^0$ and $h_n^{Q=1}$ are not mass eigenstates. From Eqs.~\eqref{goldstones}, \eqref{imag_neutral}, \eqref{sing_charged} and \eqref{real_neutral}, the mass eigenstates can be written as the following basis transformations. For real neutral fields $h_{\alpha}$ ($h_1\equiv h^0$, $h_2\equiv h_n^0$)
\begin{eqnarray}
h_{\alpha}=\sum_{\alpha ,\beta=1}^3(N_{R})_{\alpha \beta}h_{\beta}^{m0}\;,
\end{eqnarray} 
where $N_R$ denotes the $2\times 2$ orthogonal matrix and $h_{\alpha}^{m0}$ are the mass eigenstates ($h_{1(2)}^{m0}\equiv h_{1(2)}^{0}$). For imaginary neutral fields $I_{\alpha}$ ($I_1\equiv I^0$, $I_2\equiv I_n^0$)
\begin{eqnarray}
I_{\alpha}=\sum_{\alpha ,\beta=1}^3(N_{I})_{\alpha \beta}I_{\beta}^{m}\;,
\end{eqnarray}
where $N_I$ denotes the $2\times 2$ orthogonal matrix and $I_{\alpha}^{m}$ are the mass eigenstates. $I_{1}^{m}\equiv G_Z$ is the would-be Goldstone boson and $I_{2}^{m}\equiv A^0$. For singly charged fields $H_{\alpha}^+$ ($H_1^+\equiv h^+$, $H_2^+\equiv h_n^{-\ast}$, $H_3^+\equiv h_n^+$)\footnote{In general, $h_n^{|Q|-\ast}\neq h_n^{|Q|}$. The equality only holds for real representations.} 
\begin{eqnarray}
H_{\alpha}^+=\sum_{\alpha ,\beta=1}^3S_{\alpha \beta}H_{\beta}^{m+}\;,
\end{eqnarray}
where $S$ denotes the $3\times 3$ orthogonal matrix and $H_{\beta}^{m+}$ denotes the mass eigenstates. $H_1^{m+}\equiv G_{W}^+$ is the would-be Goldstone boson and the physical Higgs bosons $H_{2(3)}^{m+}\equiv h_{1(2)}^+$.

\section{Feynman rules in the general Higgs representation}

The production and detection additional Higgs boson depend on their couplings to photon, $W^\pm$ and $Z$ bosons. 
Using Eq.~\eqref{eq:kinetic}, we have the following interaction terms relevant to $A$, $W^\pm$ and $Z$ fields ($g_2\equiv g$ is the $SU(2)_L$ gauge coupling),
\begin{eqnarray}\label{higgs_int}
&&\mathcal{L}^W_{\text{int}} = i{g_2\over \sqrt{2}} \left [\sqrt{(I_n+m)(I_n - m+1)} \partial^\mu (h^Q_n)^*h^{Q-1}_n\right .\nonumber\\
&&\hspace{0.8cm}\left .  -
\sqrt{(I_n - m)(I_n +m+1)} \partial^\mu h^Q_n (h^{Q+1}_n)^*\right ]W^+_\mu+ \text{H.C.}\;,\nonumber\\
&&\mathcal{L}^{A, Z}_{\text{int}}= i\left (\partial^\mu (h^Q_n)^* h^Q_n - \partial^\mu h^Q_n (h_n^Q)^*\right ) (eQ A_\mu
+{g_2\over c_W} (m - Qs^2_W)Z_\mu )\;,\nonumber\\
&&\mathcal{L}^{WW}_{\text{int}}={g_2^2\over 2}\left [ (I_n+m)(I_n-m +1) (h^{Q-1}_n)^*h^{Q-1}_n \right .\nonumber\\
&&\hspace{0.8cm}\left .+ (I_n - m)(I_n+m+1) (h^{Q+1}_n)^* h^{Q+1}_n \right ]W^{+\mu}W^-_\mu\;,\nonumber\\
&&\hspace{0.8cm}+ \sqrt{(I_n^2-m^2)((I_n+1)^2-m^2)}\left[W^{-\mu}W_\mu^- (h^{Q-1}_n)^*h^{Q+1}_n +\text{H.C.}\right]\;,\nonumber\\
&&\mathcal{L}^{AA,ZZ,AZ}_{\text{int}}=(eQA_\mu + {g_2\over c_W}(m - Qs^2_W)Z_\mu)^2 (h^Q_n)^* h^Q_n\;,\nonumber\\
&&\mathcal{L}^{WA, WZ}_{\text{int}}={g_2\over \sqrt{2}}(eQA^\mu +{g_2\over c_W}(m-Qs^2_W)Z^\mu)
 \left [ W^-_\mu (\sqrt{(I_n+m)(I_n-m+1)}(h^{Q-1})^*h^Q_n \right .\nonumber\\
&&\hspace{0.8cm}\left . + \sqrt{(I_n-m)(I_n+m+1)}(h^Q_n)^*h^{Q +1}_n+\text{H.C.})\right]\;.
\end{eqnarray}

Substituting the physical components defined in Appendix~\ref{sec:app1} into Eq.~\eqref{higgs_int}, one can get the Feynman rules of Higgs-Gauge couplings. We list the tables of Feynman rules in the following. Note that  we have removed the would-be Goldstone bosons $G_Z$ and $G_{W}^{\pm}$ after the electroweak symmetry breaking, thus $\alpha=2,3$ for the singly charged Higgs field $H_{\alpha}^{m\pm}$.

\begin{table}[!htb]
\centering
\tabcolsep=8pt
\begin{tabular}{|c|c|}
\hline
\hline
Vertices & Coefficients \\ \hline
$W_{\mu}^{\pm}h^{m0}_{\alpha}H_{\beta}^{m\mp}$ & $i\frac{g_2}{2}((N_R)_{1\alpha}S_{1\beta}-\sqrt{(I_n-Y_n)(I_n+Y_n+1)}(N_R)_{2\alpha}S_{2\beta}$\\
 & $+\sqrt{(I_n+Y_n)(I_n-Y_n+1)}(N_R)_{2\alpha}S_{3\beta})(P_2-P_1)_{\mu}$ \\
\hline
$W_{\mu}^{\pm}A^{0}H_{\beta}^{m\mp}$  & $-\frac{g_2}{2}((N_I)_{12}S_{1\beta}+\sqrt{(I_n-Y_n)(I_n+Y_n+1)}(N_I)_{22}S_{2\beta}$\\
 & $+\sqrt{(I_n+Y_n)(I_n-Y_n+1)}(N_I)_{22}S_{3\beta})(P_2-P_1)_{\mu}$ \\
\hline
$W_{\mu}^{+}H_{\beta}^{m+}h_n^{2\ast}$ & $i\frac{g_2}{\sqrt{2}}\sqrt{(I_n-Y_n+2)(I_n+Y_n-1)}S_{3\beta}(P_2-P_1)_{\mu}$ \\
$W_{\mu}^{+}H_{\beta}^{m+}h_n^{-2}$ & $i\frac{g_2}{\sqrt{2}}\sqrt{(I_n-Y_n+2)(I_n+Y_n-1)}S_{2\beta}(P_2-P_1)_{\mu}$\\
$W_{\mu}^{+}h_n^{Q}h_n^{Q+1\ast}$, ($Q\geq 2$) & $i\frac{g_2}{\sqrt{2}}\sqrt{(I_n-Y_n+(Q+1))(I_n+Y_n-Q)}(P_2-P_1)_{\mu}$ \\
$W_{\mu}^{+}h_n^{-Q\ast}h_n^{-(Q+1)}$, ($Q\geq 2$) & $i\frac{g_2}{\sqrt{2}}\sqrt{(I_n-Y_n-Q)(I_n+Y_n+(Q+1))}(P_2-P_1)_{\mu}$ \\
\hline
\hline
$Z_{\mu}Z^{\mu}h^{m0}_{\alpha}$ & $\frac{g_2^2}{4c_W^2}((N_R)_{1\alpha}v_{H}+4Y_n^2(N_R)_{2\alpha}v_n)g_{\mu\nu}$\\
$Z_{\mu}Z^{\mu}h^{m0}_{\alpha}h^{m0}_{\alpha}$ & $\frac{g_2^2}{8c_W^2}((N_R)_{1\alpha}^2+4Y_n^2(N_R)_{2\alpha}^2)g_{\mu\nu}$\\
$Z_{\mu}Z^{\mu}A^0A^0$ & $\frac{g_2^2}{8c_W^2}((N_I)_{12}^2+4Y_n^2(N_I)_{22}^2)g_{\mu\nu}$\\
$A_{\mu}A^{\mu}H_{\alpha}^{m\pm}H_{\alpha}^{m\mp}$ & $((S_{1\alpha})^2+(S_{2\alpha})^2+(S_{3\alpha})^2)e^2g_{\mu\nu}$ \\
$A_{\mu}Z^{\mu}H_{\alpha}^{m\pm}H_{\alpha}^{m\mp}$ & $\frac{2eg_2}{c_W}((c_W^2-\frac{1}{2})(S_{1\alpha})^2+(c_W^2+Y_n)(S_{2\alpha})^2+(c_W^2-Y_n)(S_{3\alpha})^2)g_{\mu\nu}$ \\
$Z_{\mu}Z^{\mu}H_{\alpha}^{m\pm}H_{\alpha}^{m\mp}$ & $\frac{g_2^2}{c_W^2}((c_W^2-\frac{1}{2})^2(S_{1\alpha})^2+(c_W^2+Y_n)^2(S_{2\alpha})^2+(c_W^2-Y_n)^2(S_{3\alpha})^2)g_{\mu\nu}$ \\
$A_{\mu}A^{\mu}h_n^{(-)Q\ast}h_n^{(-)Q}$, ($Q\geq 2$) & $Q^2e^2g_{\mu\nu}$ \\
$A_{\mu}Z^{\mu}h_n^{(-)Q\ast}h_n^{(-)Q}$, ($Q\geq 2$) & $\frac{2Qeg_2}{c_W}(Qc_W^2-(+)Y_n)g_{\mu\nu}$ \\
$Z_{\mu}Z^{\mu}h_n^{(-)Q\ast}h_n^{(-)Q}$, ($Q\geq 2$) & $\frac{g_2^2}{c_W^2}(Qc_W^2-(+)Y_n)^2g_{\mu\nu}$ \\
\hline
\hline
$Z_{\mu}h_{\alpha}^{m0}h_{\beta}^{m0}$ & $-i\frac{g_2}{4c_W}((N_R)_{1\alpha}(N_R)_{1\beta}+2Y_n(N_R)_{2\alpha}(N_R)_{2\beta})(P_2-P_1)_{\mu}$ \\
$Z_{\mu}A^0A^0$ & $-i\frac{g_2}{4c_W}((N_I)_{12}^2+2Y_n(N_I)_{22}^2)(P_2-P_1)_{\mu}$ \\
$A_{\mu}H_{\alpha}^{m\pm}H_{\beta}^{m\mp}$ & $ie(S_{1\alpha}S_{1\beta}+S_{2\alpha}S_{2\beta}+S_{3\alpha}S_{3\beta})(P_2-P_1)_{\mu}$ \\
\hline
$Z_{\mu}H_{\alpha}^{m\pm}H_{\beta}^{m\mp}$ & $i\frac{g_2}{2c_W}((2c_W^2-1)S_{1\alpha}S_{1\beta}+2(c_W^2+Y_n)S_{2\alpha}S_{2\beta}$ \\
 & $+2(c_W^2-Y_n)S_{3\alpha}S_{3\beta})(P_2-P_1)_{\mu}$ \\
\hline
$A_{\mu}h_n^{(-)Q}h_n^{(-)Q\ast}$, ($Q\geq 2$) & $(-)iQe(P_2-P_1)_{\mu}$ \\
$Z_{\mu}h_n^{(-)Q}h_n^{(-)Q\ast}$, ($Q\geq 2$) & $(-)i\frac{g_2}{c_W}(Qc_W^2-(+)Y_n)(P_2-P_1)_{\mu}$ \\
\hline\end{tabular}
\caption{Feynman Rules. All momenta flow into the vertex.}
\label{Feynman Rules1}
\end{table}
\newpage
\begin{table}[!htb]
\centering
\tabcolsep=8pt
\begin{tabular}{|c|c|}
\hline
\hline
Vertices & Coefficients \\ \hline
\hline
$W_{\mu}^{-}W^{\mu +}h_{\alpha}^{m0}h_{\alpha}^{m0}$ & $\frac{g_2^2}{4}((N_R)_{1\alpha}^2+(I_n(I_n+1)-Y_n^2)(N_R)_{2\alpha}^2)g_{\mu\nu}$ \\
$W_{\mu}^{-}W^{\mu +}h_{\alpha}^{m0}$ & $\frac{g_2^2}{2}((N_R)_{1\alpha}v_{H}+(I_n(I_n+1)-Y_n^2)(N_R)_{2\alpha}v_n)g_{\mu\nu}$ \\
$W_{\mu}^{-}W^{\mu +}A^0A^0$ & $\frac{g_2^2}{4}((N_I)_{12}^2+(I_n(I_n+1)-Y_n^2)(N_I)_{22}^2)g_{\mu\nu}$ \\
$W_{\mu}^{-}W^{\mu +}H_{\alpha}^{m\pm}H_{\alpha}^{m\mp}$ & $\frac{g_2^2}{2}(S_{1\alpha}^2+(I_n(I_n+1)-(1+Y_n)^2)S_{2\alpha}^2+(I_n(I_n+1)-(1-Y_n)^2)S_{3\alpha}^2)g_{\mu\nu}$ \\
$W_{\mu}^{-}W^{\mu +}h_n^{Q,\ast}h_n^{Q}$ ,($\vert Q\vert\geq 2$) & $\frac{g_2^2}{2}(I_n(I_n+1)-(Q-Y_n)^2)g_{\mu\nu}$ \\
$W_{\mu}^{-}W^{\mu-}h_{\alpha}^{m0}h_n^{(-)2(\ast)}$ & $\frac{g_2^2}{2\sqrt{2}}\sqrt{(I_n^2-(1-(+)Y_n)^2)((I_n+1)^2-(1-(+)Y_n)^2)}(N_R)_{2\alpha}g_{\mu\nu}$\\
$W_{\mu}^{-}W^{\mu-}A^0h_n^{(-)2(\ast)}$ & $-(+)i\frac{g_2^2}{2\sqrt{2}}\sqrt{(I_n^2-(1-(+)Y_n)^2)((I_n+1)^2-(1-(+)Y_n)^2)}(N_I)_{22}g_{\mu\nu}$ \\
$W_{\mu}^{-}W^{\mu-}h_n^{(-)2(\ast)}$ & $\frac{g_2^2v_n}{2\sqrt{2}}\sqrt{(I_n^2-(1-(+)Y_n)^2)((I_n+1)^2-(1-(+)Y_n)^2)}g_{\mu\nu}$ \\
$W_{\mu}^{-}W^{\mu-}H_{\alpha}^{m-}h_n^{(-)3(\ast)}$ & $\frac{g_2^2}{2}\sqrt{(I_n^2-(2-(+)Y_n)^2)((I_n+1)^2-(2-(+)Y_n)^2)}S_{(2)3\alpha}g_{\mu\nu}$ \\
\hline 
$W_{\mu}^{-}W^{\mu-}h_n^{Q\ast}h_n^{Q+2}$,  & $\frac{g_2^2}{2}\sqrt{(I_n^2-((Q+1)-Y_n)^2)((I_n+1)^2-((Q+1)-Y_n)^2)}g_{\mu\nu}$\\
($Q\geq 2$ for positive Q & \\
\& $Q\leq -4$ for negative Q) & \\
\hline
\hline
\end{tabular}
\caption{Feynman Rules (continued).}
\label{Feynman Rules2}
\end{table}

\newpage
\begin{table}[!htb]
\centering
\tabcolsep=8pt
\begin{tabular}{|c|c|}
\hline
\hline
Vertices & Coefficients \\ \hline
\hline
$A_{\mu}W^{\mu\mp}h_{\alpha}^{m0}H_{\beta}^{m\pm}$ & $\frac{eg_2}{2}((N_R)_{1\alpha}S_{1\beta}-\sqrt{(I_n+Y_n+1)(I_n-Y_n)}(N_R)_{2\alpha}S_{2\beta}$\\
 & $+\sqrt{(I_n-Y_n+1)(I_n+Y_n)}(N_R)_{2\alpha}S_{3\beta})g_{\mu\nu}$ \\
\hline
$A_{\mu}W^{\mu\mp}A^0H_{\beta}^{m\pm}$ & $-i\frac{eg_2}{2}((N_I)_{12}S_{1\beta}+\sqrt{(I_n+Y_n+1)(I_n-Y_n)}(N_I)_{22}S_{2\beta}$\\
 & $+\sqrt{(I_n-Y_n+1)(I_n+Y_n)}(N_I)_{22}S_{3\beta})g_{\mu\nu}$ \\
\hline
$Z_{\mu}W^{\mu\mp}H_{\beta}^{m\pm}$ & $-\frac{g_2^2}{2c_W}( s_W^2v_{H}S_{1\beta}+(c_W^2+2Y_n)\sqrt{(I_n-Y_n)(I_n+Y_n+1)}v_nS_{2\beta}$\\
 & $-(c_W^2-2Y_n)\sqrt{(I_n+Y_n)(I_n-Y_n+1)}v_nS_{3\beta})g_{\mu\nu}$ \\
\hline
$Z_{\mu}W^{\mu\mp}h_{\alpha}^{m0}H_{\beta}^{m\pm}$ & $-\frac{g_2^2}{2c_W}( s_W^2(N_R)_{1\alpha}S_{1\beta}+(c_W^2+2Y_n)\sqrt{(I_n-Y_n)(I_n+Y_n+1)}(N_R)_{2\alpha}S_{2\beta}$\\
 & $-(c_W^2-2Y_n)\sqrt{(I_n+Y_n)(I_n-Y_n+1)}(N_R)_{2\alpha}S_{3\beta})g_{\mu\nu}$  \\
\hline 
$Z_{\mu}W^{\mu\mp}A^0H_{\beta}^{m\pm}$ & $i\frac{g_2^2}{2c_W}( s_W^2(N_I)_{12}S_{1\beta}-(c_W^2+2Y_n)\sqrt{(I_n-Y_n)(I_n+Y_n+1)}(N_I)_{22}S_{2\beta}$\\
 & $-(c_W^2-2Y_n)\sqrt{(I_n+Y_n)(I_n-Y_n+1)}(N_I)_{22}S_{3\beta})g_{\mu\nu}$  \\
\hline
$A_{\mu}W^{\mu-}h_n^{2}H_{\alpha}^{m-}$ & $3\frac{eg_2}{\sqrt{2}}\sqrt{(I_n+Y_n-1)(I_n-Y_n+2)}S_{3\alpha}g_{\mu\nu}$ \\
$A_{\mu}W^{\mu-}h_n^{-2\ast}H_{\alpha}^{m-}$ & $-3\frac{eg_2}{\sqrt{2}}\sqrt{(I_n-Y_n-1)(I_n+Y_n+2)}S_{2\alpha}g_{\mu\nu}$ \\ 
$Z_{\mu}W^{\mu-}h_n^{2}H_{\alpha}^{m-}$ & $\frac{g_2^2}{\sqrt{2}c_W}(3c_W^2-2Y_n)\sqrt{(I_n+Y_n-1)(I_n-Y_n+2)}S_{3\alpha}g_{\mu\nu}$ \\
$Z_{\mu}W^{\mu-}h_n^{-2\ast}H_{\alpha}^{m-}$ & $-\frac{g_2^2}{\sqrt{2}c_W}(3c_W^2+2Y_n)\sqrt{(I_n-Y_n-1)(I_n+Y_n+2)}S_{2\alpha}g_{\mu\nu}$ \\
\hline 
$A_{\mu}W^{\mu-}h_n^{Q\ast}h_n^{Q+1}$,  & $\frac{eg_2}{\sqrt{2}}(2Q+1)\sqrt{(I_n+Y_n-Q)(I_n-Y_n+(Q+1))}g_{\mu\nu}$ \\
($Q\geq 2$ for positive Q & \\
\& $Q\leq -3$ for negative Q) & \\
\hline  
$Z_{\mu}W^{\mu-}h_n^{Q\ast}h_n^{Q+1}$,  & $\frac{g_2^2}{\sqrt{2}c_W}((2Q+1)c_W^2-2Y_n)\sqrt{(I_n+Y_n-Q)(I_n-Y_n+(Q+1))}g_{\mu\nu}$ \\
($Q\geq 2$ for positive Q & \\
\& $Q\leq -3$ for negative Q) & \\
\hline
\hline
\end{tabular}
\caption{Feynman Rules (continued). It is noted that there is no coupling of physical singly charged Higgs boson to $\gamma W^{\pm}$. After removing the would-be Goldstone boson, the interaction of $A_{\mu}W^{\mu\pm}H_{\beta}^{m\mp}$ becomes vanishing. }
\label{Feynman Rules3}
\end{table}

\begin{acknowledgments}

We thank Yong Du, Michael Ramsey-Musolf and Yi-Lei Tang for helpful discussions and comments. The valuable correspondence with Jung Chang, Tathagata Ghosh and Takaaki Nomura is also thankfully acknowledged. 
\end{acknowledgments}

\bibliographystyle{JHEP}
\bibliography{reference}
%
\end{document}